\documentclass[11pt]{article}
\usepackage{graphicx}
\usepackage[margin=1.25in]{geometry}
\usepackage[usenames,dvipsnames]{color}
\usepackage{url}
\usepackage[colorlinks = true,
            linkcolor = blue,
            urlcolor  = blue,
            citecolor = blue,
            anchorcolor = blue]{hyperref}
\usepackage[font=itshape]{quoting}
\usepackage{graphicx,graphics,subfigure}
\usepackage{amsfonts,amsmath,amssymb,mathrsfs,bm,amsthm,nccmath}
\usepackage{cite}
\usepackage{lineno}
\usepackage{authblk}

\def\eg{{\it e.g.}}

\def\beq{\begin{equation}}
\def\eeq#1{\label{#1}\end{equation}}
\def\eeqn{\end{equation}}

\newenvironment{Eqnarray}%
   {\arraycolsep 0.14em\begin{eqnarray}}{\end{eqnarray}}
\def\beqa{\begin{Eqnarray}}
\def\eeqa#1{\label{#1}\end{Eqnarray}}
\def\eeqan{\end{Eqnarray}}

\let\bar=\overbar

\def\lsim{\mathrel{\raise.3ex\hbox{$<$\kern-.75em\lower1ex\hbox{$\sim$}}}}
\def\gsim{\mathrel{\raise.3ex\hbox{$>$\kern-.75em\lower1ex\hbox{$\sim$}}}}

\def\del{\partial}
\def\Dslash{\not{\hbox{\kern-4pt $D$}}}
\def\dslash{\not{\hbox{\kern-2pt $\del$}}}
\def\pslash{\not{\hbox{\kern-2pt $p$}}}
\def\ETmiss{\not{\hbox{\kern-4pt $E$}}_T}

\def\Dlr{\mathrel{\raise1.5ex\hbox{$\leftrightarrow$\kern-1em\lower1.5ex\hbox{$D$}}}}

\def\MSB{{\bar{M \kern -2pt S}}}
\def\msb{{\bar{\scriptsize M \kern -1pt S}}}

\def\drb{{\bar{\scriptsize D \kern -1pt R}}}

\def\GeV{{\rm GeV}}

\newcommand\snowmass{\begin{center}\rule[-0.2in]{\hsize}{0.01in}\\\rule{\hsize}{0.01in}\\
\vskip 0.1in Submitted to the  Proceedings of the US Community Study\\ 
on the Future of Particle Physics (Snowmass 2021)\\ 
\rule{\hsize}{0.01in}\\\rule[+0.2in]{\hsize}{0.01in} \end{center}}

\usepackage[usenames,dvipsnames]{xcolor}

\title{TF08 Snowmass Report: BSM Model Building\\
\bigskip
\begin{centering}
{\large {\bf Convenors:}  Patrick J.~Fox$^1$\footnote{\href{mailto:pjfox@fnal.gov}{pjfox@fnal.gov}}, Graham D.~Kribs$^2$\footnote{\href{mailto:kribs@uoregon.edu}{kribs@uoregon.edu}}, Hitoshi Murayama$^{3,4,5}$\footnote{\href{mailto:hitoshi@berkeley.edu}{hitoshi@berkeley.edu}}}
\end{centering}}

\medskip

\author[6]{Amin Aboubrahim}
\author[7]{Prateek Agrawal}
\author[8,9]{Wolfgang Altmannshofer}
\author[10]{Howard Baer}
\author[11]{Avik Banerjee}
\author[12]{Vernon Barger}
\author[13]{Brian Batell}
\author[14]{Kim V.~Berghaus}
\author[1]{Asher Berlin}
\author[15]{Nikita Blinov}
\author[16]{Diogo Buarque Franzosi}
\author[17]{Giacomo Cacciapaglia}
\author[18]{Cari Cesarotti}
\author[19]{Nathaniel Craig}
\author[20]{Csaba Cs\'aki}
\author[21]{Raffaele Tito D'Agnolo}
\author[22,23]{Jordy De Vries}
\author[17]{Aldo Deandrea}
\author[24]{Matthew J.~Dolan}
\author[25]{Patrick Draper}
\author[26]{Gilly Elor}
\author[27,28]{JiJi Fan}
\author[29]{Wan-Zhe Feng}
\author[11]{Gabriele Ferretti}
\author[30]{Thomas Flacke}
\author[31]{Benjamin Fuks}
\author[32]{Keisuke Harigaya}
\author[26]{Julia Harz}
\author[33]{Anson Hook}
\author[34]{Seyda Ipek}
\author[35]{Andreas Karch}
\author[36]{Manuel Kunkel}
\author[37]{Benjamin Lillard}
\author[13]{Matthew Low}
\author[33]{Gustavo Marques-Tavares}
\author[38]{Rashmish K.~Mishra}
\author[39]{Pran Nath}
\author[40]{Ethan T.~Neil}
\author[41]{Luca Panizzi}
\author[36]{Werner Porod}
\author[38]{Lisa Randall}
\author[38]{Matthew Reece}
\author[4]{Tom Rudelius}
\author[10]{Shadman Salam}
\author[36]{Leonard Schwarze}
\author[42]{Dibyashree Sengupta}
\author[43]{Bibhushan Shakya}
\author[37]{Jessie Shelton}
\author[33]{Raman Sundrum}
\author[32]{Daniele Teresi}
\author[44]{Christopher B.~Verhaaren}
\author[39]{Zhu-Yao Wang}
\author[45]{Jure Zupan}
\medskip
\affil[1]{Theory Division, Fermi National Accelerator Laboratory, Batavia, IL 60510, USA}
\affil[2]{Institute for Fundamental Science and Department of Physics, University of Oregon, Eugene, OR, 97403 USA}
\affil[3]{Kavli Institute for the Physics and Mathematics of the Universe (WPI), University of Tokyo Institutes for Advanced Study, University of Tokyo, Kashiwa 277-8583, Japan}
\affil[4]{Department of Physics, University of California, Berkeley, CA 94720, USA}
\affil[5]{Ernest Orlando Lawrence Berkeley National Laboratory, University of California, Berkeley, CA 94720, USA}
\affil[6]{Institut f\"ur Theoretische Physik, Westf\"alische Wilhelms-Universit\"at M\"unster, Wilhelm-Klemm-Stra{\ss}e 9, 48149 M\"unster, Germany}
\affil[7]{Rudolf Peierls Centre for Theoretical Physics, University of Oxford, Parks Road, Oxford OX1 3PU, United Kingdom}
\affil[8]{Department of Physics, University of California Santa Cruz, CA 95064, USA}
\affil[9]{Santa Cruz Institute for Particle Physics, 1156 High St., Santa Cruz, CA 95064, USA}
\affil[10]{Homer L. Dodge Department of Physics and Astronomy, University of Oklahoma, Norman, OK 73019, USA}
\affil[11]{Department of Physics, Chalmers University of Technology, Fysikg\aa rden, 41296 G\"oteborg, Sweden}
\affil[12]{Department of Physics, University of Wisconsin, Madison, WI 53706 USA}
\affil[13]{Pittsburgh Particle Physics, Astrophysics, and Cosmology Center, Department of Physics and Astronomy, University of Pittsburgh, Pittsburgh, USA}
\affil[14]{C.N. Yang Institute for Theoretical Physics, Stony Brook University, NY 11794, USA}
\affil[15]{Department of Physics and Astronomy, University of Victoria, Victoria, BC V8P 5C2, Canada}
\affil[16]{Stockholm University, Department of Physics, 106 91 Stockholm, Sweden}
\affil[17]{Univ. Lyon, Universit{\' e} Claude Bernard Lyon 1, CNRS/IN2P3,IP2I UMR5822, F-69622, Villeurbanne, France}
\affil[18]{Center for Theoretical Physics, Massachusetts Institute of Technology, Cambridge, MA 02139, USA}
\affil[19]{Department of Physics, University of California, Santa Barbara, CA 93106, USA}
\affil[20]{Department of Physics, LEPP, Cornell University, Ithaca, NY 14853, USA}
\affil[21]{Universit\'e Paris-Saclay, CEA, Institut de Physique Th\'eorique, 91191, Gif-sur-Yvette, France}
\affil[22]{Institute for Theoretical Physics Amsterdam and Delta Institute for Theoretical Physics, University of Amsterdam, Science Park 904, 1098 XH Amsterdam, The Netherlands}
\affil[23]{Nikhef, Theory Group, Science Park 105, 1098 XG, Amsterdam, The Netherlands}
\affil[24]{ARC Centre of Excellence for Dark Matter Particle Physics, School of Physics, University of Melbourne, Victoria 3010, Australia}
\affil[25]{Department of Physics and the Illinois Center for the Advanced Study of the Universe, University of Illinois, Urbana, IL 61801}
\affil[26]{PRISMA$^+$ Cluster of Excellence $\&$ Mainz Institute for Theoretical Physics Johannes Gutenberg University, 55099 Mainz, Germany}
\affil[27]{Department of Physics, Brown University, Providence, RI, 02912, USA}
\affil[28]{Brown Theoretical Physics Center, Brown University, Providence, RI, 02912, U.S.A.}
\affil[29]{Center for Joint Quantum Studies and Department of Physics, School of Science, Tianjin University, Tianjin 300350, PR. China}
\affil[30]{Center for AI and Natural Sciences, KIAS, Seoul 02455, Korea}
\affil[31]{Laboratoire de Physique Th\'eorique et Hautes Energies (LPTHE), UMR 7589, Sorbonne Universit\'e et CNRS, 4 place Jussieu, 75252 Paris Cedex 05, France}
\affil[32]{CERN, Theoretical Physics Department, 1211 Geneva 23, Switzerland}
\affil[33]{Maryland Center for Fundamental Physics, University of Maryland, College Park, MD 20742, USA}
\affil[34]{Carleton University 1125 Colonel By Drive, Ottawa, Ontario K1S 5B6, Canada}
\affil[35]{Theory Group, Department of Physics, University of Texas, Austin, TX 78712, USA}
\affil[36]{Institut f\"ur Theoretische Physik und Astrophysik,  Uni W\"urzburg, Emil-Hilb-Weg 22, D-97074 W\"urzburg, Germany}
\affil[37]{Department of Physics and Illinois Center for the Advanced Study of the Universe, University of Illinois, Urbana, IL 61801, USA}
\affil[38]{Harvard University, 17 Oxford Street, Cambridge, MA, 02139, USA}
\affil[39]{Department of Physics, Northeastern University, Boston, MA 02115-5000, USA}
\affil[40]{Department of Physics, University of Colorado, Boulder, Colorado 80309, USA}
\affil[41]{Department of Physics and Astronomy, Uppsala University, Box 516, SE-751 20 Uppsala, Sweden}
\affil[42]{Department of Physics, National Taiwan University, Taipei, Taiwan 10617, R.O.C.}
\affil[43]{Deutsches Elektronen-Synchrotron DESY, Notkestrasse 85, 22607 Hamburg, Germany}
\affil[44]{Department of Physics and Astronomy, Brigham Young University, Provo, UT, 84602 USA}
\affil[45]{Department of Physics, University of Cincinnati, Cincinnati, Ohio 45221, USA}\renewcommand\Affilfont{\itshape\footnotesize}

\date{\small{(\today)}}

\begin{document}

\renewcommand\Affilfont{\itshape\footnotesize}
\setcounter{footnote}{0}
\maketitle
\vspace{-0.6in}
\snowmass
\begin{abstract}
  We summarize the state of Beyond the Standard Model (BSM) model building
  in particle physics for Snowmass 2021,
  focusing mainly on several whitepaper contributions to BSM model building (TF08)
  and closely related areas.
\end{abstract}

\section{Executive Summary}

Despite its phenomenal successes, the Standard Model (SM) can only be considered a low-energy, effective field theory (EFT) description of particle physics which leaves many unanswered questions about the nature of reality at distance scales shorter than $\sim\mathrm{TeV}^{-1}$.  Among these so-far unanswered questions are the origin of the neutrino masses, an explanation for the quark and lepton flavor structures, the absence of measurable CP violation in QCD, and why the scales associated with gravity and weak interactions are so disparate.  The SM also does not explain the observed asymmetry between matter and antimatter in the Universe. It is known that the cosmological dark matter cannot be composed of SM fields and the origin of the periods of accelerated expansion of the Universe are a mystery.  Any explanation for these above mentioned puzzles must involve physics beyond the SM.

Using these clues, and others, the goal of beyond the Standard Model (BSM) model building is to uncover the next set of principles which determine how Nature behaves at the shortest distance scales and to build the ``next'' Standard Model
that addresses one or more of these puzzles.  At some level BSM model building can be thought of as turning ``why?'' questions into ``how?'' questions.
While all BSM models follow the principle of \emph{primum non nocere}, there are a plethora of creative, testable ideas that have been
proposed that use a broad range of theoretical approaches.   Some use concepts familiar
from the SM while others introduce completely new ideas and techniques. Many lead to unique
%
%
experimentally testable predictions, across experiments/observations that span a vast range of scales.  Model building is an integral part of many other aspects of HEP, both taking inspiration from, and providing inspiration to, other areas~\cite{TF01, TF02, TF03,TF04,TF05,TF06,TF07,TF09,TF10,TF11}.
Through BSM model building, theorists have also been instrumental in leading the case for new
experimental programs, as well as motivating new search strategies at existing experiments \cite{Essig:2022yzw}.  The success of the theoretical efforts has been strongly advanced by the semi-automated event generation pipeline of FeynRules, the UFO file format, and event generators (for additional details see Sec.~2.8 of \cite{Campbell:2022qmc}).

Understanding the origin and stability of the electroweak scale has long been pursued by BSM model builders.  The so-far null results of searches for new particles from LEP through to LHC have demonstrated that many elegant models are constrained, requiring some to substantial fine-tuning, or are simply not viable. This is not a failure -- on the contrary, this demonstrates the remarkable success of the synergy between theory and experiment where ideas are proposed, models are built, particles are predicted, parameters are constrained, and then ideas/models are ultimately ruled out.  Most BSM theories addressing the hierarchy problem are only midway through this shakeout process with experiment.  Model-building remains an extremely active area to address a range of the puzzles of SM, and continues to push the envelope of new experimental probes to find or exclude the BSM ideas.

Flavor probes are highly sensitive to new physics, and the existing flavor 
constraints are very stringent, with a handful of hints of deviations
beyond the SM\@. 
Many solutions to the hierarchy problem, such as composite Higgs models, models with extra dimensions, and models with low energy supersymmetry, all require a non-trivial flavor structure in order to avoid experimental bounds on BSM flavor-violation.
The origin of such non-trivial new physics flavor structure is an open question, as is the origin of the hierarchies in the spectrum of the SM quarks and leptons, the so-called SM flavor puzzle. An important aspect of flavor model building is to construct mechanisms that address such open issues, implement them in new physics models, and derive phenomenological consequences.  The recent hint for BSM physics associated with flavor, from the muon anomalous magnetic moment and the $B$ anomalies, could point to an exciting future ahead.

Cosmology and astrophysics have deep implications for BSM model building beyond providing evidence of dark matter and dark energy.  The early universe may be considered as a ``laboratory'' equipped with a high-temperature bath.  Its dynamics allows us to probe particle physics models in a way that is impossible at the present time. For example, particles that are too heavy and/or too feebly interacting to be abundantly produced today can be easily produced at much earlier times, potentially leading to observable signatures (e.g., dark radiation in many extensions to the SM) or constraints on new particle physics models.  In addition to understanding the origin of the baryon asymmetry, 
the cosmos is also filled with exotic objects such as neutron stars and supernov\ae~that provide unique constraints on BSM models.
Developments within observational cosmology play an increasingly important role in guiding model-building efforts, \eg\ \cite{Dvorkin:2022jyg,TF09,CF1,CF2,CF3,CF4,CF5,CF6,CF7}.  With the impressive constraints from dark matter direct detection experiments, along with the apparent excess of photons from the Galactic center and the discrepancy between early and late measurements of the Hubble constant, the area of dark matter/dark sector model building has seen substantial growth in the last decade. 

There has also been an ever increasing need to study the structure of models using methods beyond perturbative calculations.  For example, a 
recurring theme in dark sector model building, and many other areas of BSM model building, is extended gauge sectors in which a new gauge group becomes strongly-coupled at scales relevant to the model.
While effective field theory techniques and ``scaled up QCD'' can provide some insights, definitive results require dedicated lattice gauge theory simulations to study the non-perturbative dynamics.  As detailed in \cite{TF05}, there is a strong synergy between effective field theory and lattice gauge theory that continues to inform and strengthen our understanding of gauge theory with applications to model-building beyond the SM.  The list of applications is extensive: determining the properties of a QCD axion requires input from lattice QCD simulations; lattice simulations of the non-perturbative hadronic corrections to $(g-2)_\mu$; lattice simulations of theories of strongly-coupled dark matter to determine the strength of the self-interactions as well as between dark matter and the SM.


Since the last Snowmass the landscape of experimental results has grown substantially \eg\ many of the Higgs boson's properties have been measured and confirmed to be SM like while there is no clear evidence for new particles at the LHC after Run 2, $(g-2)_\mu$ appears to be 
discrepant from the SM prediction, there are numerous flavor anomalies involving $b$-quarks, and the search for direct detection of dark matter has tightly constrained dark matter's properties.  Alongside this evolution, the theory landscape has also substantially altered, which we outline in more details below, based upon contributions to TF08  \cite{Aboubrahim:2021dei,Baer:2022naw,Batell:2022pzc,Elor:2022hpa,Blinov:2022tfy,Windchime:2022whs,Carney:2022gse,Draper:2022pvk,Dutta:2022qvn,Altmannshofer:2022aml,Agrawal:2022yvu,PeterCameron,DagnoloTeresi,Craig:2022uua,Agrawal:2022rqd,Banerjee:2022xmu,Asadi:2022njl}.


\begin{figure}[t]
\begin{center}
\includegraphics[width=0.8\textwidth]{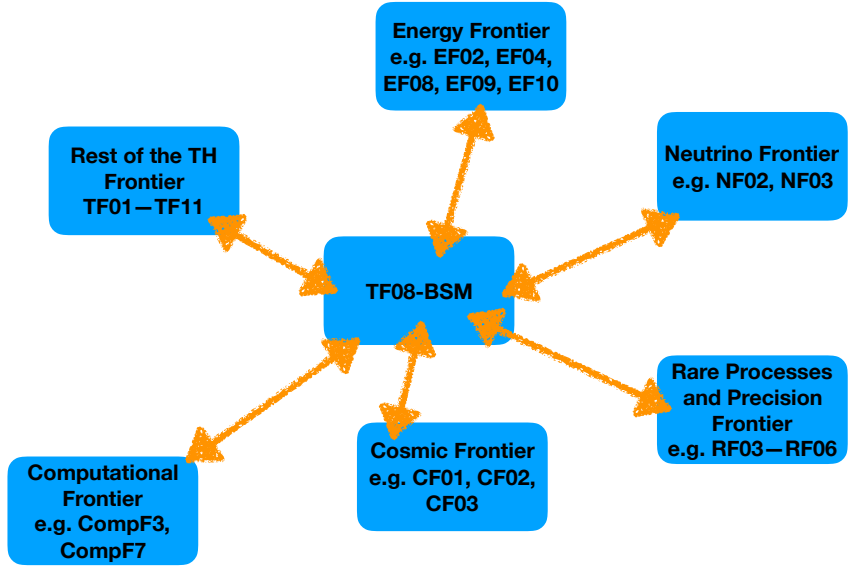}
\caption{Model Building's connections to other facets of the Snowmass process.}
\label{default}
\end{center}
\end{figure}

\section{Naturalness}

Naturalness has been one of the guiding principles in the development of
Beyond the Standard Model (BSM) physics over the last several decades.
Many of the naturalness puzzles of the Standard Model arise in trying
to understand
why a dimensionless number is small rather than $O(1)$, which might
be viewed as the generic expectation. These include the famous electroweak
hierarchy problem ($m_h^2/M_\mathrm{Pl}^2 \ll 1$); the cosmological constant
problem ($\rho_\Lambda/M_\mathrm{Pl}^4 \ll 1$); the strong CP problem
(effective ${\theta} \ll 1$); and the flavor problem ($y_f \ll 1$,
or more broadly, the peculiar structure of fermion masses and mixings).

The scope of ideas to address the hiearchy puzzles
has grown substantially.
This includes such well-known scenarios as low-scale supersymmetry, composite Higgs, and extra dimensions as standard frameworks for BSM physics (for a recent review \cite{Craig:2022uua}).  However, the lack of evidence for new states at the LHC has widened the playing field to new ideas, such as
neutral naturalness, that may be able to address part of the
hierarchy problem while avoiding the substantial constraints
from experiment.
These various proposals are discussed in turn below.
Moreover, models that solve the naturalness problems of the Standard Model often
have one or more dark matter candidates, having the potential to
solve multiple problems within one framework.  In this section,
we focus on the naturalness, while in Sec.~\ref{sec:darkmatter},
we discuss dark matter models and mechanisms.


\subsection{Supersymmetry}

Supersymmetry is perhaps the best known solution to the electroweak hierarchy problem, and proceeds by making the mass of a scalar proportional to that of a fermion, which is itself protected by chiral symmetry. This symmetry must be softly broken in order to be consistent with the non-observation of degenerate superpartners. For excellent reviews, see e.g.~\cite{Martin:1997ns,Baer:2006rs}.
One of the conceptual virtues of supersymmetry, beyond the fact that supersymmetry is the most general space-time symmetry allowed by Nature, is that it provides a very concrete theory in which quadratic divergences are absent. In supersymmetric extensions of the Standard Model, the quadratic divergences indeed vanish and are replaced by the mass splittings between Standard Model particles and their superpartners.  

The status of weak scale supersymmetry, in light of LHC data, is still far from settled.  At this point it is fair to say that supersymmetry has not yet appeared where the simplest implementations in a weak scale model would have led us to expect it.
In its original incarnation, the MSSM now has to exist in moderately fine-tuned \cite{Barbieri:1987fn,Baer:2021tta} regions of its parameter space; searches for Higgsinos are of particular importance as a way of gauging the level of tuning.
That is not to say that supersymmetry could not appear above the TeV scale (e.g., see \cite{Baer:2022naw}), but this leaves a fair bit of daylight between Nature and naturalness expectations.
It is also possible that the superpartners most easily accessible at the LHC are beyond reach, as can occur in variants of supersymmetry such as split-SUSY\cite{Wells:2003tf,Arkani-Hamed:2004ymt}, or that the structure of weak scale supersymmetry is qualitatively different from the MSSM \cite{Fox:2002bu,Kribs:2007ac,Martin:2007gf,Fan:2011yu}.




\subsection{Warped Extra Dimensions and Composite Higgs Theories}

Warped extra dimensions (reviewed in \cite{Agrawal:2022rqd}) have provided 
a major driving force for many model building and  phenomenological activities in high energy physics in the last few decades.
The initial work in the late 90's on Randall-Sundrum (RS) models, based on warped extra-dimensional geometry, was a new Beyond Standard Model (BSM) framework to maintain the EW hierarchy and as an added feature could naturally account for fermion hierarchies, making it one of the few of the electroweak theories to naturally account for both hierarchies with a single underlying mechanism.
The experimental efforts to look for new physics in this framework, both by direct production, or by precision measurements, has been one of the drivers of our experimental programs--often in ways applicable to other BSM scenarios.
In another work with the same broad motivation, ref.~\cite{Csaki:2018kxb} constructed a 5D model where the KK states form a continuum beyond a certain scale--a qualitatively different spectrum of new physics states.


Models of electroweak symmetry breaking by a composite Higgs~\cite{Kaplan:1983fs} with partial compositeness~\cite{Kaplan:1991dc} in the top-quark sector give promising solutions to the hierarchy problem~\cite{Agashe:2004rs,Contino:2010rs,Bellazzini:2014yua,Panico:2015jxa,Cacciapaglia:2020kgq} of the Standard Model, extensively reviewed in \cite{Banerjee:2022xmu}. At the effective field theory level,  they can be described by specifying the pattern of symmetry breaking involved. Generically these models predict the existence of light scalars in addition to the Higgs boson, all emerging as pseudo Nambu-Goldstone bosons. Partial compositeness also implies the presence of vector-like quarks. This proliferation of new particles requires a systematic approach to study this class of models and to facilitate a seamless transition between theory, data, and simulation codes.

There are many approaches to strongly-coupled electroweak symmetry breaking scenarios~\cite{Contino:2010rs,Bellazzini:2014yua,Panico:2015jxa,Cacciapaglia:2020kgq}. The most investigated ones include, e.g., 5D holographic theories~\cite{Contino:2003ve} and multi-site deconstructed models like Little Higgs models \cite{Schmaltz:2005ky}. There are also a class of composite Higgs models based on 4D asymptotically free (hypercolor) gauge theories. The assumption is that the hypercolor theory, after going through a near conformal running \cite{Holdom:1981rm,Cohen:1988sq}, confines at the multi-TeV scale. To include top partial compositeness, its fundamental degrees of freedom contain fermionic matter in two inequivalent irreducible representations of hypercolor~\cite{Barnard:2013zea,Ferretti:2013kya}, chosen in order to sequester the EW coset from the composite states carrying QCD color.
There are two major types of experimental signatures in the low energy range. First, the pNGB nature of the Higgs boson implies modifications of its couplings with the other SM particles \cite{Montull:2013mla,Sanz:2017tco,Liu:2017dsz,Banerjee:2017wmg,Liu:2018qtb}. The other kind of signature involves additional particles beyond the SM, predicted by these models, for example, exotic electroweak pNGBs \cite{Arbey:2015exa,Ferretti:2016upr,Belyaev:2016ftv,Agugliaro:2018vsu}, colored pNGBs \cite{Cacciapaglia:2015eqa,Belyaev:2016ftv}, vector-like quarks \cite{Bizot:2018tds,Cacciapaglia:2019zmj,Xie:2019gya,Matsedonskyi:2014lla,Corcella:2021mdl,Dasgupta:2021fzw,Chala:2017xgc} and other colored fermionic states in non-triplet irreps \cite{Cacciapaglia:2021uqh}, vector resonances \cite{Azatov:2015xqa,BuarqueFranzosi:2016ooy,Yepes:2018dlw,Dasgupta:2019yjm,Banerjee:2021efl}, and axion-like particles \cite{Cacciapaglia:2019bqz,BuarqueFranzosi:2021kky}.

These theories can also address the flavor structure of the SM, naturally explaining neutrino masses and large mixing angles in the lepton sector alongside small mixing angles in the quark sector.  The scale of new physics need not be exceptionally high, for example in RS models this arises due a RS GIM mechanism~\cite{Agashe:2004cp,Cacciapaglia:2007fw,Fitzpatrick:2007sa,Perez:2008ee}. Nevertheless, a multi-TeV (or higher) scale can pose challenges, but also in principle makes the framework  more readily testable than many flavor models--although signals can still be challenging.  More details on the flavor structure of the models is discussed in Sec.~\ref{sec:flavor}.


\subsection{Neutral Naturalness}

Higgs naturalness could arise by positing a new symmetry which relates the SM quarks to colorless partners, which has become known as neutral naturalness
(reviewed in \cite{Batell:2022pzc}).
These are primarily bottom-up constructions that characterize the new partner particles and interactions only up to the few TeV scale and require completion at high energies, solving the so-called 
``little'' hierarchy problem.   Typically, these completions include new colored states with masses of order a few TeV, motivating future high-energy colliders to thoroughly test Higgs naturalness due to new symmetries. 

Realizations of neutral naturalness typically include a hidden, dark, or secluded sector of particles that are related to at least some of the SM fields by a discrete symmetry.  Because at least some of the hidden-sector particles are essential to explaining the mass of the Higgs boson, the Higgs is a robust link between the sectors. Signals of these connections can appear in deviations from the SM Higgs couplings and exotic Higgs decay modes that can be probed at future Higgs factories.
Examining the Higgs for indications of these structures is highly motivated. The Higgs often acquires new, or exotic, decay modes.
The strengths of these couplings and the mass of the twin fermions are both controlled by the ratio of the SM Higgs VEV to the twin Higgs VEV, $v/f$. This quantity is also directly proportional to the tuning of the model, making both Higgs coupling deviations and the invisible Higgs width direct probes of naturalness.

While there are a variety of realizations of the idea of neutral naturalness, the twin Higgs framework has received extensive analysis.  This pertains to higher-energy (UV) completions, collider phenomenology, dark matter candidates, cosmological and astrophysical signals, and intersections with neutrino and flavor physics. That this narrower focus has led to such a rich variety of signals and overlaps with other aspects of beyond the SM physics is impressive, but also points the way to further exploration through other neutral naturalness models.
%
For example, the so-called fraternal model~\cite{Craig:2015pha} makes $\mathbb{Z}_2$ breaking a main feature, only twinning the SM's third generation and allowing for modest deviations from the twin equality in some couplings. In effect, the model keeps only the minimal ingredients required for a twin-Higgs-like protection of the Higgs mass. In a similar vein, if the twin top quark is taken to be vector-like then even the third generation leptons can be removed without generating anomalies~\cite{Craig:2016kue}. Rather than removing entire generations from the mirror twin model, simply allowing significant $\mathbb{Z}_2$ breaking between the fermion Yukawa couplings (other than the top quark) is enough to eliminate some cosmological concerns~\cite{Barbieri:2016zxn}. In weakly-coupled UV completions, a hard breaking of the $\mathbb{Z}_2$ in the quartic terms of the scalar potential can further reduce the need for fine-tuning~\cite{Katz:2016wtw}.




Neutral naturalness models typically include a confining gauge group in the hidden sector. In the case of a mirror twin Higgs model, this includes states at or below the hidden confining scale, but this appears to be more the exception than the rule.  Without light quarks, which can be pair produced to fragment tubes of confining color flux, states much heavier than the confinement scale evolve as though connected by a string of constant tension. These quirky bound states must shed energy until they reach low angular momentum configurations, after which they can decay efficiently.  In addition to decays into SM vectors, the quirks or squirks have a significant branching fraction into hidden gluons. Thus, heavy quirks can produce showers of glueballs~\cite{Burdman:2018ehe,Curtin:2022tou}. Some fraction of these glueballs, the $0^{++}$ states, have displaced decays back into the SM, leading to striking signatures at current and upcoming colliders~\cite{Chacko:2015fbc}.

\subsection{Cosmological Selection}

The past decade has seen the emergence of several entirely new approaches,
in some cases inspired by proposals for the strong CP and cosmological
constant problems. Chief among these is relaxation of the weak scale.
Aspects of these approaches can be found in several whitepapers, including
\cite{Asadi:2022njl,DagnoloTeresi,Craig:2022uua}. 

The original incarnation is the {\it relaxion} \cite{Graham:2015cka},
with a QCD axion-like particle $\phi$ coupled to the Standard Model
with an additional inflationary sector whose properties are
necessarily somewhat special.
The scalar field $\phi$ must transit very large field values so it is a non-compact field.
Below the QCD confinement scale, the coupling between $\phi$ and the
gluon field  strength gives rise to the familiar periodic axion potential with
the height of the cosine potential typically
$
\Lambda^4 \sim f_\pi^2 m_\pi^2 \sim yv f_\pi^3
$.
The barrier height is linearly proportional to the
Higgs VEV since $m_\pi^2$ changes linearly with the quark masses.

The idea is thus: starting at values of $\phi$ such that the
total Higgs mass is large and positive, and assuming the slope
of the $\phi$ potential causes it to evolve in a direction that
lowers the Higgs mass, the $\phi$ potential will initially be
completely dominated by the potential terms, until the point at which
the total Higgs mass-squared goes from positive to negative and the
Higgs acquires a vacuum expectation value. At this point the wiggles
due to the quark masses grow linearly in the Higgs VEV,
and generically $\phi$ will stop when the slope of the QCD-induced
wiggles matches the slope of the potential of $\phi$. This classical
stopping point leads to a small electroweak scale relative to
moderately large cutoff scale (perhaps $10^7$ GeV or more).

There are challenges to protecting the shift symmetry of the relaxion
over the vastly trans-Planckian excursions in field space required to
explain the value of the weak scale, as enumerated by the
Swampland program (see e.g.~\cite{Harlow:2022gzl, Draper:2022pvk}).

An alternative cosmological selection approach is to put many copies of the Standard Model in the
same universe, but explain why one copy acquires the dominant
energy density, called $N$-Naturalness \cite{Arkani-Hamed:2016rle}.
$N$ sectors, e.g., $N$ copies of the Standard Model, are taken where
from copy to copy, the Higgs mass parameters are distributed in
some range from $-\Lambda_H^2$ to $\Lambda_H^2$ according to some
probability distribution. For a wide range of distributions, the generic
expectation is that some sectors have accidentally small Higgs masses,
$m_H^2 \sim \Lambda_H^2 / N$.  In a universe with many sectors,
the universe is populated by whatever sectors are abundant.
It is necessary that the cosmological mechanism preferentially
reheats sectors with smaller scales.  The simplest way to
accomplish this is to imagine an inflationary epoch, followed by
reheating due to the decay of some reheaton.



Other ideas involve ``trigger'' operators~\cite{Arkani-Hamed:2020yna} that play a privileged role in cosmological solutions to the hierarchy problem. Only three such operators are known and each leads to distinctive phenomenological predictions that are completely different compared to traditional solutions to the hierarchy problem. Triggers are therefore our best shot at discovering cosmological selection of the weak scale, either through axion-like signatures ($G \widetilde G$) a new (quite special) Higgs doublet at HL-LHC~\cite{Arkani-Hamed:2020yna} ($H_1 H_2$) or new vector-like leptons at the HL-LHC~\cite{Beauchesne:2017ukw}.

\subsection{Strong CP and Axions}

The strong CP problem -- why the effective $\theta \ll 1$ --
remains one of the vexing naturalness puzzles in the SM\@.
There are numerous approaches to solving the
the strong CP problem.
Perhaps the most famous solution posits a new pseudoscalar field, the QCD axion,
that dynamically adjusts its vev to cancel the strong CP phase~\cite{Peccei:1977ur, Peccei:1977hh, Weinberg:1977ma,Wilczek:1977pj,Kim:1979if,Shifman:1979if,Zhitnitsky:1980tq,Dine:1981rt}.
Other approaches to the strong CP problem are based on the observation that $\theta$ can violate certain discrete symmetries and the renormalization of $\theta$ in the Standard Model alone is minuscule
\cite{ellisgaillard,nelsoncp,barrcp,barrcp2,Beg:1978mt,mohapatrasenjanovic,Georgi:1978xz,Babu:1989rb,barrsenjanovic}. 



Solutions to the strong CP problem involving axions have a wide range of phenomenological and cosmological applications \cite{Agrawal:2022yvu,Blinov:2022tfy}.
One of the most important axion interactions is its coupling to photons. 
Measuring this coupling could give us information about the far UV of the SM.  A thriving experimental program is underway to detect axions, with a lot of synergy between theory and experiment, and a number of new experimental proposals on the horizon \cite{Lawson:2019brd,BRASS,
Liu:2021pei,DMRadio,Michimura:2019qxr,
Baryakhtar:2018doz,Beurthey:2020yuq,
Alesini:2017ifp,McAllister:2017lkb,
Schutte-Engel:2021bqm,Zhang:2021bpa,
Berlin:2020vrk, Meyer:2016wrm,Thorpe-Morgan:2020rwc,
Dekker:2021bos,Dolan:2021rya,Foster:2021ngm,Cadamuro:2011fd,Depta:2020wmr,
Ortiz:2020tgs, Shilon:2013xma,Ge:2020zww}.  
The broad array of employed technologies, and their associated range of sensitivities in mass-coupling plane, has motivated a rejuvenation of axion model building where the axion's properties are varied away from the na\"ive expectations \cite{Baer:2018avn,DiLuzio:2020wdo,Choi:2020rgn}.

Axions can also play a role in large-field inflation. The periodic shift symmetry $a \rightarrow a + 2 \pi f_a$ naturally protects axions from Planck-suppressed operators, which generically spoil large-field inflationary models. Axions acquire a periodic potential from instantons of the form $V(a) = \Lambda^4 \sin(a/f_a)$, which for $f_a > M_{\textrm{Pl}}$ leads to a large-field model of inflation called natural inflation \cite{Freese:1990rb}, which can be distinguished from other models of inflation by the spectral tilt and tensor-to-scalar ratio of the primordial power spectrum.
However, quantum gravity appears to censor such natural inflation models. Within string theory, axion decay constants are constrained in all known examples to satisfy $f_a < M_{\textrm{Pl}}$ \cite{Banks:2003sx}, and $f_a < M_{\textrm{Pl}}$ is a generic prediction of the axion version of the ``Weak Gravity Conjecture'' \cite{ArkaniHamed:2006dz}. Models of natural inflation involving multiple axions \cite{Liddle:1998jc, Dimopoulos:2005ac, Kim:2004rp} are similarly constrained by the Weak Gravity Conjecture \cite{Rudelius:2014wla, Rudelius:2015xta, Montero:2015ofa, Heidenreich:2015wga, Brown:2015iha, Heidenreich:2019bjd}. Although loopholes to these constraints exist \cite{delaFuente:2014aca,Rudelius:2015xta,Brown:2015iha, Bachlechner:2015qja, Hebecker:2015rya}, it is unclear whether or not these loopholes can be successfully threaded in a UV complete theory of quantum gravity \cite{Brown:2015lia}. Indeed, surveys of axion landscapes in string theory have so far found them to be barren of large-field natural inflation \cite{Rudelius:2014wla, Conlon:2016aea, Long:2016jvd}.

For the axion to produce a sufficiently small effective value of $\theta$, the Peccei-Quinn symmetry must be very nearly conserved, with no sources of explicit $U(1)_\text{PQ}$ violation other than QCD itself, otherwise the minimum of the axion potential shifts away from the desired value of $\theta(a) = 0$. This is the axion quality problem: the $10^{-10}$ fine tuning associated with $\theta$ has only been shifted into a different (and more severe) fine-tuning in the corrections to the axion potential.  This requirement may be in tension with general expectations about quantum gravity since gravitational effects are not expected to respect global symmetries and thus provides an example of insights from high scale physics affecting BSM physics, more of this will be discussed in the next section.
Several solutions have been proposed and recently there have been developments both in field- and string-theoretic model building. For instance, in ``accidental'' axion models the PQ-charged fields are also charged under some locally conserved symmetries, so that all gauge-invariant PQ-charged operators are necessarily higher-dimensional, and the PQ breaking $\sim (f_a / M_p)^n$ is suppressed by multiple powers of the Planck mass $M_p$.  The local symmetry may be discrete~\cite{Chun:1992bn,Carpenter:2009zs,Harigaya:2013vja,Chen:2021haa} or continuous~\cite{Cheng:2001ys,Hill:2002kq,Fukuda:2017ylt,Lee:2018yak,Darme:2021cxx,Nakai:2021nyf}. In composite axion models~\cite{Randall:1992ut, DiLuzio:2017tjx, Lillard:2017cwx, Lillard:2018fdt,Lee:2021slp,Lillard:2021dxd} the axion is a baryon-like particle, composed of quarks charged under a strongly coupled non-Abelian gauge group. In these models the confining dynamics often instigate the spontaneous breaking of $U(1)_\text{PQ}$, so that the scale $f_a$ is generated dynamically. 



\subsection{Quantum Gravity Implications: Swampland}


From the absence of global symmetries to the sub-extensive entropy of black holes, there is unambiguous evidence that the rules of effective field theory (EFT) consistency as currently understood are not enough to describe even the low energy regime of theories that admit a gravitational UV completion
(reviewed in \cite{Draper:2022pvk}).
Quantum gravity provides us with powerful reasons for seeking theories that are truly natural, in the sense that they explain small dimensionless numbers in terms of $O(1)$ inputs.

The expectation that there are no global symmetries in quantum  gravity, and that at the UV cutoff scale there are not even {\em approximate} ones, provides a strong impetus, beyond simple aesthetic considerations, for the traditional model-building goal of explaining small numbers of the  theory in terms of reasonable  inputs. Technical naturalness can be a useful guiding principle for IR effective field theories,  but a complete theory must go beyond it. This is not to say that a number cannot be small simply because of an accidental  cancelation, although quantum gravity may also constrain how many such fine-tunings can occur~\cite{Heckman:2019bzm}. It does mean that hierarchies that are not technically unnatural, like the flavor hierarchies, are important puzzles that could shed light on UV physics.

A general consequence of approximate global symmetries is a lowering of the cutoff scale as $\sim M_\text{Pl} / \sqrt{\log x^{-1}}$, with $x$ some parameter such that the limit $x \rightarrow 0$ restores the corresponding symmetry. For example, $x = y_f$ for a fermion chiral symmetry, or $x = \mu^4 / M_\text{Pl}^4$ for the shift-symmetry of an axion, with $\mu^4$ the contribution to the axion potential.  This conclusion seems a reasonable lower bound on the ``cost'' of realizing an approximate global symmetry, but it is not the whole story.

An alternative realization of an approximate global symmetry can be obtained by taking the weak coupling limit of a gauge theory. In this case, the obstruction to recovering an unbroken global symmetry is embodied by the Weak Gravity Conjecture (WGC) \cite{ArkaniHamed:2006dz} (see e.g.~\cite{Harlow:2022gzl} for a recent review). Loosely speaking, the WGC implies the presence of ``super-extremal'' degrees of freedom, whose charge-to-mass ratio is larger than that corresponding to an extremal black hole. Perhaps the most notable consequence of this statement is the existence of an upper bound on the cutoff scale of a $U(1)$ gauge theory given by $\Lambda \lesssim g M_\text{Pl}$, related to the presence of new dynamics linked to the existence of magnetic monopoles that satisfy the super-extremality bound \cite{ArkaniHamed:2006dz}. A more sophisticated version of this statement, but that nevertheless seems to be satisfied in all known examples, suggests that a full \emph{tower} of super-extremal states should be part of the theory \cite{Heidenreich:2015nta,Heidenreich:2016aqi,Montero:2016tif,Andriolo:2018lvp}. In combination with the species bound (the expectation that gravity becomes strong at the scale $\Lambda_\mathrm{QG} \sim M_\mathrm{Pl}/\sqrt{N}$~\cite{Veneziano:2001ah, Dvali:2007hz} in a theory with $N$ degrees of freedom), this translates into $\Lambda_\text{QG} \sim g^{1/3} M_\text{Pl}$~\cite{Heidenreich:2016aqi,Heidenreich:2017sim}. (Remarkably, the same $g^{1/3}$ scaling on the cutoff appears from an entirely different argument involving massive Stueckelberg gauge fields~\cite{Craig:2019zkf}.) In other words, the cost of approaching a global symmetry by taking the $g \rightarrow 0$ limit of a gauge theory appears to be power-law rather than logarithmic. This is a much more severe cost, and a stronger constraint on model-building.

\section{Dark Matter}
\label{sec:darkmatter}

The existence of DM is one of the most compelling pieces of evidence for physics beyond the SM. The most detailed understanding of DM arises from its gravitational imprints on cosmological observables, most notably from its effects on the acoustic peaks of the CMB angular anisotropies. Such observations have informed us of the abundance and general characteristics of DM when the universe was only $O(10^5)$ years old. While its general particle physics description currently remains unknown, any DM candidate must be consistent with a plethora of terrestrial, astrophysical, and cosmological probes. At a minimum, any theory of DM must posit something that is non-relativistic, essentially collisionless, feebly-coupled to normal matter, and stable. Even with these restrictions, the theory space is incredibly vast. Over the past decade the broadening of theory priors has enlarged the exploration of this theoretical landscape. This is due in part to: 1) the null results at large underground direct detection experiments and high-energy colliders and 2) the realization that the strong empirical evidence for DM motivates an examination of viable models beyond just those tied to other top-down motivations of new physics.

Continued development of experimental technologies has also driven theoretical investigations into the nature of DM. For instance, low-threshold detectors sensitive to the sub-eV energy deposition of sub-MeV DM have fueled the exploration of cosmologically viable and detectable light DM candidates, while developments in precision sensors for ultralight bosonic DM across a wide range of mass scales have led to the identification of new cosmological targets within the larger parameter space \cite{CF1,CF2,CF3}.

\subsection{Interaction Mechanisms}
\label{sec:mechanisms}

In classic \emph{freeze-out}, the DM abundance is depleted through a two-to-two scattering process until its rate drops below the Hubble expansion rate, at which time its co-moving number density is (again) conserved. To get an abundance in agreement with observations requires a thermally-averaged annihilation cross-section, $\langle \sigma v \rangle \sim 10^{-26} \; {\rm cm}^3/s$.
For $O(1)$ coupling sizes, this famously points to a DM mass around
$0.1$-$1$~TeV\@. As described, the classic freeze-out scenario is highly predictive, though it can be significantly altered through various modifications to the framework above.  There are a host of ``exceptions'' to thermal
freezeout that began with the classic work from
Griest and Seckel \cite{Griest:1990kh}.

DM may not annihilate substantially with the visible sector; instead, it might annihilate or be converted into a secondary state ($ \chi ' $) which itself has interactions with the SM. The evolution of DM depends sensitively on the mass of the secondary state. If $m_\chi\gg m_{\chi'} $, dark matter is in the {\em secluded} regime~\cite{Pospelov:2007mp}, where the annihilations proceed as in classic freeze-out requiring $ \left\langle \sigma v \right\rangle \sim  10 ^{ - 26 }~ {\rm cm} ^3 /{\rm sec} $ while the direct detection signal can be negligible.

If $ m _{ \chi } \simeq m _{ \chi ' } $, thermal freeze-out of both $ \chi $ and $ \chi ' $ must be studied in parallel and the dynamics depend on whether $ \chi ' $ annihilates or decays to deplete its abundance. If $\chi'$ annihilates, and the DM annihilation cross-section is similar to classic freeze-out (though to dark states), the process is termed {\em coannihilation}~\cite{Griest:1990kh}). Coannihilating DM can have direct detection and indirect detection rates comparable to that of classic freeze-out suppressed by a loop factor.

\emph{Forbidden DM}~\cite{Griest:1990kh,D'Agnolo:2015koa} is the same $ 2 \rightarrow 2 $ process of classic freeze-out, but with the mass of the DM slightly below that of the outgoing SM state. This process, forbidden at zero temperature, has an exponentially suppressed rate proportional to $ e ^{ - \Delta E  / T } $, where $ \Delta E $ is the energy threshold for the process.

DM densities can be substantially influenced by $ 3 \rightarrow 2 $ or $ 4 \rightarrow 2 $ interactions, also known as a phase of {\em cannibalism}~\cite{Carlson:1992fn}. The dynamics during a cannibal phase depend on whether DM is chemically coupled to the SM, kinetically coupled to the SM, or neither. If DM is both chemically and kinetically coupled, its bulk properties are unchanged over time. If it is only kinetically coupled, its number density drops, but its energy is transferred to the SM as is the case for SIMPs~\cite{Hochberg:2014dra}, co-SIMPs~\cite{Smirnov:2020zwf}, and ELDERs~\cite{Kuflik:2015isi}.

The dark sector may be populated by the leakage of energy from the visible sector through sub-Hubble rate annihilations or decays of SM particles over time \cite{Hall:2009bx,Bernal:2017kxu,Chu:2013jja}. This production mechanism is known as \textit{freeze-in} because in many respects it is the opposite of freeze-out; the DM in non-thermal.  Many freeze-out processes can be time-reversed to provide a channel for freeze-in, and the relationship between couplings and DM abundances have opposite behaviors for freeze-out and freeze-in. Due to the smallness of coupling strengths relevant for freeze-in, the DM candidate in such models is sometimes called a \textit{Feebly Interacting Massive Particle} (FIMP). The freeze-in parameter space is an extreme limiting case because it represents the smallest relevant coupling the DM can have with the SM plasma that affects early-universe observables in any meaningful way. Historically, some classic examples of DM candidates with freeze-in production mechanisms include sterile neutrinos~\cite{Dodelson:1993je,Kusenko:2006rh,Petraki:2007gq,Shakya:2015xnx}, singlet scalars~\cite{McDonald:2001vt}, and various particles in supersymmetric models including sneutrinos~\cite{Asaka:2005cn,Asaka:2006fs,Gopalakrishna:2006kr,Page:2007sh}, axinos and neutralinos~\cite{Covi:2002vw,Cheung:2011mg,Bae:2014rfa, Co:2015pka}, gravitinos~\cite{Moroi:1993mb, Choi:2005vq, Cheung:2011nn,Hall:2012zp,Co:2016fln,Benakli:2017whb}, goldstinos~\cite{Monteux:2015qqa}, and photinos~\cite{Kolda:2014ppa}.

\emph{Asymmetric dark matter} posits
a common origin for the baryon content of the Universe and DM (see Ref.~\cite{Hut:1979xw} for an early suggestion in this direction).  Since the former is set by an asymmetry, based on various direct and implicit empirical arguments, the DM abundance would also be taken to correspond to an asymmetry of DM over its antiparticle.
A generic requirement of such models is the presence of sufficiently strong interactions that can lead to efficient elimination of the symmetric populations in both the visible and the dark sectors.  For baryons, this is achieved by the presence of strong interactions in the low energy regime of QCD, while the requirement that the same occurs for asymmetric dark matter places restrictions on the new physics.

There is a multitude of models of asymmetric dark matter (for reviews see
Refs.~\cite{Davoudiasl:2012uw,Petraki:2013wwa,Zurek:2013wia}).  There is no clear way to assign various asymmetric dark matter models to a few distinct classes.  Nevertheless, there are some general features that broadly define different scenarios: (A)  where a quantum number has an asymmetry that gets shared between the dark and visible sectors and (B) those that postulate a generalized ``baryon number'' that is preserved, with equal and opposite baryon and dark asymmetries.

\emph{Sommerfeld enhancement} \cite{Hisano:2006nn,Arkani-Hamed:2008hhe}
appears whenever the dark sector system is such that effectively long-range forces arise in the non-relativistic limit. Here, at non-relativistic velocities, a potential between the incoming DM particles leads to the formation of off-shell bound states, and significantly affects the interaction amplitude. This can lead to a strongly enhanced late-time annihilation signal and needs to be taken into account for the freeze-out calculation, as well as present-day indirect detection observables.  

On-shell bound-state formation is a related process that has been investigated more recently~\cite{vonHarling:2014kha,An:2016gad,Asadi:2016ybp}.  Its effect on the DM abundance~\cite{Mitridate:2017izz,Binder:2020efn} and late-time signatures~\cite{Smirnov:2019ngs} become increasingly important with growing dark-sector coupling. A particularly interesting case is DM with EW charges. In particular for $SU(2)$ representations larger than 3, the effective coupling strength is large enough to support effective bound state formation~\cite{Smirnov:2019ngs,Bottaro:2021snn}.

\emph{Inelastic dark matter} (e.g., \cite{Hall:1997ah,Tucker-Smith:2001myb,Tucker-Smith:2004mxa,Chang:2008gd,March-Russell:2008rkh,Cui:2009xq,Alves:2009nf,Chang:2010en,Barello:2014uda,Bramante:2016rdh,Berlin:2018jbm}) posits that 
the predominant interaction of dark matter with Standard Model particles is mediated by an interaction $X_1 X_2 \mathcal{O}$, where $\mathcal{O}$ is an operator built from standard model field(s).  There are numerous models
of ineleastic dark matter, where the inelastic interaction 
could be dimension-4, e.g., interacting with the $Z$ or Higgs boson, or a higher dimensional interaction.
The initial kinetic energy of the $X_1$-nuclear system must be greater than the mass difference ($\delta \equiv m_{\rm X_2} - m_{\rm X_1}$) between $X_1$ and $X_2$ in order for scattering to take place. The fact that \emph{only} DM with sufficiently large kinetic energy can scatter has important consequences.
First, because inelastic dark matter must impinge with sufficiently large kinetic energy to scatter with a direct detection target nucleus, the available kinematic phase space for DM-nuclear scattering is reduced, and the effective DM-nuclear scattering rate is suppressed. The amount of suppression will depend on what fraction of dark matter in the Galactic halo has enough kinetic energy to overcome the inelastic scattering energy threshold.
Next, the minimum required energy for inelastic DM-nuclear collisions implies a minimum recoil energy in the detector, $E_{\rm R}^{\rm min}$. Traditional dark matter searches have optimized sensitivity to \emph{elastic} DM-nuclei collisions by focusing on the limit $E_{\rm R}^{\rm min} \rightarrow 0$, and pushing the observed recoil energy window as low as backgrounds and detector sensitivities allow. For inelastic dark matter with a sizable mass splitting, a low maximum recoil energy reduces detector sensitivity. At best, a low maximum recoil energy will be sub-optimal for detecting inelastic dark matter. At worst, if the dark matter's \emph{minimum} inelastic recoil energy lies above the window of recoil energies, considered in a direct detection analysis, the experiment is {\em insensitive} to inelastic dark matter.

\subsection{Models}

\emph{Supersymmetric dark matter} appears in 
R-parity conserving weak scale supersymmetric theories, where the lightest superpartner is expected to be absolutely stable. The minimal content of a supersymmetric theory includes superpartners of all the SM particles, the Minimal Supersymmetric Standard Model~(MSSM)~(see for eg. Refs.~\cite{Martin:1997ns, Chung:2003fi}). Further, unlike the SM, supersymmetry requires at least two Higgs doublets, one coupling to all the up-type fermions and the other to down-type fermions. Therefore, at a minimum, supersymmetry would include fermionic superpartners of these Higgs bosons and the SM gauge bosons. In the presence of non-minimal field content, there may be additional neutral fermionic states. A popular example is the Next-to-Minimal Supersymmetric SM~(NMSSM)~(see for e.g., Ref.~\cite{Ellwanger:2009dp}) which includes an additional singlet superfield coupling only to the Higgs superfields. Any one of the weakly charged, electrically neutral set of these {\it neutralinos} can be excellent thermal WIMP DM candidates. A key difference between a SUSY neutralino dark matter candidate and a ``random stable fermion'' is the expected presence of a set of correlated states in SUSY.

Since the couplings of the winos and Higgsinos are large and they are further accompanied by almost mass-degenerate charged states~(the charginos), consistent relic density is obtained for DM masses of the order of the TeV scale~\cite{Roszkowski:2017nbc, Delgado:2020url, Kowalska:2018toh}. However, weak scale bino or singlino DM candidates may realize an observationally consistent relic density without necessitating TeV scale winos and Higgsinos. An observationally consistent relic density for either  binos or  singlinos may be obtained by a wide variety of processes~\cite{Han:2013gba, Cabrera:2016wwr, Baum:2017enm}.

\emph{Dark matter in composite Higgs theories} 
can be easily obtained in holographic composite Higgs scenarios by introducing a suitable discrete symmetry that makes some of the massive KK modes stable.\footnote{From the 4-dimensional effective description perspective, this corresponds to identify the DM candidate with a heavy massive resonance coming from the composite dynamics.}
Two types of symmetries are often considered in the literature: discrete exchange symmetries that relate different copies of the bulk fields~\cite{Panico:2006em}, and geometrical parity symmetries\footnote{This type of symmetries are analogous to KK parity in universal extra dimension models (see for instance~\cite{Servant:2002aq}).} connected to the $S_1/Z_2$ orbifold structure representing the extra spatial dimension (see for instance~\cite{Haba:2009xu}).

Since all the fields propagating in the bulk of the extra dimension give rise to KK modes, there are different options for the DM candidate. A natural option is to identify the DM with a $\mathbb{Z}_2$-odd vector KK mode associated to a 5-dimensional gauge field. Gauge singlet candidates are easily obtained if the 5-dimensional gauge symmetry of the model (corresponding to the global symmetry of the composite sector) contains ${\rm U}(1)$ subgroups. This happens, for instance in the minimal models with custodial symmetry, based on the coset ${\rm SO}(5)\times{\rm U}(1)_\textsc{x}/{\rm SO}(4)\times{\rm U}(1)_\textsc{x}$~\cite{Agashe:2004rs}. Models featuring vector DM states have been constructed both on warped space~\cite{Panico:2008bx,Maru:2018ocf} and on flat space~\cite{Panico:2006em,Regis:2006hc}.

\emph{Strong-coupled composite dark matter:}
It is also possible that the DM is a stable bound state of a confining dark sector. Depending on the details of the model, different symmetries can guarantee the stability of various DM candidates \cite{Kribs:2009fy,Appelquist:2015yfa,Antipin:2015xia,Kribs:2016cew,Dondi:2019olm}. 
Given their rich dynamics, the confining dark sectors can give rise to many interesting observable phenomena. Below we review different DM candidates in such sectors and some intriguing dynamics that can happen therein, see Refs.~\cite{Kribs:2016cew,Garani:2021zrr,Cline:2021itd} for recent reviews of such models.

Dark pions can become stable kinematically or through various symmetries, giving rise to dark meson DM candidates, see for instance Refs.~\cite{Alves:2009nf,Hambye:2009fg,SpierMoreiraAlves:2010err,Antipin:2014qva,Hochberg:2014kqa,Antipin:2015xia,Carmona:2015haa,Lonsdale:2017mzg,DeLuca:2018mzn,Kribs:2018oad,Tsai:2020vpi,Garani:2021zrr,Cheng:2021kjg}.  These dark mesons are composed of a dark quark and a dark anti-quark and are always bosons, regardless of the spin of dark quarks.

A more natural possibility, in direct analog with the SM, is that $N$ dark quarks are charged under a dark confining $SU(N)$ form a color-neutral and stable baryon that can account for the observed DM abundance today, see for instance \cite{Kribs:2009fy,Buckley:2012ky,LatticeStrongDynamicsLSD:2013elk,Antipin:2014qva,Appelquist:2015yfa,Antipin:2015xia,Cline:2016nab,Lonsdale:2017mzg,Mitridate:2017oky,Morrison:2020yeg,Garani:2021zrr}. Depending on $N$, such baryons can be either fermions or bosons. Stability of such candidates can be guaranteed in many different ways, including by conservation of dark baryon number.

Finally, yet another potential DM candidate from confining dark sectors are glueballs, see for instance Refs.~\cite{Faraggi:2000pv,Juknevich:2009ji,Juknevich:2009gg,Soni:2016gzf,Forestell:2016qhc}. The spectrum of glueballs in pure confining gauge groups has been studied extensively in the literature, see for instance Refs.~\cite{Morningstar:1999rf,Mathieu:2008me}.
It is shown that many glueball states with different parity and charge-conjugation properties can exist in the spectrum of any confining gauge theory. 
Interactions of various glueball states within a pure Yang-Mills theory have been studied in Ref.~\cite{Forestell:2016qhc}.

\emph{Atomic/Mirror dark matter} posits that 
DM could exist in atom-like bound states. Such dark atoms have a long history (see Refs.~\cite{Blinnikov:1983gh,Goldberg:1986nk}) and occur naturally in mirror twin Higgs models (see e.g.~Refs~\cite{Chacko:2005pe,Chacko:2005vw,Chacko:2005un,Barbieri:2005ri,Craig:2013fga,Craig:2015pha,GarciaGarcia:2015fol,Craig:2015xla,Farina:2015uea,Farina:2016ndq,Prilepina:2016rlq,Barbieri:2016zxn,Craig:2016lyx,Berger:2016vxi,Chacko:2016hvu,Csaki:2017spo,Chacko:2018vss,Elor:2018xku,Hochberg:2018vdo,Francis:2018xjd,Harigaya:2019shz,Ibe:2019ena,Dunsky:2019upk,Csaki:2019qgb,Koren:2019iuv,Terning:2019hgj,Johns:2020rtp,Roux:2020wkp,Ritter:2021hgu,Curtin:2021alk,Curtin:2021spx}). In its simplest implementation, atomic DM \cite{Foot:2002iy,Foot:2003jt,Foot:2004pa,Foot:2004wz,Khlopov:2008ty,Kaplan:2009de,Kaplan:2011yj,Behbahani:2010xa,Cline:2012is,Cyr-Racine:2012tfp,Cline:2013pca,Pearce:2015zca,Choquette:2015mca,Petraki:2014uza,Cirelli:2016rnw,Petraki:2016cnz,Ciarcelluti:2004ik,Ciarcelluti:2004ip,Ciarcelluti:2008vs,Ciarcelluti:2010zz,Ciarcelluti:2012zz,Ciarcelluti:2014scd,Cudell:2014wca} is made of two fermions of different masses oppositely charged under a dark $U(1)_{\rm D}$ gauge symmetry \cite{HOLDOM198665,HOLDOM1986196}. More complex scenarios in which one or both constituents of the dark atoms are themselves composite particles (such as dark nucleons) have also been considered. Similar to the visible sector, the presence of atomic DM generally requires a matter-antimatter asymmetry in the dark sector to set its relic abundance, although it is also possible for a symmetric component to survive \cite{Agrawal:2017rvu}, resulting in a mixture of darkly-charged DM \cite{Ackerman:2008gi,Feng:2009mn,Agrawal:2016quu} and dark atoms. 

\emph{Light dark matter} is an entire class of models in which dark matter
has sub-GeV mass, and is invisible to ton-scale nuclear recoil detectors
(reviewed in \cite{Kahn:2021ttr,Mitridate:2022tnv}). 
As $m_\chi$ decreases, $n_\chi$ increases, so experiments with relatively small targets (gram-scale rather than ton-scale) can have comparable discovery prospects for the same thermally-motivated cross sections, if the energy threshold can be reduced.
Furthermore, from the point of view of maximizing the DM signal, it is optimal to have systems with available excitations that match the low energies and momentum transfers associated with DM masses in the keV--GeV range. Since the DM mass is much lower than a nucleus mass in this regime, nuclear recoils are a poor kinematic match, but the wide range of available excitations in condensed matter systems offers a promising way forward.

There are several mechanisms that light dark matter can employ to get the correct relic abundance, see Sec.~\ref{sec:mechanisms}.
The thermal histories for sub-GeV dark matter require, at a minimum, one additional ingredient: a new force which mediates the thermal contact between the dark matter and the SM. Indeed, dark matter cannot interact with the SM through the strong force (otherwise it would not be ``dark'' with respect to baryons), and neither can it be the weak force, which has too small of an annihilation cross section to generate the correct relic abundance of sub-GeV dark matter \cite{PhysRevLett.39.165}. In principle, it could be the photon if dark matter had a small enough electric charge to be cosmologically ``dark'', but the CMB excludes this possibility for freeze-out because such a small charge would not lead to sufficient annihilation and would yield an overabundance of DM unless other annihilation channels are introduced~\cite{McDermott:2010pa}.
Hence, a benchmark model of such a new force is a \emph{dark photon}  \cite{Holdom:1985ag,Okun:1982xi}. In this model, dark matter has a charge $g_{D}$ under a ``dark'' version of electromagnetism, but unlike electromagnetism, the dark photon is massive, in many models comparable to the dark matter mass itself.  The experimental and observational program to search for dark photons, motivated in part by developments in BSM model building, is one of the major developments over the past decade 
\cite{Battaglieri:2017aum}.

\emph{Axion and wave-like dark matter:}
The existence of dwarf galaxies means that Dark Matter whose mass is below $\sim \mathrm{keV}$ scale must be bosonic and furthermore that such dark matter must be heavier than $10^{-22} \; \mathrm{eV}$.  Such ultra-light dark matter has a high number density in our galaxy and is best understood in terms of classical waves.  The (QCD) axion is the quintessential example, and there has been a resurgence in interest over the last decade as direct searches for particle dark matter become ever more constraining.

The non-thermal production of the axion is closely related to when and how the $U(1)_{PQ}$ symmetry is spontaneously broken in the early universe.
If such breaking occurs after inflation, the radial mode is initially trapped at the origin and later rolls towards the potential minimum, spontaneously breaking the symmetry. At this time, the axion obtains random field values in different patches of the universe.
There is an irreducible abundance from the misalignment mechanism~\cite{Preskill:1982cy, Dine:1982ah,Abbott:1982af} and the domain-wall network. Since the axion field is randomized, the axion field must have a spatial average of the misalignment angle.
To explain the observed DM abundance, this misalignment contribution alone predicts an axion mass of order $m_a \simeq 30 ~\mu{\rm eV}$ (or $f_a \simeq 2 \times 10^{11} \GeV$) for the QCD axion and $m_a \simeq 600 ~\mu{\rm eV}~(10^{12} \GeV /f_a)^4$ for ALPs.
In addition to axion misalignment, topological defects~\cite{Vilenkin:1984ib,Davis:1986xc}, such as axion strings and domain walls, will also form. The decay of the axion-string and domain-wall network radiates axions, which also contribute to the DM abundance. An accurate determination of the total axion abundance in this scenario requires lattice simulations because of the complex dynamics involved from the $U(1)$ symmetry breaking to the late-time oscillations. There have been extensive efforts dedicated to this numerical study~\cite{Hiramatsu:2012gg,Kawasaki:2014sqa,Fleury:2015aca,Klaer:2017ond,Gorghetto:2018myk,Vaquero:2018tib,Buschmann:2019icd,Hindmarsh:2019csc,Gorghetto:2020qws,Hindmarsh:2021vih,Buschmann:2021sdq}. In the case where the domain wall number is larger than unity, the domain walls are stable and will overclose the universe. This issue is avoided if explicit $U(1)$ breaking is introduced so that the domain walls decay and make an additional contribution to the axion abundance~\cite{Sikivie:1982qv,Chang:1998tb,Hiramatsu:2010yn,Hiramatsu:2012sc,Kawasaki:2014sqa,Ringwald:2015dsf,Harigaya:2018ooc,Caputo:2019wsd}, which allows for a smaller $f_a$ to still explain DM.

If the symmetry is instead broken before or during inflation, the field values of both modes will be homogenized by inflation up to possible quantum fluctuations
and the misalignment angle $\theta_i$ is a constant throughout the observable universe. The axion abundance, therefore, depends on the value of $\theta_i$. The observed DM abundance is reproduced by $m_a \simeq 10 ~\mu{\rm eV} \times \theta_i^{12/7}$ (or $f_a \simeq 6 \times 10^{11} \GeV / \theta_i^{12/7}$) for the QCD axion and $m_a \simeq 7 \; {\rm meV} /\theta_i^4 \times (10^{12} \; {\rm GeV} /f_a)^4$ for ALPs. Thus, a large decay constant, as motivated by string theory or grand unification of the gauge and $U(1)$ symmetries~\cite{Nilles:1981py,Hall:1995eq}, needs a small $\theta_i$ to avoid overabundance. Similarly, a small decay constant calls for an angle very close to the hilltop of the cosine potential, $\theta_i \rightarrow \pi$, in order to exploit the inharmonicity. In the simplistic scenario, a very small/large $\theta_i$ has to come from a tuned initial condition. However, a small angle may also result from the early relaxation of the axion field during or after inflation (a large angle is similarly achieved with a further phase shift of $\pi$~\cite{Co:2018mho,Takahashi:2019pqf,Huang:2020etx}) if the axion mass is larger in the early universe~\cite{Dvali:1995ce,Banks:1996ea,Choi:1996fs,Co:2018phi} or if inflation lasts a very long time~\cite{Dimopoulos:1988pw,Graham:2018jyp,Takahashi:2018tdu,Kitajima:2019ibn}.

\emph{Other models of dark matter:}
There are a plethora of other models of dark matter with
intriguing properties and signals that we can only briefly
mention due solely to space limitations of this summary report.
This includes
\emph{sterile neutrino dark matter} (reviewed in \cite{Asadi:2022njl}), 
\emph{ultraheavy dark matter} (reviewed in \cite{Carney:2022gse}), 
\emph{dynamical dark matter} (reviewed in \cite{Dienes:2022zbh}), and
\emph{hidden sectors and a multi-temperature universe}
(reviewed in \cite{Aboubrahim:2021dei}).


%

\section{Baryogenesis}
\label{sec:baryogenesis}

There is more matter than antimatter in the Universe. This asymmetry, quantified as the ratio of baryon density to photon density, is measured at the time of Big Bang Nucleosynthesis (BBN) and the Cosmic Microwave Background (CMB) to be
$
 (n_b-n_{\bar{b}})/n_\gamma = n_b/n_\gamma = \left(6.10\pm 0.4\right)\times 10^{-10}\,
$~\cite{planck}.
Inflation dictates that such an asymmetry must be dynamically generated after reheating, necessitating a mechanism of \emph{baryogenesis}. 

In order to produce a matter--antimatter asymmetry, a model of particle physics must satisfy the so-called Sakharov conditions \cite{sakharov}. These are: (i) Baryon number ($B$) violation, (ii) $C$ and $CP$ violation, and (iii) departure from thermodynamic equilibrium. 
In the Standard Model  (SM), (i) Baryon number is anomalously violated in the weak interactions of the SM. Although the rate of $B$-violating  \emph{sphaleron} processes is exponentially suppressed at zero temperature, sphalerons are very efficient at temperatures at which electroweak symmetry is restored, $T\gtrsim 130$~GeV \cite{Arnold:1987mh,Arnold:1996dy}. (ii) There is $CP$ violation in the CKM matrix, and possibly in the PMNS matrix~\cite{Abe:2019vii}. It has been argued that the CKM phase is not sufficient (in fact orders of magnitude too small) for producing the baryon asymmetry of the Universe. Within the SM there is no process to employ the $CP$ violation in the PMNS matrix to produce the asymmetry. (iii) There are many ways a process could occur out of thermal equilibrium, such as particle decay at temperatures below its mass, or a first-order phase transition. There is no process in the SM that goes out of thermal equilibrium in the early Universe. These shortcomings of the SM are a clear sign of BSM physics. By the nature of the problem, these observations and the related new physics have strong implications for early Universe cosmology. Beyond-the-Standard Model models that seek to explain the baryon asymmetry invoke certain ingredients to satisfy the Sakharov conditions. 

The question of generating the baryon asymmetry of the universe has been around for several decades.  Leptogenesis is perhaps the best known model \cite{Fukugita:1986hr}, where a lepton asymmetry is generated by right-handed neutrino decays and transferred to a baryon symmetry though electroweak sphalerons.  Electroweak baryogenesis (reviewed in \cite{Cline:2006ts,Morrissey:2012db,Konstandin:2013caa,Garbrecht:2018mrp,Bodeker:2020ghk}) is also a well-known mechanism where the baryon asymmetry is generated during the electroweak phase transition.
Nevertheless, there continue to be new ways to address this mystery, inspired by new theoretical ideas and observational opportunities.
In the reviews \cite{Elor:2022hpa,Asadi:2022njl},
numerous baryogenesis mechanisms have been discussed, with a focus on those developed in the last decade. A salient aspect of many of these contributions is experimental testability: while the vast majority of traditional baryogenesis models have involved high-scale physics and hence are difficult to probe experimentally, many new BSM models produce the baryon asymmetry at low scales and involve low-scale new physics, and are therefore experimentally observable. Many of these new models are expected to produce signals at multiple experiments, allowing for a multi-prong search for new physics (see \cite{Elor:2022hpa} for details).

\section{Flavor Model Building}
\label{sec:flavor}

Flavor violating processes, in particular those involving flavor changing neutral currents (FCNC) have exquisite sensitivity to new sources of flavor and CP violation beyond those of the SM \cite{Altmannshofer:2022aml}. This high sensitivity to new physics has its origin in the small amount of flavor breaking that is present in the SM. 
In the SM, the only sources of flavor violation are the hierarchical Yukawa couplings of the Higgs.  The origin of the SM arrangement of the various quark and charged lepton masses, the hierarchical structure of the CKM matrix, and the absence of visible hierarchies in the PMNS matrix is often referred to as the SM flavor puzzle.  Various classes of ideas exist to solve this puzzle: horizontal flavor symmetries  \cite{Froggatt:1978nt}, warped extra dimentions \cite{Randall:1999vf,Randall:1999ee}, partial compositeness \cite{Kaplan:1991dc,Agashe:2004rs,Contino:2003ve,Contino:2006qr}, and radiative fermion masses~\cite{Weinberg:1972ws}.
In the SM, the quark FCNCs are suppressed by a loop factor and by small CKM matrix elements. As long as theoretical uncertainties in the SM predictions are under control, quark flavor violating processes can indirectly explore very high mass scales, in some cases far beyond the direct reach of collider experiments.
In the lepton sector, SM predictions for FCNCs are suppressed by the tiny neutrino masses and below any imaginable experimental sensitivities.  Electroweak contributions to electric dipole moments are also predicted to be strongly suppressed in the SM, several orders of magnitude below the current bounds. Charged lepton flavor violation and electric dipole moments are thus null tests of the SM. Any observation of such processes would be an unambiguous sign of new physics.

In the SM, the Yukawa couplings of the Higgs to the fermions are the only sources of flavor violation. Therefore, the Higgs might be the window into understanding flavor, with the precision Higgs program at the LHC, and in the future also at a Higgs factory, able to provide valuable inputs. In particular, Higgs decays involving tau, $h \to \tau \mu$ and $h \to \tau e$, are cases where the direct searches at the LHC are the most sensitive probes.

In addition to long-standing puzzles, in the last several years a number of ``flavor anomalies''  have created considerable excitement in the community. Discrepancies between SM predictions and experimental measurements are seen in $B$ decays as well as in the anomalous magnetic moment of the muon.
If the new physics origin for these experimental anomalies could be established, it would have a transformative impact on the field. First and foremost, such an indirect sign of new physics would establish a new mass scale in particle physics. This scale could become the next target for direct exploration at future high-energy colliders. With sufficient energy, discoveries would, at least in principle, be guaranteed. 
Second, the couplings of the new physics constitute new sources of flavor violation beyond the SM Yukawa couplings. Existing low energy constraints suggest that such new physics couplings have a hierarchical flavor structure. This provides a new perspective on the Standard Model flavor puzzle and invites the construction of flavor models that link the structure of the SM and BSM sources of flavor violation.
At present, the global analyses point towards a small consistent set of dimension-6 effective operators  ($C_9$ and/or $C_{10}$) to explain the B-physics anomalies. The leading candidate UV models generating these operators involve $Z'$ (e.g.~$L_\mu-L_\tau$) gauge bosons or leptoquarks.

\section*{Acknowledgments}

The conveners thank the authors of all whitepapers submitted to TF08 and those who contributed to the development of this document in numerous other ways.
P.J.F. was supported by Fermi Research Alliance, LLC under Contract
DE-AC02-07CH11359 with the U.S. Department of Energy.
G.D.K. was supported in part by the U.S. Department of Energy
under Grant Number DE-SC0011640.
H.M. was supported in part by 
Beyond AI Institute at the University of Tokyo; 
Director, Office of Science, Office of High Energy Physics of the U.S.
Department of Energy under the Contract DE-AC02-05CH11231;
NSF grant PHY-1915314;
JSPS Grant-in-Aid for Scientific Research JP20K03942;
MEXT Grant-in-Aid for Transformative Research Areas (A) JP20H05850, JP20A203;
Hamamatsu Photonics, K.K.;
and World Premier International Research Center Initiative
(WPI) MEXT, Japan.


\bibliographystyle{JHEP}
\bibliography{TF08refs,Agrawalrefs,Altmannshoferrefs,Attanasiorefs,Baerrefs,Blinovrefs,Carneyrefs,Duttarefs,Elorrefs,Batellrefs,Teresirefs,Draperrefs,Berlinrefs,RandallSundrumrefs,Dvorkinrefs,LDM,Ferrettirefs}

\providecommand{\href}[2]{#2}\begingroup\raggedright\begin{thebibliography}{100}

\bibitem{TF01}
D.~Harlow, S.~Kachru and J.~Maldacena, \emph{{TF01 Snowmass Report: String
  Theory, Quantum Gravity, Black Holes}},  2022.

\bibitem{TF02}
P.~Draper and I.~Rothstein, \emph{{TF02 Snowmass Report: Effective Field Theory
  Techniques}},  2022.

\bibitem{TF03}
D.~Poland and L.~Rastelli, \emph{{TF03 Snowmass Report: CFT and Formal QFT}},
  2022.

\bibitem{TF04}
Z.~Bern and J.~Trnka, \emph{{TF04 Snowmass Report: Scattering Amplitudes}},
  2022.

\bibitem{TF05}
Z.~Davoudi, T.~Izubuchi and E.~Neil, \emph{{TF05 Snowmass Report: Lattice Gauge
  Theory}},  2022.

\bibitem{TF06}
R.~Boughezal and Z.~Ligeti, \emph{{TF06 Snowmass Report: Theory Techniques for
  Precision Physics}},  2022.

\bibitem{TF07}
F.~Maltoni, S.~Su and J.~Thaler, \emph{{TF07 Snowmass Report: Collider
  Phenomenology}},  2022.

\bibitem{TF09}
D.~Green, J.T.~Ruderman, B.R.~Safdi and J.~Shelton, \emph{{TF09 Snowmass
  Report: Astrophysics and Cosmology}},  2022.

\bibitem{TF10}
S.~Catterall, R.~Harnik and V.~Hubeny, \emph{{TF10 Snowmass Report: Quantum
  Information Science}},  2022.

\bibitem{TF11}
A.~de~Gouv\^{e}a, I.~Mocioiu, S.~Pastore and L.~Strigari, \emph{{TF11 Snowmass
  Report: Theory of Neutrino Physics}},  2022.

\bibitem{Essig:2022yzw}
R.~Essig, Y.~Kahn, S.~Knapen, A.~Ringwald and N.~Toro, \emph{{Snowmass2021
  Theory Frontier: Theory Meets the Lab}},  in \emph{{2022 Snowmass Summer
  Study}}, 3, 2022 [\href{https://arxiv.org/abs/2203.10089}{{\ttfamily
  2203.10089}}].

\bibitem{Campbell:2022qmc}
J.M.~Campbell et~al., \emph{{Event Generators for High-Energy Physics
  Experiments}},  in \emph{{2022 Snowmass Summer Study}}, 3, 2022
  [\href{https://arxiv.org/abs/2203.11110}{{\ttfamily 2203.11110}}].

\bibitem{Dvorkin:2022jyg}
C.~Dvorkin et~al., \emph{{The Physics of Light Relics}},  in \emph{{2022
  Snowmass Summer Study}}, 3, 2022
  [\href{https://arxiv.org/abs/2203.07943}{{\ttfamily 2203.07943}}].

\bibitem{CF1}
J.~Cooley, T.~Lin, H.~Lippincott and T.~Slatyer, \emph{{CF1 Snowmass Report --
  Dark Matter: Particle-like}},  2022.

\bibitem{CF2}
J.~Jaeckel, G.~Rybka and L.~Winslow, \emph{{CF2 Snowmass Report -- Dark Matter:
  Wave-like}},  2022.

\bibitem{CF3}
A.~Drlica-Wagner, C.~Prescod-Weinstein and H.~bo~Yu, \emph{{CF3 Snowmass Report
  -- Dark Matter: Cosmic Probes}},  2022.

\bibitem{CF4}
J.~Annis, J.~Newman and A.~Slosar, \emph{{CF4 Snowmass Report -- Dark Energy
  and Cosmic Acceleration: The Modern Universe}},  2022.

\bibitem{CF5}
C.~Chang, L.~Newburgh and D.~Shoemaker, \emph{{CF5 Snowmass Report -- Dark
  Energy and Cosmic Acceleration: Cosmic Dawn and Before}},  2022.

\bibitem{CF6}
V.~Miranda, B.~Flaugher and D.~Schlegel, \emph{{CF6 Snowmass Report -- Dark
  Energy and Cosmic Acceleration: Complementarity of Probes and New
  Facilities}},  2022.

\bibitem{CF7}
R.~Adhikari, L.~Anchordoqui, K.~Fang, B.~Sathyaprakash and K.~Tollefson,
  \emph{{CF7 Snowmass Report -- Cosmic Probes of Fundamental Physics}},  2022.

\bibitem{Aboubrahim:2021dei}
A.~Aboubrahim, W.-Z.~Feng, P.~Nath and Z.-Y.~Wang, \emph{{Hidden sectors and a
  multi-temperature universe}},  in \emph{{2022 Snowmass Summer Study}}, 6,
  2021 [\href{https://arxiv.org/abs/2106.06494}{{\ttfamily 2106.06494}}].

\bibitem{Baer:2022naw}
H.~Baer, V.~Barger, S.~Salam and D.~Sengupta, \emph{{Mini-review: Expectations
  for supersymmetry from the string landscape}},  in \emph{{2022 Snowmass
  Summer Study}}, 2, 2022 [\href{https://arxiv.org/abs/2202.11578}{{\ttfamily
  2202.11578}}].

\bibitem{Batell:2022pzc}
B.~Batell, M.~Low, E.T.~Neil and C.B.~Verhaaren, \emph{{Review of Neutral
  Naturalness}},  in \emph{{2022 Snowmass Summer Study}}, 3, 2022
  [\href{https://arxiv.org/abs/2203.05531}{{\ttfamily 2203.05531}}].

\bibitem{Elor:2022hpa}
G.~Elor et~al., \emph{{New Ideas in Baryogenesis: A Snowmass White Paper}},  in
  \emph{{2022 Snowmass Summer Study}}, 3, 2022
  [\href{https://arxiv.org/abs/2203.05010}{{\ttfamily 2203.05010}}].

\bibitem{Blinov:2022tfy}
N.~Blinov, N.~Craig, M.J.~Dolan, J.~de~Vries, P.~Draper, I.G.~Garcia et~al.,
  \emph{{Snowmass White Paper: Strong CP Beyond Axion Direct Detection}},  in
  \emph{{2022 Snowmass Summer Study}}, 3, 2022
  [\href{https://arxiv.org/abs/2203.07218}{{\ttfamily 2203.07218}}].

\bibitem{Windchime:2022whs}
{\scshape Windchime} collaboration, \emph{{Snowmass 2021 White Paper: The
  Windchime Project}},  in \emph{{2022 Snowmass Summer Study}}, 3, 2022
  [\href{https://arxiv.org/abs/2203.07242}{{\ttfamily 2203.07242}}].

\bibitem{Carney:2022gse}
D.~Carney et~al., \emph{{Snowmass2021 Cosmic Frontier White Paper: Ultraheavy
  particle dark matter}},  \href{https://arxiv.org/abs/2203.06508}{{\ttfamily
  2203.06508}}.

\bibitem{Draper:2022pvk}
P.~Draper, I.G.~Garcia and M.~Reece, \emph{{Snowmass White Paper: Implications
  of Quantum Gravity for Particle Physics}},  in \emph{{2022 Snowmass Summer
  Study}}, 3, 2022 [\href{https://arxiv.org/abs/2203.07624}{{\ttfamily
  2203.07624}}].

\bibitem{Dutta:2022qvn}
B.~Dutta, S.~Ghosh and J.~Kumar, \emph{{$U(1)_{T3R}$ Extension of Standard
  Model: A Sub-GeV Dark Matter Model}},  in \emph{{2022 Snowmass Summer
  Study}}, 3, 2022 [\href{https://arxiv.org/abs/2203.07786}{{\ttfamily
  2203.07786}}].

\bibitem{Altmannshofer:2022aml}
W.~Altmannshofer and J.~Zupan, \emph{{Snowmass White Paper: Flavor Model
  Building}},  in \emph{{2022 Snowmass Summer Study}}, 3, 2022
  [\href{https://arxiv.org/abs/2203.07726}{{\ttfamily 2203.07726}}].

\bibitem{Agrawal:2022yvu}
P.~Agrawal, K.V.~Berghaus, J.~Fan, A.~Hook, G.~Marques-Tavares and T.~Rudelius,
  \emph{{Some open questions in axion theory}},  in \emph{{2022 Snowmass Summer
  Study}}, 3, 2022 [\href{https://arxiv.org/abs/2203.08026}{{\ttfamily
  2203.08026}}].

\bibitem{PeterCameron}
P.~Cameron, \emph{{Bootstrapping the Muon Collider: Massless Neutrinos in the
  $g-2$ Delivery Ring}},  2022.

\bibitem{DagnoloTeresi}
R.~Tito~D'Agnolo and D.~Teresi, \emph{{Cosmological Selection of The Weak
  Scale}},  2022.

\bibitem{Craig:2022uua}
N.~Craig, \emph{{Naturalness: A Snowmass White Paper}},  in \emph{{2022
  Snowmass Summer Study}}, 5, 2022
  [\href{https://arxiv.org/abs/2205.05708}{{\ttfamily 2205.05708}}].

\bibitem{Agrawal:2022rqd}
P.~Agrawal, C.~Cesarotti, A.~Karch, R.K.~Mishra, L.~Randall and R.~Sundrum,
  \emph{{Warped Compactifications in Particle Physics, Cosmology and Quantum
  Gravity}},  in \emph{{2022 Snowmass Summer Study}}, 3, 2022
  [\href{https://arxiv.org/abs/2203.07533}{{\ttfamily 2203.07533}}].

\bibitem{Banerjee:2022xmu}
A.~Banerjee et~al., \emph{{Phenomenological aspects of composite Higgs
  scenarios: exotic scalars and vector-like quarks}},
  \href{https://arxiv.org/abs/2203.07270}{{\ttfamily 2203.07270}}.

\bibitem{Asadi:2022njl}
P.~Asadi et~al., \emph{{Early-Universe Model Building}},
  \href{https://arxiv.org/abs/2203.06680}{{\ttfamily 2203.06680}}.

\bibitem{Martin:1997ns}
S.P.~Martin, \emph{{A Supersymmetry primer}},
  \href{https://doi.org/10.1142/9789812839657_0001}{\emph{Adv. Ser. Direct.
  High Energy Phys.} {\bfseries 18} (1998) 1}
  [\href{https://arxiv.org/abs/hep-ph/9709356}{{\ttfamily hep-ph/9709356}}].

\bibitem{Baer:2006rs}
H.~Baer and X.~Tata, \emph{{Weak scale supersymmetry: From superfields to
  scattering events}}, Cambridge University Press (5, 2006).

\bibitem{Barbieri:1987fn}
R.~Barbieri and G.F.~Giudice, \emph{{Upper Bounds on Supersymmetric Particle
  Masses}}, \href{https://doi.org/10.1016/0550-3213(88)90171-X}{\emph{Nucl.
  Phys. B} {\bfseries 306} (1988) 63}.

\bibitem{Baer:2021tta}
H.~Baer, V.~Barger and D.~Martinez, \emph{{Comparison of SUSY spectra
  generators for natural SUSY and string landscape predictions}},
  \href{https://doi.org/10.1140/epjc/s10052-022-10141-2}{\emph{Eur. Phys. J. C}
  {\bfseries 82} (2022) 172}
  [\href{https://arxiv.org/abs/2111.03096}{{\ttfamily 2111.03096}}].

\bibitem{Wells:2003tf}
J.D.~Wells, \emph{{Implications of supersymmetry breaking with a little
  hierarchy between gauginos and scalars}},  in \emph{{11th International
  Conference on Supersymmetry and the Unification of Fundamental
  Interactions}}, 6, 2003
  [\href{https://arxiv.org/abs/hep-ph/0306127}{{\ttfamily hep-ph/0306127}}].

\bibitem{Arkani-Hamed:2004ymt}
N.~Arkani-Hamed and S.~Dimopoulos, \emph{{Supersymmetric unification without
  low energy supersymmetry and signatures for fine-tuning at the LHC}},
  \href{https://doi.org/10.1088/1126-6708/2005/06/073}{\emph{JHEP} {\bfseries
  06} (2005) 073} [\href{https://arxiv.org/abs/hep-th/0405159}{{\ttfamily
  hep-th/0405159}}].

\bibitem{Fox:2002bu}
P.J.~Fox, A.E.~Nelson and N.~Weiner, \emph{{Dirac gaugino masses and supersoft
  supersymmetry breaking}},
  \href{https://doi.org/10.1088/1126-6708/2002/08/035}{\emph{JHEP} {\bfseries
  08} (2002) 035} [\href{https://arxiv.org/abs/hep-ph/0206096}{{\ttfamily
  hep-ph/0206096}}].

\bibitem{Kribs:2007ac}
G.D.~Kribs, E.~Poppitz and N.~Weiner, \emph{{Flavor in supersymmetry with an
  extended R-symmetry}},
  \href{https://doi.org/10.1103/PhysRevD.78.055010}{\emph{Phys. Rev.}
  {\bfseries D78} (2008) 055010}
  [\href{https://arxiv.org/abs/0712.2039}{{\ttfamily 0712.2039}}].

\bibitem{Martin:2007gf}
S.P.~Martin, \emph{{Compressed supersymmetry and natural neutralino dark matter
  from top squark-mediated annihilation to top quarks}},
  \href{https://doi.org/10.1103/PhysRevD.75.115005}{\emph{Phys. Rev. D}
  {\bfseries 75} (2007) 115005}
  [\href{https://arxiv.org/abs/hep-ph/0703097}{{\ttfamily hep-ph/0703097}}].

\bibitem{Fan:2011yu}
J.~Fan, M.~Reece and J.T.~Ruderman, \emph{{Stealth Supersymmetry}},
  \href{https://doi.org/10.1007/JHEP11(2011)012}{\emph{JHEP} {\bfseries 11}
  (2011) 012} [\href{https://arxiv.org/abs/1105.5135}{{\ttfamily 1105.5135}}].

\bibitem{Csaki:2018kxb}
C.~Cs\'aki, G.~Lee, S.J.~Lee, S.~Lombardo and O.~Telem, \emph{{Continuum
  Naturalness}}, \href{https://doi.org/10.1007/JHEP03(2019)142}{\emph{JHEP}
  {\bfseries 03} (2019) 142}
  [\href{https://arxiv.org/abs/1811.06019}{{\ttfamily 1811.06019}}].

\bibitem{Kaplan:1983fs}
D.B.~Kaplan and H.~Georgi, \emph{{SU(2) x U(1) Breaking by Vacuum
  Misalignment}},
  \href{https://doi.org/10.1016/0370-2693(84)91177-8}{\emph{Phys. Lett.}
  {\bfseries 136B} (1984) 183}.

\bibitem{Kaplan:1991dc}
D.B.~Kaplan, \emph{{Flavor at SSC energies: A New mechanism for dynamically
  generated fermion masses}},
  \href{https://doi.org/10.1016/S0550-3213(05)80021-5}{\emph{Nucl. Phys. B}
  {\bfseries 365} (1991) 259}.

\bibitem{Agashe:2004rs}
K.~Agashe, R.~Contino and A.~Pomarol, \emph{{The Minimal composite Higgs
  model}}, \href{https://doi.org/10.1016/j.nuclphysb.2005.04.035}{\emph{Nucl.
  Phys. B} {\bfseries 719} (2005) 165}
  [\href{https://arxiv.org/abs/hep-ph/0412089}{{\ttfamily hep-ph/0412089}}].

\bibitem{Contino:2010rs}
R.~Contino, \emph{{The Higgs as a Composite Nambu-Goldstone Boson}},  in
  \emph{{Theoretical Advanced Study Institute in Elementary Particle Physics}:
  {Physics of the Large and the Small}}, pp.~235--306, 2011,
  \href{https://doi.org/10.1142/9789814327183_0005}{DOI}
  [\href{https://arxiv.org/abs/1005.4269}{{\ttfamily 1005.4269}}].

\bibitem{Bellazzini:2014yua}
B.~Bellazzini, C.~Csáki and J.~Serra, \emph{{Composite Higgses}},
  \href{https://doi.org/10.1140/epjc/s10052-014-2766-x}{\emph{Eur. Phys. J. C}
  {\bfseries 74} (2014) 2766}
  [\href{https://arxiv.org/abs/1401.2457}{{\ttfamily 1401.2457}}].

\bibitem{Panico:2015jxa}
G.~Panico and A.~Wulzer, \emph{{The Composite Nambu-Goldstone Higgs}},
  vol.~913, Springer (2016),
  \href{https://doi.org/10.1007/978-3-319-22617-0}{10.1007/978-3-319-22617-0},
  [\href{https://arxiv.org/abs/1506.01961}{{\ttfamily 1506.01961}}].

\bibitem{Cacciapaglia:2020kgq}
G.~Cacciapaglia, C.~Pica and F.~Sannino, \emph{{Fundamental Composite Dynamics:
  A Review}}, \href{https://doi.org/10.1016/j.physrep.2020.07.002}{\emph{Phys.
  Rept.} {\bfseries 877} (2020) 1}
  [\href{https://arxiv.org/abs/2002.04914}{{\ttfamily 2002.04914}}].

\bibitem{Contino:2003ve}
R.~Contino, Y.~Nomura and A.~Pomarol, \emph{{Higgs as a holographic
  pseudoGoldstone boson}},
  \href{https://doi.org/10.1016/j.nuclphysb.2003.08.027}{\emph{Nucl. Phys. B}
  {\bfseries 671} (2003) 148}
  [\href{https://arxiv.org/abs/hep-ph/0306259}{{\ttfamily hep-ph/0306259}}].

\bibitem{Schmaltz:2005ky}
M.~Schmaltz and D.~Tucker-Smith, \emph{{Little Higgs review}},
  \href{https://doi.org/10.1146/annurev.nucl.55.090704.151502}{\emph{Ann. Rev.
  Nucl. Part. Sci.} {\bfseries 55} (2005) 229}
  [\href{https://arxiv.org/abs/hep-ph/0502182}{{\ttfamily hep-ph/0502182}}].

\bibitem{Holdom:1981rm}
B.~Holdom, \emph{{Raising the Sideways Scale}},
  \href{https://doi.org/10.1103/PhysRevD.24.1441}{\emph{Phys. Rev. D}
  {\bfseries 24} (1981) 1441}.

\bibitem{Cohen:1988sq}
A.G.~Cohen and H.~Georgi, \emph{{Walking Beyond the Rainbow}},
  \href{https://doi.org/10.1016/0550-3213(89)90109-0}{\emph{Nucl. Phys. B}
  {\bfseries 314} (1989) 7}.

\bibitem{Barnard:2013zea}
J.~Barnard, T.~Gherghetta and T.S.~Ray, \emph{{UV descriptions of composite
  Higgs models without elementary scalars}},
  \href{https://doi.org/10.1007/JHEP02(2014)002}{\emph{JHEP} {\bfseries 02}
  (2014) 002} [\href{https://arxiv.org/abs/1311.6562}{{\ttfamily 1311.6562}}].

\bibitem{Ferretti:2013kya}
G.~Ferretti and D.~Karateev, \emph{{Fermionic UV completions of Composite Higgs
  models}}, \href{https://doi.org/10.1007/JHEP03(2014)077}{\emph{JHEP}
  {\bfseries 03} (2014) 077} [\href{https://arxiv.org/abs/1312.5330}{{\ttfamily
  1312.5330}}].

\bibitem{Montull:2013mla}
M.~Montull, F.~Riva, E.~Salvioni and R.~Torre, \emph{{Higgs Couplings in
  Composite Models}},
  \href{https://doi.org/10.1103/PhysRevD.88.095006}{\emph{Phys. Rev. D}
  {\bfseries 88} (2013) 095006}
  [\href{https://arxiv.org/abs/1308.0559}{{\ttfamily 1308.0559}}].

\bibitem{Sanz:2017tco}
V.~Sanz and J.~Setford, \emph{{Composite Higgs Models after Run 2}},
  \href{https://doi.org/10.1155/2018/7168480}{\emph{Adv. High Energy Phys.}
  {\bfseries 2018} (2018) 7168480}
  [\href{https://arxiv.org/abs/1703.10190}{{\ttfamily 1703.10190}}].

\bibitem{Liu:2017dsz}
D.~Liu, I.~Low and C.E.M.~Wagner, \emph{{Modification of Higgs Couplings in
  Minimal Composite Models}},
  \href{https://doi.org/10.1103/PhysRevD.96.035013}{\emph{Phys. Rev. D}
  {\bfseries 96} (2017) 035013}
  [\href{https://arxiv.org/abs/1703.07791}{{\ttfamily 1703.07791}}].

\bibitem{Banerjee:2017wmg}
A.~Banerjee, G.~Bhattacharyya, N.~Kumar and T.S.~Ray, \emph{{Constraining
  Composite Higgs Models using LHC data}},
  \href{https://doi.org/10.1007/JHEP03(2018)062}{\emph{JHEP} {\bfseries 03}
  (2018) 062} [\href{https://arxiv.org/abs/1712.07494}{{\ttfamily
  1712.07494}}].

\bibitem{Liu:2018qtb}
D.~Liu, I.~Low and Z.~Yin, \emph{{Universal Relations in Composite Higgs
  Models}}, \href{https://doi.org/10.1007/JHEP05(2019)170}{\emph{JHEP}
  {\bfseries 05} (2019) 170}
  [\href{https://arxiv.org/abs/1809.09126}{{\ttfamily 1809.09126}}].

\bibitem{Arbey:2015exa}
A.~Arbey, G.~Cacciapaglia, H.~Cai, A.~Deandrea, S.~Le~Corre and F.~Sannino,
  \emph{{Fundamental Composite Electroweak Dynamics: Status at the LHC}},
  \href{https://doi.org/10.1103/PhysRevD.95.015028}{\emph{Phys. Rev. D}
  {\bfseries 95} (2017) 015028}
  [\href{https://arxiv.org/abs/1502.04718}{{\ttfamily 1502.04718}}].

\bibitem{Ferretti:2016upr}
G.~Ferretti, \emph{{Gauge theories of Partial Compositeness: Scenarios for
  Run-II of the LHC}},
  \href{https://doi.org/10.1007/JHEP06(2016)107}{\emph{JHEP} {\bfseries 06}
  (2016) 107} [\href{https://arxiv.org/abs/1604.06467}{{\ttfamily
  1604.06467}}].

\bibitem{Belyaev:2016ftv}
A.~Belyaev, G.~Cacciapaglia, H.~Cai, G.~Ferretti, T.~Flacke, A.~Parolini
  et~al., \emph{{Di-boson signatures as Standard Candles for Partial
  Compositeness}}, \href{https://doi.org/10.1007/JHEP01(2017)094}{\emph{JHEP}
  {\bfseries 01} (2017) 094}
  [\href{https://arxiv.org/abs/1610.06591}{{\ttfamily 1610.06591}}].

\bibitem{Agugliaro:2018vsu}
A.~Agugliaro, G.~Cacciapaglia, A.~Deandrea and S.~De~Curtis, \emph{{Vacuum
  misalignment and pattern of scalar masses in the SU(5)/SO(5) composite Higgs
  model}}, \href{https://doi.org/10.1007/JHEP02(2019)089}{\emph{JHEP}
  {\bfseries 02} (2019) 089}
  [\href{https://arxiv.org/abs/1808.10175}{{\ttfamily 1808.10175}}].

\bibitem{Cacciapaglia:2015eqa}
G.~Cacciapaglia, H.~Cai, A.~Deandrea, T.~Flacke, S.J.~Lee and A.~Parolini,
  \emph{{Composite scalars at the LHC: the Higgs, the Sextet and the Octet}},
  \href{https://doi.org/10.1007/JHEP11(2015)201}{\emph{JHEP} {\bfseries 11}
  (2015) 201} [\href{https://arxiv.org/abs/1507.02283}{{\ttfamily
  1507.02283}}].

\bibitem{Bizot:2018tds}
N.~Bizot, G.~Cacciapaglia and T.~Flacke, \emph{{Common exotic decays of top
  partners}}, \href{https://doi.org/10.1007/JHEP06(2018)065}{\emph{JHEP}
  {\bfseries 06} (2018) 065}
  [\href{https://arxiv.org/abs/1803.00021}{{\ttfamily 1803.00021}}].

\bibitem{Cacciapaglia:2019zmj}
G.~Cacciapaglia, T.~Flacke, M.~Park and M.~Zhang, \emph{{Exotic decays of top
  partners: mind the search gap}},
  \href{https://doi.org/10.1016/j.physletb.2019.135015}{\emph{Phys. Lett. B}
  {\bfseries 798} (2019) 135015}
  [\href{https://arxiv.org/abs/1908.07524}{{\ttfamily 1908.07524}}].

\bibitem{Xie:2019gya}
K.-P.~Xie, G.~Cacciapaglia and T.~Flacke, \emph{{Exotic decays of top partners
  with charge 5/3: bounds and opportunities}},
  \href{https://doi.org/10.1007/JHEP10(2019)134}{\emph{JHEP} {\bfseries 10}
  (2019) 134} [\href{https://arxiv.org/abs/1907.05894}{{\ttfamily
  1907.05894}}].

\bibitem{Matsedonskyi:2014lla}
O.~Matsedonskyi, F.~Riva and T.~Vantalon, \emph{{Composite Charge 8/3
  Resonances at the LHC}},
  \href{https://doi.org/10.1007/JHEP04(2014)059}{\emph{JHEP} {\bfseries 04}
  (2014) 059} [\href{https://arxiv.org/abs/1401.3740}{{\ttfamily 1401.3740}}].

\bibitem{Corcella:2021mdl}
G.~Corcella, A.~Costantini, M.~Ghezzi, L.~Panizzi, G.M.~Pruna and J.~\v{S}alko,
  \emph{{Vector-like quarks decaying into singly and doubly charged bosons at
  LHC}}, \href{https://doi.org/10.1007/JHEP10(2021)108}{\emph{JHEP} {\bfseries
  10} (2021) 108} [\href{https://arxiv.org/abs/2107.07426}{{\ttfamily
  2107.07426}}].

\bibitem{Dasgupta:2021fzw}
S.~Dasgupta, R.~Pramanick and T.S.~Ray, \emph{{Broad top-like vector quarks at
  LHC and HL-LHC}},  \href{https://arxiv.org/abs/2112.03742}{{\ttfamily
  2112.03742}}.

\bibitem{Chala:2017xgc}
M.~Chala, \emph{{Direct bounds on heavy toplike quarks with standard and exotic
  decays}}, \href{https://doi.org/10.1103/PhysRevD.96.015028}{\emph{Phys. Rev.
  D} {\bfseries 96} (2017) 015028}
  [\href{https://arxiv.org/abs/1705.03013}{{\ttfamily 1705.03013}}].

\bibitem{Cacciapaglia:2021uqh}
G.~Cacciapaglia, T.~Flacke, M.~Kunkel and W.~Porod, \emph{{Phenomenology of
  unusual top partners in composite Higgs models}},
  \href{https://doi.org/10.1007/JHEP02(2022)208}{\emph{JHEP} {\bfseries 02}
  (2022) 208} [\href{https://arxiv.org/abs/2112.00019}{{\ttfamily
  2112.00019}}].

\bibitem{Azatov:2015xqa}
A.~Azatov, D.~Chowdhury, D.~Ghosh and T.S.~Ray, \emph{{Same sign di-lepton
  candles of the composite gluons}},
  \href{https://doi.org/10.1007/JHEP08(2015)140}{\emph{JHEP} {\bfseries 08}
  (2015) 140} [\href{https://arxiv.org/abs/1505.01506}{{\ttfamily
  1505.01506}}].

\bibitem{BuarqueFranzosi:2016ooy}
D.~Buarque~Franzosi, G.~Cacciapaglia, H.~Cai, A.~Deandrea and M.~Frandsen,
  \emph{{Vector and Axial-vector resonances in composite models of the Higgs
  boson}}, \href{https://doi.org/10.1007/JHEP11(2016)076}{\emph{JHEP}
  {\bfseries 11} (2016) 076}
  [\href{https://arxiv.org/abs/1605.01363}{{\ttfamily 1605.01363}}].

\bibitem{Yepes:2018dlw}
J.~Yepes and A.~Zerwekh, \emph{{Modelling top partner-vector resonance
  phenomenology}},
  \href{https://doi.org/10.1016/j.nuclphysb.2019.02.028}{\emph{Nucl. Phys. B}
  {\bfseries 941} (2019) 560}
  [\href{https://arxiv.org/abs/1806.06694}{{\ttfamily 1806.06694}}].

\bibitem{Dasgupta:2019yjm}
S.~Dasgupta, S.K.~Rai and T.S.~Ray, \emph{{Impact of a colored vector resonance
  on the collider constraints for top-like top partner}},
  \href{https://doi.org/10.1103/PhysRevD.102.115014}{\emph{Phys. Rev. D}
  {\bfseries 102} (2020) 115014}
  [\href{https://arxiv.org/abs/1912.13022}{{\ttfamily 1912.13022}}].

\bibitem{Banerjee:2021efl}
A.~Banerjee, S.~Dasgupta and T.S.~Ray, \emph{{Probing composite Higgs boson
  substructure at the HL-LHC}},
  \href{https://doi.org/10.1103/PhysRevD.104.095021}{\emph{Phys. Rev. D}
  {\bfseries 104} (2021) 095021}
  [\href{https://arxiv.org/abs/2105.01093}{{\ttfamily 2105.01093}}].

\bibitem{Cacciapaglia:2019bqz}
G.~Cacciapaglia, G.~Ferretti, T.~Flacke and H.~Ser\^odio, \emph{{Light scalars
  in composite Higgs models}},
  \href{https://doi.org/10.3389/fphy.2019.00022}{\emph{Front. in Phys.}
  {\bfseries 7} (2019) 22} [\href{https://arxiv.org/abs/1902.06890}{{\ttfamily
  1902.06890}}].

\bibitem{BuarqueFranzosi:2021kky}
D.~Buarque~Franzosi, G.~Cacciapaglia, X.~Cid~Vidal, G.~Ferretti, T.~Flacke and
  C.~V\'azquez~Sierra, \emph{{Exploring new possibilities to discover a light
  pseudo-scalar at LHCb}},
  \href{https://doi.org/10.1140/epjc/s10052-021-09930-y}{\emph{Eur. Phys. J. C}
  {\bfseries 82} (2022) 3} [\href{https://arxiv.org/abs/2106.12615}{{\ttfamily
  2106.12615}}].

\bibitem{Agashe:2004cp}
K.~Agashe, G.~Perez and A.~Soni, \emph{{Flavor structure of warped extra
  dimension models}},
  \href{https://doi.org/10.1103/PhysRevD.71.016002}{\emph{Phys. Rev. D}
  {\bfseries 71} (2005) 016002}
  [\href{https://arxiv.org/abs/hep-ph/0408134}{{\ttfamily hep-ph/0408134}}].

\bibitem{Cacciapaglia:2007fw}
G.~Cacciapaglia, C.~Csaki, J.~Galloway, G.~Marandella, J.~Terning and
  A.~Weiler, \emph{{A GIM Mechanism from Extra Dimensions}},
  \href{https://doi.org/10.1088/1126-6708/2008/04/006}{\emph{JHEP} {\bfseries
  04} (2008) 006} [\href{https://arxiv.org/abs/0709.1714}{{\ttfamily
  0709.1714}}].

\bibitem{Fitzpatrick:2007sa}
A.L.~Fitzpatrick, G.~Perez and L.~Randall, \emph{{Flavor anarchy in a
  Randall-Sundrum model with 5D minimal flavor violation and a low Kaluza-Klein
  scale}}, \href{https://doi.org/10.1103/PhysRevLett.100.171604}{\emph{Phys.
  Rev. Lett.} {\bfseries 100} (2008) 171604}
  [\href{https://arxiv.org/abs/0710.1869}{{\ttfamily 0710.1869}}].

\bibitem{Perez:2008ee}
G.~Perez and L.~Randall, \emph{{Natural Neutrino Masses and Mixings from Warped
  Geometry}}, \href{https://doi.org/10.1088/1126-6708/2009/01/077}{\emph{JHEP}
  {\bfseries 01} (2009) 077} [\href{https://arxiv.org/abs/0805.4652}{{\ttfamily
  0805.4652}}].

\bibitem{Craig:2015pha}
N.~Craig, A.~Katz, M.~Strassler and R.~Sundrum, \emph{{Naturalness in the Dark
  at the Lhc}}, \href{https://doi.org/10.1007/JHEP07(2015)105}{\emph{JHEP}
  {\bfseries 07} (2015) 105}
  [\href{https://arxiv.org/abs/1501.05310}{{\ttfamily 1501.05310}}].

\bibitem{Craig:2016kue}
N.~Craig, S.~Knapen, P.~Longhi and M.~Strassler, \emph{{The Vector-Like Twin
  Higgs}}, \href{https://doi.org/10.1007/JHEP07(2016)002}{\emph{JHEP}
  {\bfseries 07} (2016) 002}
  [\href{https://arxiv.org/abs/1601.07181}{{\ttfamily 1601.07181}}].

\bibitem{Barbieri:2016zxn}
R.~Barbieri, L.J.~Hall and K.~Harigaya, \emph{{Minimal Mirror Twin Higgs}},
  \href{https://doi.org/10.1007/JHEP11(2016)172}{\emph{JHEP} {\bfseries 11}
  (2016) 172} [\href{https://arxiv.org/abs/1609.05589}{{\ttfamily
  1609.05589}}].

\bibitem{Katz:2016wtw}
A.~Katz, A.~Mariotti, S.~Pokorski, D.~Redigolo and R.~Ziegler, \emph{{SUSY
  Meets Her Twin}}, \href{https://doi.org/10.1007/JHEP01(2017)142}{\emph{JHEP}
  {\bfseries 01} (2017) 142}
  [\href{https://arxiv.org/abs/1611.08615}{{\ttfamily 1611.08615}}].

\bibitem{Burdman:2018ehe}
G.~Burdman and G.~Lichtenstein, \emph{{Displaced Vertices from Hidden Glue}},
  \href{https://doi.org/10.1007/JHEP08(2018)146}{\emph{JHEP} {\bfseries 08}
  (2018) 146} [\href{https://arxiv.org/abs/1807.03801}{{\ttfamily
  1807.03801}}].

\bibitem{Curtin:2022tou}
D.~Curtin, C.~Gemmell and C.B.~Verhaaren, \emph{{Simulating Glueball Production
  in $N_f = 0$ QCD}},  \href{https://arxiv.org/abs/2202.12899}{{\ttfamily
  2202.12899}}.

\bibitem{Chacko:2015fbc}
Z.~Chacko, D.~Curtin and C.B.~Verhaaren, \emph{{A Quirky Probe of Neutral
  Naturalness}}, \href{https://doi.org/10.1103/PhysRevD.94.011504}{\emph{Phys.
  Rev. D} {\bfseries 94} (2016) 011504}
  [\href{https://arxiv.org/abs/1512.05782}{{\ttfamily 1512.05782}}].

\bibitem{Graham:2015cka}
P.W.~Graham, D.E.~Kaplan and S.~Rajendran, \emph{{Cosmological Relaxation of
  the Electroweak Scale}},
  \href{https://doi.org/10.1103/PhysRevLett.115.221801}{\emph{Phys. Rev. Lett.}
  {\bfseries 115} (2015) 221801}
  [\href{https://arxiv.org/abs/1504.07551}{{\ttfamily 1504.07551}}].

\bibitem{Harlow:2022gzl}
D.~Harlow, B.~Heidenreich, M.~Reece and T.~Rudelius, \emph{{The Weak Gravity
  Conjecture: A Review}},  \href{https://arxiv.org/abs/2201.08380}{{\ttfamily
  2201.08380}}.

\bibitem{Arkani-Hamed:2016rle}
N.~Arkani-Hamed, T.~Cohen, R.T.~D'Agnolo, A.~Hook, H.D.~Kim and D.~Pinner,
  \emph{{Solving the Hierarchy Problem at Reheating with a Large Number of
  Degrees of Freedom}},
  \href{https://doi.org/10.1103/PhysRevLett.117.251801}{\emph{Phys. Rev. Lett.}
  {\bfseries 117} (2016) 251801}
  [\href{https://arxiv.org/abs/1607.06821}{{\ttfamily 1607.06821}}].

\bibitem{Arkani-Hamed:2020yna}
N.~Arkani-Hamed, R.T.~{ D'Agnolo} and H.D.~Kim, \emph{{The Weak Scale as a
  Trigger}},  \href{https://arxiv.org/abs/2012.04652}{{\ttfamily 2012.04652}}.

\bibitem{Beauchesne:2017ukw}
H.~Beauchesne, E.~Bertuzzo and G.~Grilli~di Cortona, \emph{{Constraints on the
  relaxion mechanism with strongly interacting vector-fermions}},
  \href{https://doi.org/10.1007/JHEP08(2017)093}{\emph{JHEP} {\bfseries 08}
  (2017) 093} [\href{https://arxiv.org/abs/1705.06325}{{\ttfamily
  1705.06325}}].

\bibitem{Peccei:1977ur}
R.D.~Peccei and H.R.~Quinn, \emph{{Constraints Imposed by CP Conservation in
  the Presence of Instantons}},
  \href{https://doi.org/10.1103/PhysRevD.16.1791}{\emph{Phys. Rev. D}
  {\bfseries 16} (1977) 1791}.

\bibitem{Peccei:1977hh}
R.D.~Peccei and H.R.~Quinn, \emph{{CP Conservation in the Presence of
  Instantons}}, \href{https://doi.org/10.1103/PhysRevLett.38.1440}{\emph{Phys.
  Rev. Lett.} {\bfseries 38} (1977) 1440}.

\bibitem{Weinberg:1977ma}
S.~Weinberg, \emph{{A New Light Boson?}},
  \href{https://doi.org/10.1103/PhysRevLett.40.223}{\emph{Phys. Rev. Lett.}
  {\bfseries 40} (1978) 223}.

\bibitem{Wilczek:1977pj}
F.~Wilczek, \emph{{Problem of Strong $P$ and $T$ Invariance in the Presence of
  Instantons}}, \href{https://doi.org/10.1103/PhysRevLett.40.279}{\emph{Phys.
  Rev. Lett.} {\bfseries 40} (1978) 279}.

\bibitem{Kim:1979if}
J.E.~Kim, \emph{{Weak Interaction Singlet and Strong CP Invariance}},
  \href{https://doi.org/10.1103/PhysRevLett.43.103}{\emph{Phys. Rev. Lett.}
  {\bfseries 43} (1979) 103}.

\bibitem{Shifman:1979if}
M.A.~Shifman, A.I.~Vainshtein and V.I.~Zakharov, \emph{{Can Confinement Ensure
  Natural CP Invariance of Strong Interactions?}},
  \href{https://doi.org/10.1016/0550-3213(80)90209-6}{\emph{Nucl. Phys. B}
  {\bfseries 166} (1980) 493}.

\bibitem{Zhitnitsky:1980tq}
A.R.~Zhitnitsky, \emph{{On Possible Suppression of the Axion Hadron
  Interactions. (In Russian)}}, {\emph{Sov. J. Nucl. Phys.} {\bfseries 31}
  (1980) 260}.

\bibitem{Dine:1981rt}
M.~Dine, W.~Fischler and M.~Srednicki, \emph{{A Simple Solution to the Strong
  CP Problem with a Harmless Axion}},
  \href{https://doi.org/10.1016/0370-2693(81)90590-6}{\emph{Phys. Lett. B}
  {\bfseries 104} (1981) 199}.

\bibitem{ellisgaillard}
J.R.~Ellis and M.K.~Gaillard, \emph{{Strong and Weak CP Violation}},
  \href{https://doi.org/10.1016/0550-3213(79)90297-9}{\emph{Nucl. Phys.}
  {\bfseries B150} (1979) 141}.

\bibitem{nelsoncp}
A.E.~Nelson, \emph{{Naturally Weak CP Violation}},
  \href{https://doi.org/10.1016/0370-2693(84)92025-2}{\emph{Phys.Lett.}
  {\bfseries B136} (1984) 387}.

\bibitem{barrcp}
S.M.~Barr, \emph{{Solving the Strong CP Problem Without the Peccei-Quinn
  Symmetry}},
  \href{https://doi.org/10.1103/PhysRevLett.53.329}{\emph{Phys.Rev.Lett.}
  {\bfseries 53} (1984) 329}.

\bibitem{barrcp2}
S.M.~Barr, \emph{{A Natural Class of Nonpeccei-quinn Models}},
  \href{https://doi.org/10.1103/PhysRevD.30.1805}{\emph{Phys.Rev.} {\bfseries
  D30} (1984) 1805}.

\bibitem{Beg:1978mt}
M.~Beg and H.-S.~Tsao, \emph{{Strong P, T Noninvariances in a Superweak
  Theory}},
  \href{https://doi.org/10.1103/PhysRevLett.41.278}{\emph{Phys.Rev.Lett.}
  {\bfseries 41} (1978) 278}.

\bibitem{mohapatrasenjanovic}
R.N.~Mohapatra and G.~Senjanovic, \emph{{Natural Suppression of Strong p and t
  Noninvariance}},
  \href{https://doi.org/10.1016/0370-2693(78)90243-5}{\emph{Phys.Lett.}
  {\bfseries B79} (1978) 283}.

\bibitem{Georgi:1978xz}
H.~Georgi, \emph{{A Model of Soft CP Violation}}, {\emph{Hadronic J.}
  {\bfseries 1} (1978) 155}.

\bibitem{Babu:1989rb}
K.S.~Babu and R.N.~Mohapatra, \emph{{A Solution to the Strong {CP} Problem
  Without an Axion}},
  \href{https://doi.org/10.1103/PhysRevD.41.1286}{\emph{Phys. Rev.} {\bfseries
  D41} (1990) 1286}.

\bibitem{barrsenjanovic}
S.M.~Barr, D.~Chang and G.~Senjanovic, \emph{{Strong CP problem and parity}},
  \href{https://doi.org/10.1103/PhysRevLett.67.2765}{\emph{Phys. Rev. Lett.}
  {\bfseries 67} (1991) 2765}.

\bibitem{Lawson:2019brd}
M.~Lawson, A.J.~Millar, M.~Pancaldi, E.~Vitagliano and F.~Wilczek,
  \emph{{Tunable axion plasma haloscopes}},
  \href{https://doi.org/10.1103/PhysRevLett.123.141802}{\emph{Phys. Rev. Lett.}
  {\bfseries 123} (2019) 141802}
  [\href{https://arxiv.org/abs/1904.11872}{{\ttfamily 1904.11872}}].

\bibitem{BRASS}
``Brass.''
  \url{https://www1.physik.uni-hamburg.de/iexp/gruppe-horns/forschung/brass.html}.

\bibitem{Liu:2021pei}
J.~Liu et~al., \emph{{Broadband solenoidal haloscope for terahertz axion
  detection}},  \href{https://arxiv.org/abs/2111.12103}{{\ttfamily
  2111.12103}}.

\bibitem{DMRadio}
``Dmradio.''
  \url{https://indico.mit.edu/event/151/contributions/295/attachments/96/172/Dark%20Matter%20Radio_CambridgeAxions2021.pdf}.

\bibitem{Michimura:2019qxr}
Y.~Michimura, Y.~Oshima, T.~Watanabe, T.~Kawasaki, H.~Takeda, M.~Ando et~al.,
  \emph{{DANCE: Dark matter Axion search with riNg Cavity Experiment}},
  \href{https://doi.org/10.1088/1742-6596/1468/1/012032}{\emph{J. Phys. Conf.
  Ser.} {\bfseries 1468} (2020) 012032}
  [\href{https://arxiv.org/abs/1911.05196}{{\ttfamily 1911.05196}}].

\bibitem{Baryakhtar:2018doz}
M.~Baryakhtar, J.~Huang and R.~Lasenby, \emph{{Axion and hidden photon dark
  matter detection with multilayer optical haloscopes}},
  \href{https://doi.org/10.1103/PhysRevD.98.035006}{\emph{Phys. Rev. D}
  {\bfseries 98} (2018) 035006}
  [\href{https://arxiv.org/abs/1803.11455}{{\ttfamily 1803.11455}}].

\bibitem{Beurthey:2020yuq}
S.~Beurthey et~al., \emph{{MADMAX Status Report}},
  \href{https://arxiv.org/abs/2003.10894}{{\ttfamily 2003.10894}}.

\bibitem{Alesini:2017ifp}
D.~Alesini, D.~Babusci, D.~Di~Gioacchino, C.~Gatti, G.~Lamanna and C.~Ligi,
  \emph{{The KLASH Proposal}},
  \href{https://arxiv.org/abs/1707.06010}{{\ttfamily 1707.06010}}.

\bibitem{McAllister:2017lkb}
B.T.~McAllister, G.~Flower, E.N.~Ivanov, M.~Goryachev, J.~Bourhill and
  M.E.~Tobar, \emph{{The ORGAN Experiment: An axion haloscope above 15 GHz}},
  \href{https://doi.org/10.1016/j.dark.2017.09.010}{\emph{Phys. Dark Univ.}
  {\bfseries 18} (2017) 67} [\href{https://arxiv.org/abs/1706.00209}{{\ttfamily
  1706.00209}}].

\bibitem{Schutte-Engel:2021bqm}
J.~Sch\"utte-Engel, D.J.E.~Marsh, A.J.~Millar, A.~Sekine, F.~Chadha-Day,
  S.~Hoof et~al., \emph{{Axion quasiparticles for axion dark matter
  detection}}, \href{https://doi.org/10.1088/1475-7516/2021/08/066}{\emph{JCAP}
  {\bfseries 08} (2021) 066}
  [\href{https://arxiv.org/abs/2102.05366}{{\ttfamily 2102.05366}}].

\bibitem{Zhang:2021bpa}
Z.~Zhang, O.~Ghosh and D.~Horns, \emph{{WISPLC: Search for Dark Matter with LC
  Circuit}},  \href{https://arxiv.org/abs/2111.04541}{{\ttfamily 2111.04541}}.

\bibitem{Berlin:2020vrk}
A.~Berlin, R.T.~D'Agnolo, S.A.R.~Ellis and K.~Zhou, \emph{{Heterodyne broadband
  detection of axion dark matter}},
  \href{https://doi.org/10.1103/PhysRevD.104.L111701}{\emph{Phys. Rev. D}
  {\bfseries 104} (2021) L111701}
  [\href{https://arxiv.org/abs/2007.15656}{{\ttfamily 2007.15656}}].

\bibitem{Meyer:2016wrm}
M.~Meyer, M.~Giannotti, A.~Mirizzi, J.~Conrad and M.A.~S\'anchez-Conde,
  \emph{{Fermi Large Area Telescope as a Galactic Supernovae Axionscope}},
  \href{https://doi.org/10.1103/PhysRevLett.118.011103}{\emph{Phys. Rev. Lett.}
  {\bfseries 118} (2017) 011103}
  [\href{https://arxiv.org/abs/1609.02350}{{\ttfamily 1609.02350}}].

\bibitem{Thorpe-Morgan:2020rwc}
C.~Thorpe-Morgan, D.~Malyshev, A.~Santangelo, J.~Jochum, B.~J\"ager, M.~Sasaki
  et~al., \emph{{THESEUS insights into axionlike particles, dark photon, and
  sterile neutrino dark matter}},
  \href{https://doi.org/10.1103/PhysRevD.102.123003}{\emph{Phys. Rev. D}
  {\bfseries 102} (2020) 123003}
  [\href{https://arxiv.org/abs/2008.08306}{{\ttfamily 2008.08306}}].

\bibitem{Dekker:2021bos}
A.~Dekker, E.~Peerbooms, F.~Zimmer, K.C.Y.~Ng and S.~Ando, \emph{{Searches for
  sterile neutrinos and axionlike particles from the Galactic halo with
  eROSITA}}, \href{https://doi.org/10.1103/PhysRevD.104.023021}{\emph{Phys.
  Rev. D} {\bfseries 104} (2021) 023021}
  [\href{https://arxiv.org/abs/2103.13241}{{\ttfamily 2103.13241}}].

\bibitem{Dolan:2021rya}
M.J.~Dolan, F.J.~Hiskens and R.R.~Volkas, \emph{{Constraining axion-like
  particles using the white dwarf initial-final mass relation}},
  \href{https://doi.org/10.1088/1475-7516/2021/09/010}{\emph{JCAP} {\bfseries
  09} (2021) 010} [\href{https://arxiv.org/abs/2102.00379}{{\ttfamily
  2102.00379}}].

\bibitem{Foster:2021ngm}
J.W.~Foster, M.~Kongsore, C.~Dessert, Y.~Park, N.L.~Rodd, K.~Cranmer et~al.,
  \emph{{Deep Search for Decaying Dark Matter with XMM-Newton Blank-Sky
  Observations}},
  \href{https://doi.org/10.1103/PhysRevLett.127.051101}{\emph{Phys. Rev. Lett.}
  {\bfseries 127} (2021) 051101}
  [\href{https://arxiv.org/abs/2102.02207}{{\ttfamily 2102.02207}}].

\bibitem{Cadamuro:2011fd}
D.~Cadamuro and J.~Redondo, \emph{{Cosmological bounds on pseudo
  Nambu-Goldstone bosons}},
  \href{https://doi.org/10.1088/1475-7516/2012/02/032}{\emph{JCAP} {\bfseries
  02} (2012) 032} [\href{https://arxiv.org/abs/1110.2895}{{\ttfamily
  1110.2895}}].

\bibitem{Depta:2020wmr}
P.F.~Depta, M.~Hufnagel and K.~Schmidt-Hoberg, \emph{{Robust cosmological
  constraints on axion-like particles}},
  \href{https://doi.org/10.1088/1475-7516/2020/05/009}{\emph{JCAP} {\bfseries
  05} (2020) 009} [\href{https://arxiv.org/abs/2002.08370}{{\ttfamily
  2002.08370}}].

\bibitem{Ortiz:2020tgs}
M.D.~Ortiz et~al., \emph{{Design of the ALPS II Optical System}},
  \href{https://arxiv.org/abs/2009.14294}{{\ttfamily 2009.14294}}.

\bibitem{Shilon:2013xma}
I.~Shilon, A.~Dudarev, H.~Silva, U.~Wagner and H.H.J.~ten Kate, \emph{{The
  Superconducting Toroid for the New International AXion Observatory (IAXO)}},
  \href{https://doi.org/10.1109/TASC.2013.2280654}{\emph{IEEE Trans. Appl.
  Supercond.} {\bfseries 24} (2014) 4500104}
  [\href{https://arxiv.org/abs/1309.2117}{{\ttfamily 1309.2117}}].

\bibitem{Ge:2020zww}
S.-F.~Ge, K.~Hamaguchi, K.~Ichimura, K.~Ishidoshiro, Y.~Kanazawa, Y.~Kishimoto
  et~al., \emph{{Supernova-scope for the Direct Search of Supernova Axions}},
  \href{https://doi.org/10.1088/1475-7516/2020/11/059}{\emph{JCAP} {\bfseries
  11} (2020) 059} [\href{https://arxiv.org/abs/2008.03924}{{\ttfamily
  2008.03924}}].

\bibitem{Baer:2018avn}
H.~Baer, V.~Barger and D.~Sengupta, \emph{{Gravity safe, electroweak natural
  axionic solution to strong $CP$ and SUSY $\mu$ problems}},
  \href{https://doi.org/10.1016/j.physletb.2019.01.007}{\emph{Phys. Lett. B}
  {\bfseries 790} (2019) 58}
  [\href{https://arxiv.org/abs/1810.03713}{{\ttfamily 1810.03713}}].

\bibitem{DiLuzio:2020wdo}
L.~Di~Luzio, M.~Giannotti, E.~Nardi and L.~Visinelli, \emph{{The landscape of
  QCD axion models}},
  \href{https://doi.org/10.1016/j.physrep.2020.06.002}{\emph{Phys. Rept.}
  {\bfseries 870} (2020) 1} [\href{https://arxiv.org/abs/2003.01100}{{\ttfamily
  2003.01100}}].

\bibitem{Choi:2020rgn}
K.~Choi, S.H.~Im and C.~Sub~Shin, \emph{{Recent Progress in the Physics of
  Axions and Axion-Like Particles}},
  \href{https://doi.org/10.1146/annurev-nucl-120720-031147}{\emph{Ann. Rev.
  Nucl. Part. Sci.} {\bfseries 71} (2021) 225}
  [\href{https://arxiv.org/abs/2012.05029}{{\ttfamily 2012.05029}}].

\bibitem{Freese:1990rb}
K.~Freese, J.A.~Frieman and A.V.~Olinto, \emph{{Natural inflation with pseudo -
  Nambu-Goldstone bosons}},
  \href{https://doi.org/10.1103/PhysRevLett.65.3233}{\emph{Phys.Rev.Lett.}
  {\bfseries 65} (1990) 3233}.

\bibitem{Banks:2003sx}
T.~Banks, M.~Dine, P.J.~Fox and E.~Gorbatov, \emph{{On the possibility of large
  axion decay constants}},
  \href{https://doi.org/10.1088/1475-7516/2003/06/001}{\emph{JCAP} {\bfseries
  0306} (2003) 001} [\href{https://arxiv.org/abs/hep-th/0303252}{{\ttfamily
  hep-th/0303252}}].

\bibitem{ArkaniHamed:2006dz}
N.~Arkani-Hamed, L.~Motl, A.~Nicolis and C.~Vafa, \emph{{The String landscape,
  black holes and gravity as the weakest force}},
  \href{https://doi.org/10.1088/1126-6708/2007/06/060}{\emph{JHEP} {\bfseries
  0706} (2007) 060} [\href{https://arxiv.org/abs/hep-th/0601001}{{\ttfamily
  hep-th/0601001}}].

\bibitem{Liddle:1998jc}
A.R.~Liddle, A.~Mazumdar and F.E.~Schunck, \emph{{Assisted inflation}},
  \href{https://doi.org/10.1103/PhysRevD.58.061301}{\emph{Phys. Rev.}
  {\bfseries D58} (1998) 061301}
  [\href{https://arxiv.org/abs/astro-ph/9804177}{{\ttfamily
  astro-ph/9804177}}].

\bibitem{Dimopoulos:2005ac}
S.~Dimopoulos, S.~Kachru, J.~McGreevy and J.G.~Wacker, \emph{{N-flation}},
  \href{https://doi.org/10.1088/1475-7516/2008/08/003}{\emph{JCAP} {\bfseries
  0808} (2008) 003} [\href{https://arxiv.org/abs/hep-th/0507205}{{\ttfamily
  hep-th/0507205}}].

\bibitem{Kim:2004rp}
J.E.~Kim, H.P.~Nilles and M.~Peloso, \emph{{Completing natural inflation}},
  \href{https://doi.org/10.1088/1475-7516/2005/01/005}{\emph{JCAP} {\bfseries
  01} (2005) 005} [\href{https://arxiv.org/abs/hep-ph/0409138}{{\ttfamily
  hep-ph/0409138}}].

\bibitem{Rudelius:2014wla}
T.~Rudelius, \emph{{On the Possibility of Large Axion Moduli Spaces}},
  \href{https://doi.org/10.1088/1475-7516/2015/04/049}{\emph{JCAP} {\bfseries
  1504} (2015) 049} [\href{https://arxiv.org/abs/1409.5793}{{\ttfamily
  1409.5793}}].

\bibitem{Rudelius:2015xta}
T.~Rudelius, \emph{{Constraints on Axion Inflation from the Weak Gravity
  Conjecture}}, \href{https://doi.org/10.1088/1475-7516/2015/9/020}{\emph{JCAP}
  {\bfseries 09} (2015) 020}
  [\href{https://arxiv.org/abs/1503.00795}{{\ttfamily 1503.00795}}].

\bibitem{Montero:2015ofa}
M.~Montero, A.M.~Uranga and I.~Valenzuela, \emph{{Transplanckian axions!?}},
  \href{https://doi.org/10.1007/JHEP08(2015)032}{\emph{JHEP} {\bfseries 08}
  (2015) 032} [\href{https://arxiv.org/abs/1503.03886}{{\ttfamily
  1503.03886}}].

\bibitem{Heidenreich:2015wga}
B.~Heidenreich, M.~Reece and T.~Rudelius, \emph{{Weak Gravity Strongly
  Constrains Large-Field Axion Inflation}},
  \href{https://doi.org/10.1007/JHEP12(2015)108}{\emph{JHEP} {\bfseries 12}
  (2015) 108} [\href{https://arxiv.org/abs/1506.03447}{{\ttfamily
  1506.03447}}].

\bibitem{Brown:2015iha}
J.~Brown, W.~Cottrell, G.~Shiu and P.~Soler, \emph{{Fencing in the Swampland:
  Quantum Gravity Constraints on Large Field Inflation}},
  \href{https://doi.org/10.1007/JHEP10(2015)023}{\emph{JHEP} {\bfseries 10}
  (2015) 023} [\href{https://arxiv.org/abs/1503.04783}{{\ttfamily
  1503.04783}}].

\bibitem{Heidenreich:2019bjd}
B.~Heidenreich, C.~Long, L.~McAllister, T.~Rudelius and J.~Stout,
  \emph{{Instanton Resummation and the Weak Gravity Conjecture}},
  \href{https://doi.org/10.1007/JHEP11(2020)166}{\emph{JHEP} {\bfseries 11}
  (2020) 166} [\href{https://arxiv.org/abs/1910.14053}{{\ttfamily
  1910.14053}}].

\bibitem{delaFuente:2014aca}
A.~de~la Fuente, P.~Saraswat and R.~Sundrum, \emph{{Natural Inflation and
  Quantum Gravity}},
  \href{https://doi.org/10.1103/PhysRevLett.114.151303}{\emph{Phys. Rev. Lett.}
  {\bfseries 114} (2015) 151303}
  [\href{https://arxiv.org/abs/1412.3457}{{\ttfamily 1412.3457}}].

\bibitem{Bachlechner:2015qja}
T.C.~Bachlechner, C.~Long and L.~McAllister, \emph{{Planckian Axions and the
  Weak Gravity Conjecture}},
  \href{https://doi.org/10.1007/JHEP01(2016)091}{\emph{JHEP} {\bfseries 01}
  (2016) 091} [\href{https://arxiv.org/abs/1503.07853}{{\ttfamily
  1503.07853}}].

\bibitem{Hebecker:2015rya}
A.~Hebecker, P.~Mangat, F.~Rompineve and L.T.~Witkowski, \emph{{Winding out of
  the Swamp: Evading the Weak Gravity Conjecture with F-term Winding
  Inflation?}},
  \href{https://doi.org/10.1016/j.physletb.2015.07.026}{\emph{Phys. Lett. B}
  {\bfseries 748} (2015) 455}
  [\href{https://arxiv.org/abs/1503.07912}{{\ttfamily 1503.07912}}].

\bibitem{Brown:2015lia}
J.~Brown, W.~Cottrell, G.~Shiu and P.~Soler, \emph{{On Axionic Field Ranges,
  Loopholes and the Weak Gravity Conjecture}},
  \href{https://doi.org/10.1007/JHEP04(2016)017}{\emph{JHEP} {\bfseries 04}
  (2016) 017} [\href{https://arxiv.org/abs/1504.00659}{{\ttfamily
  1504.00659}}].

\bibitem{Conlon:2016aea}
J.P.~Conlon and S.~Krippendorf, \emph{{Axion decay constants away from the
  lamppost}}, \href{https://doi.org/10.1007/JHEP04(2016)085}{\emph{JHEP}
  {\bfseries 04} (2016) 085}
  [\href{https://arxiv.org/abs/1601.00647}{{\ttfamily 1601.00647}}].

\bibitem{Long:2016jvd}
C.~Long, L.~McAllister and J.~Stout, \emph{{Systematics of Axion Inflation in
  Calabi-Yau Hypersurfaces}},
  \href{https://doi.org/10.1007/JHEP02(2017)014}{\emph{JHEP} {\bfseries 02}
  (2017) 014} [\href{https://arxiv.org/abs/1603.01259}{{\ttfamily
  1603.01259}}].

\bibitem{Chun:1992bn}
E.J.~Chun and A.~Lukas, \emph{{Discrete gauge symmetries in axionic extensions
  of the SSM}}, \href{https://doi.org/10.1016/0370-2693(92)91266-C}{\emph{Phys.
  Lett. B} {\bfseries 297} (1992) 298}
  [\href{https://arxiv.org/abs/hep-ph/9209208}{{\ttfamily hep-ph/9209208}}].

\bibitem{Carpenter:2009zs}
L.M.~Carpenter, M.~Dine and G.~Festuccia, \emph{{Dynamics of the Peccei Quinn
  Scale}}, \href{https://doi.org/10.1103/PhysRevD.80.125017}{\emph{Phys. Rev.
  D} {\bfseries 80} (2009) 125017}
  [\href{https://arxiv.org/abs/0906.1273}{{\ttfamily 0906.1273}}].

\bibitem{Harigaya:2013vja}
K.~Harigaya, M.~Ibe, K.~Schmitz and T.T.~Yanagida, \emph{{Peccei-Quinn symmetry
  from a gauged discrete R symmetry}},
  \href{https://doi.org/10.1103/PhysRevD.88.075022}{\emph{Phys. Rev. D}
  {\bfseries 88} (2013) 075022}
  [\href{https://arxiv.org/abs/1308.1227}{{\ttfamily 1308.1227}}].

\bibitem{Chen:2021haa}
N.~Chen, Y.~Liu and Z.~Teng, \emph{{Axion model with the SU(6) unification}},
  \href{https://doi.org/10.1103/PhysRevD.104.115011}{\emph{Phys. Rev. D}
  {\bfseries 104} (2021) 115011}
  [\href{https://arxiv.org/abs/2106.00223}{{\ttfamily 2106.00223}}].

\bibitem{Cheng:2001ys}
H.-C.~Cheng and D.E.~Kaplan, \emph{{Axions and a gauged Peccei-Quinn
  symmetry}},  \href{https://arxiv.org/abs/hep-ph/0103346}{{\ttfamily
  hep-ph/0103346}}.

\bibitem{Hill:2002kq}
C.T.~Hill and A.K.~Leibovich, \emph{{Natural Theories of Ultralow Mass PNGB's:
  Axions and Quintessence}},
  \href{https://doi.org/10.1103/PhysRevD.66.075010}{\emph{Phys. Rev. D}
  {\bfseries 66} (2002) 075010}
  [\href{https://arxiv.org/abs/hep-ph/0205237}{{\ttfamily hep-ph/0205237}}].

\bibitem{Fukuda:2017ylt}
H.~Fukuda, M.~Ibe, M.~Suzuki and T.T.~Yanagida, \emph{{A ''gauged'' $U(1)$
  Peccei\textendash{}Quinn symmetry}},
  \href{https://doi.org/10.1016/j.physletb.2017.05.071}{\emph{Phys. Lett. B}
  {\bfseries 771} (2017) 327}
  [\href{https://arxiv.org/abs/1703.01112}{{\ttfamily 1703.01112}}].

\bibitem{Lee:2018yak}
H.-S.~Lee and W.~Yin, \emph{{Peccei-Quinn symmetry from a hidden gauge group
  structure}}, \href{https://doi.org/10.1103/PhysRevD.99.015041}{\emph{Phys.
  Rev. D} {\bfseries 99} (2019) 015041}
  [\href{https://arxiv.org/abs/1811.04039}{{\ttfamily 1811.04039}}].

\bibitem{Darme:2021cxx}
L.~Darm\'e and E.~Nardi, \emph{{Exact accidental U(1) symmetries for the
  axion}}, \href{https://doi.org/10.1103/PhysRevD.104.055013}{\emph{Phys. Rev.
  D} {\bfseries 104} (2021) 055013}
  [\href{https://arxiv.org/abs/2102.05055}{{\ttfamily 2102.05055}}].

\bibitem{Nakai:2021nyf}
Y.~Nakai and M.~Suzuki, \emph{{Axion Quality from Superconformal Dynamics}},
  \href{https://doi.org/10.1016/j.physletb.2021.136239}{\emph{Phys. Lett. B}
  {\bfseries 816} (2021) 136239}
  [\href{https://arxiv.org/abs/2102.01329}{{\ttfamily 2102.01329}}].

\bibitem{Randall:1992ut}
L.~Randall, \emph{{Composite axion models and Planck scale physics}},
  \href{https://doi.org/10.1016/0370-2693(92)91928-3}{\emph{Phys. Lett. B}
  {\bfseries 284} (1992) 77}.

\bibitem{DiLuzio:2017tjx}
L.~Di~Luzio, E.~Nardi and L.~Ubaldi, \emph{{Accidental Peccei-Quinn symmetry
  protected to arbitrary order}},
  \href{https://doi.org/10.1103/PhysRevLett.119.011801}{\emph{Phys. Rev. Lett.}
  {\bfseries 119} (2017) 011801}
  [\href{https://arxiv.org/abs/1704.01122}{{\ttfamily 1704.01122}}].

\bibitem{Lillard:2017cwx}
B.~Lillard and T.M.P.~Tait, \emph{{A Composite Axion from a Supersymmetric
  Product Group}}, \href{https://doi.org/10.1007/JHEP11(2017)005}{\emph{JHEP}
  {\bfseries 11} (2017) 005}
  [\href{https://arxiv.org/abs/1707.04261}{{\ttfamily 1707.04261}}].

\bibitem{Lillard:2018fdt}
B.~Lillard and T.M.P.~Tait, \emph{{A High Quality Composite Axion}},
  \href{https://doi.org/10.1007/JHEP11(2018)199}{\emph{JHEP} {\bfseries 11}
  (2018) 199} [\href{https://arxiv.org/abs/1811.03089}{{\ttfamily
  1811.03089}}].

\bibitem{Lee:2021slp}
S.J.~Lee, Y.~Nakai and M.~Suzuki, \emph{{High quality axion via a doubly
  composite dynamics}},
  \href{https://doi.org/10.1007/JHEP03(2022)038}{\emph{JHEP} {\bfseries 03}
  (2022) 038} [\href{https://arxiv.org/abs/2112.08083}{{\ttfamily
  2112.08083}}].

\bibitem{Lillard:2021dxd}
B.~Lillard, \emph{{Confinement On the Moose Lattice}},
  \href{https://arxiv.org/abs/2112.13828}{{\ttfamily 2112.13828}}.

\bibitem{Heckman:2019bzm}
J.J.~Heckman and C.~Vafa, \emph{{Fine Tuning, Sequestering, and the
  Swampland}},
  \href{https://doi.org/10.1016/j.physletb.2019.135004}{\emph{Phys. Lett. B}
  {\bfseries 798} (2019) 135004}
  [\href{https://arxiv.org/abs/1905.06342}{{\ttfamily 1905.06342}}].

\bibitem{Heidenreich:2015nta}
B.~Heidenreich, M.~Reece and T.~Rudelius, \emph{{Sharpening the Weak Gravity
  Conjecture with Dimensional Reduction}},
  \href{https://doi.org/10.1007/JHEP02(2016)140}{\emph{JHEP} {\bfseries 02}
  (2016) 140} [\href{https://arxiv.org/abs/1509.06374}{{\ttfamily
  1509.06374}}].

\bibitem{Heidenreich:2016aqi}
B.~Heidenreich, M.~Reece and T.~Rudelius, \emph{{Evidence for a sublattice weak
  gravity conjecture}},
  \href{https://doi.org/10.1007/JHEP08(2017)025}{\emph{JHEP} {\bfseries 08}
  (2017) 025} [\href{https://arxiv.org/abs/1606.08437}{{\ttfamily
  1606.08437}}].

\bibitem{Montero:2016tif}
M.~Montero, G.~Shiu and P.~Soler, \emph{{The Weak Gravity Conjecture in three
  dimensions}}, \href{https://doi.org/10.1007/JHEP10(2016)159}{\emph{JHEP}
  {\bfseries 10} (2016) 159}
  [\href{https://arxiv.org/abs/1606.08438}{{\ttfamily 1606.08438}}].

\bibitem{Andriolo:2018lvp}
S.~Andriolo, D.~Junghans, T.~Noumi and G.~Shiu, \emph{{A Tower Weak Gravity
  Conjecture from Infrared Consistency}},
  \href{https://doi.org/10.1002/prop.201800020}{\emph{Fortsch. Phys.}
  {\bfseries 66} (2018) 1800020}
  [\href{https://arxiv.org/abs/1802.04287}{{\ttfamily 1802.04287}}].

\bibitem{Veneziano:2001ah}
G.~Veneziano, \emph{{Large N bounds on, and compositeness limit of, gauge and
  gravitational interactions}},
  \href{https://doi.org/10.1088/1126-6708/2002/06/051}{\emph{JHEP} {\bfseries
  06} (2002) 051} [\href{https://arxiv.org/abs/hep-th/0110129}{{\ttfamily
  hep-th/0110129}}].

\bibitem{Dvali:2007hz}
G.~Dvali, \emph{{Black Holes and Large N Species Solution to the Hierarchy
  Problem}}, \href{https://doi.org/10.1002/prop.201000009}{\emph{Fortsch.
  Phys.} {\bfseries 58} (2010) 528}
  [\href{https://arxiv.org/abs/0706.2050}{{\ttfamily 0706.2050}}].

\bibitem{Heidenreich:2017sim}
B.~Heidenreich, M.~Reece and T.~Rudelius, \emph{{The Weak Gravity Conjecture
  and Emergence from an Ultraviolet Cutoff}},
  \href{https://doi.org/10.1140/epjc/s10052-018-5811-3}{\emph{Eur. Phys. J. C}
  {\bfseries 78} (2018) 337}
  [\href{https://arxiv.org/abs/1712.01868}{{\ttfamily 1712.01868}}].

\bibitem{Craig:2019zkf}
N.~Craig, I.~Garcia~Garcia and G.D.~Kribs, \emph{{The UV fate of anomalous
  U(1)s and the Swampland}},
  \href{https://doi.org/10.1007/JHEP11(2020)063}{\emph{JHEP} {\bfseries 11}
  (2020) 063} [\href{https://arxiv.org/abs/1912.10054}{{\ttfamily
  1912.10054}}].

\bibitem{Griest:1990kh}
K.~Griest and D.~Seckel, \emph{{Three exceptions in the calculation of relic
  abundances}}, \href{https://doi.org/10.1103/PhysRevD.43.3191}{\emph{Phys.
  Rev.} {\bfseries D43} (1991) 3191}.

\bibitem{Pospelov:2007mp}
M.~Pospelov, A.~Ritz and M.B.~Voloshin, \emph{{Secluded WIMP Dark Matter}},
  \href{https://doi.org/10.1016/j.physletb.2008.02.052}{\emph{Phys. Lett.}
  {\bfseries B662} (2008) 53}
  [\href{https://arxiv.org/abs/0711.4866}{{\ttfamily 0711.4866}}].

\bibitem{D'Agnolo:2015koa}
R.T.~D'Agnolo and J.T.~Ruderman, \emph{{Light Dark Matter from Forbidden
  Channels}}, \href{https://doi.org/10.1103/PhysRevLett.115.061301}{\emph{Phys.
  Rev. Lett.} {\bfseries 115} (2015) 061301}
  [\href{https://arxiv.org/abs/1505.07107}{{\ttfamily 1505.07107}}].

\bibitem{Carlson:1992fn}
E.D.~Carlson, M.E.~Machacek and L.J.~Hall, \emph{{Self-interacting dark
  matter}}, \href{https://doi.org/10.1086/171833}{\emph{Astrophys. J.}
  {\bfseries 398} (1992) 43}.

\bibitem{Hochberg:2014dra}
Y.~Hochberg, E.~Kuflik, T.~Volansky and J.G.~Wacker, \emph{{Mechanism for
  Thermal Relic Dark Matter of Strongly Interacting Massive Particles}},
  \href{https://doi.org/10.1103/PhysRevLett.113.171301}{\emph{Phys. Rev. Lett.}
  {\bfseries 113} (2014) 171301}
  [\href{https://arxiv.org/abs/1402.5143}{{\ttfamily 1402.5143}}].

\bibitem{Smirnov:2020zwf}
J.~Smirnov and J.F.~Beacom, \emph{{New Freezeout Mechanism for Strongly
  Interacting Dark Matter}},
  \href{https://doi.org/10.1103/PhysRevLett.125.131301}{\emph{Phys. Rev. Lett.}
  {\bfseries 125} (2020) 131301}
  [\href{https://arxiv.org/abs/2002.04038}{{\ttfamily 2002.04038}}].

\bibitem{Kuflik:2015isi}
E.~Kuflik, M.~Perelstein, N.R.-L.~Lorier and Y.-D.~Tsai, \emph{{Elastically
  Decoupling Dark Matter}},
  \href{https://doi.org/10.1103/PhysRevLett.116.221302}{\emph{Phys. Rev. Lett.}
  {\bfseries 116} (2016) 221302}
  [\href{https://arxiv.org/abs/1512.04545}{{\ttfamily 1512.04545}}].

\bibitem{Hall:2009bx}
L.J.~Hall, K.~Jedamzik, J.~March-Russell and S.M.~West, \emph{{Freeze-In
  Production of FIMP Dark Matter}},
  \href{https://doi.org/10.1007/JHEP03(2010)080}{\emph{JHEP} {\bfseries 03}
  (2010) 080} [\href{https://arxiv.org/abs/0911.1120}{{\ttfamily 0911.1120}}].

\bibitem{Bernal:2017kxu}
N.~Bernal, M.~Heikinheimo, T.~Tenkanen, K.~Tuominen and V.~Vaskonen, \emph{{The
  Dawn of FIMP Dark Matter: A Review of Models and Constraints}},
  \href{https://doi.org/10.1142/S0217751X1730023X}{\emph{Int. J. Mod. Phys. A}
  {\bfseries 32} (2017) 1730023}
  [\href{https://arxiv.org/abs/1706.07442}{{\ttfamily 1706.07442}}].

\bibitem{Chu:2013jja}
X.~Chu, Y.~Mambrini, J.~Quevillon and B.~Zaldivar, \emph{{Thermal and
  non-thermal production of dark matter via Z'-portal(s)}},
  \href{https://doi.org/10.1088/1475-7516/2014/01/034}{\emph{JCAP} {\bfseries
  01} (2014) 034} [\href{https://arxiv.org/abs/1306.4677}{{\ttfamily
  1306.4677}}].

\bibitem{Dodelson:1993je}
S.~Dodelson and L.M.~Widrow, \emph{{Sterile-neutrinos as dark matter}},
  \href{https://doi.org/10.1103/PhysRevLett.72.17}{\emph{Phys. Rev. Lett.}
  {\bfseries 72} (1994) 17}
  [\href{https://arxiv.org/abs/hep-ph/9303287}{{\ttfamily hep-ph/9303287}}].

\bibitem{Kusenko:2006rh}
A.~Kusenko, \emph{{Sterile neutrinos, dark matter, and the pulsar velocities in
  models with a Higgs singlet}},
  \href{https://doi.org/10.1103/PhysRevLett.97.241301}{\emph{Phys. Rev. Lett.}
  {\bfseries 97} (2006) 241301}
  [\href{https://arxiv.org/abs/hep-ph/0609081}{{\ttfamily hep-ph/0609081}}].

\bibitem{Petraki:2007gq}
K.~Petraki and A.~Kusenko, \emph{{Dark-matter sterile neutrinos in models with
  a gauge singlet in the Higgs sector}},
  \href{https://doi.org/10.1103/PhysRevD.77.065014}{\emph{Phys. Rev. D}
  {\bfseries 77} (2008) 065014}
  [\href{https://arxiv.org/abs/0711.4646}{{\ttfamily 0711.4646}}].

\bibitem{Shakya:2015xnx}
B.~Shakya, \emph{{Sterile Neutrino Dark Matter from Freeze-In}},
  \href{https://doi.org/10.1142/S0217732316300056}{\emph{Mod. Phys. Lett. A}
  {\bfseries 31} (2016) 1630005}
  [\href{https://arxiv.org/abs/1512.02751}{{\ttfamily 1512.02751}}].

\bibitem{McDonald:2001vt}
J.~McDonald, \emph{{Thermally generated gauge singlet scalars as
  selfinteracting dark matter}},
  \href{https://doi.org/10.1103/PhysRevLett.88.091304}{\emph{Phys. Rev. Lett.}
  {\bfseries 88} (2002) 091304}
  [\href{https://arxiv.org/abs/hep-ph/0106249}{{\ttfamily hep-ph/0106249}}].

\bibitem{Asaka:2005cn}
T.~Asaka, K.~Ishiwata and T.~Moroi, \emph{{Right-handed sneutrino as cold dark
  matter}}, \href{https://doi.org/10.1103/PhysRevD.73.051301}{\emph{Phys. Rev.
  D} {\bfseries 73} (2006) 051301}
  [\href{https://arxiv.org/abs/hep-ph/0512118}{{\ttfamily hep-ph/0512118}}].

\bibitem{Asaka:2006fs}
T.~Asaka, K.~Ishiwata and T.~Moroi, \emph{{Right-handed sneutrino as cold dark
  matter of the universe}},
  \href{https://doi.org/10.1103/PhysRevD.75.065001}{\emph{Phys. Rev. D}
  {\bfseries 75} (2007) 065001}
  [\href{https://arxiv.org/abs/hep-ph/0612211}{{\ttfamily hep-ph/0612211}}].

\bibitem{Gopalakrishna:2006kr}
S.~Gopalakrishna, A.~de~Gouvea and W.~Porod, \emph{{Right-handed sneutrinos as
  nonthermal dark matter}},
  \href{https://doi.org/10.1088/1475-7516/2006/05/005}{\emph{JCAP} {\bfseries
  05} (2006) 005} [\href{https://arxiv.org/abs/hep-ph/0602027}{{\ttfamily
  hep-ph/0602027}}].

\bibitem{Page:2007sh}
V.~Page, \emph{{Non-thermal right-handed sneutrino dark matter and the
  Omega(DM)/Omega(b) problem}},
  \href{https://doi.org/10.1088/1126-6708/2007/04/021}{\emph{JHEP} {\bfseries
  04} (2007) 021} [\href{https://arxiv.org/abs/hep-ph/0701266}{{\ttfamily
  hep-ph/0701266}}].

\bibitem{Covi:2002vw}
L.~Covi, L.~Roszkowski and M.~Small, \emph{{Effects of squark processes on the
  axino CDM abundance}},
  \href{https://doi.org/10.1088/1126-6708/2002/07/023}{\emph{JHEP} {\bfseries
  07} (2002) 023} [\href{https://arxiv.org/abs/hep-ph/0206119}{{\ttfamily
  hep-ph/0206119}}].

\bibitem{Cheung:2011mg}
C.~Cheung, G.~Elor and L.J.~Hall, \emph{{The Cosmological Axino Problem}},
  \href{https://doi.org/10.1103/PhysRevD.85.015008}{\emph{Phys. Rev. D}
  {\bfseries 85} (2012) 015008}
  [\href{https://arxiv.org/abs/1104.0692}{{\ttfamily 1104.0692}}].

\bibitem{Bae:2014rfa}
K.J.~Bae, H.~Baer, A.~Lessa and H.~Serce, \emph{{Coupled Boltzmann computation
  of mixed axion neutralino dark matter in the SUSY DFSZ axion model}},
  \href{https://doi.org/10.1088/1475-7516/2014/10/082}{\emph{JCAP} {\bfseries
  10} (2014) 082} [\href{https://arxiv.org/abs/1406.4138}{{\ttfamily
  1406.4138}}].

\bibitem{Co:2015pka}
R.T.~Co, F.~D'Eramo, L.J.~Hall and D.~Pappadopulo, \emph{{Freeze-In Dark Matter
  with Displaced Signatures at Colliders}},
  \href{https://doi.org/10.1088/1475-7516/2015/12/024}{\emph{JCAP} {\bfseries
  12} (2015) 024} [\href{https://arxiv.org/abs/1506.07532}{{\ttfamily
  1506.07532}}].

\bibitem{Moroi:1993mb}
T.~Moroi, H.~Murayama and M.~Yamaguchi, \emph{{Cosmological constraints on the
  light stable gravitino}},
  \href{https://doi.org/10.1016/0370-2693(93)91434-O}{\emph{Phys. Lett.}
  {\bfseries B303} (1993) 289}.

\bibitem{Choi:2005vq}
K.-Y.~Choi and L.~Roszkowski, \emph{{E-WIMPs}},
  \href{https://doi.org/10.1063/1.2149672}{\emph{AIP Conf. Proc.} {\bfseries
  805} (2005) 30} [\href{https://arxiv.org/abs/hep-ph/0511003}{{\ttfamily
  hep-ph/0511003}}].

\bibitem{Cheung:2011nn}
C.~Cheung, G.~Elor and L.~Hall, \emph{{Gravitino Freeze-In}},
  \href{https://doi.org/10.1103/PhysRevD.84.115021}{\emph{Phys. Rev.}
  {\bfseries D84} (2011) 115021}
  [\href{https://arxiv.org/abs/1103.4394}{{\ttfamily 1103.4394}}].

\bibitem{Hall:2012zp}
L.J.~Hall, Y.~Nomura and S.~Shirai, \emph{{Spread Supersymmetry with Wino LSP:
  Gluino and Dark Matter Signals}},
  \href{https://doi.org/10.1007/JHEP01(2013)036}{\emph{JHEP} {\bfseries 01}
  (2013) 036} [\href{https://arxiv.org/abs/1210.2395}{{\ttfamily 1210.2395}}].

\bibitem{Co:2016fln}
R.T.~Co, F.~D'Eramo and L.J.~Hall, \emph{{Gravitino or Axino Dark Matter with
  Reheat Temperature as high as $10^{16}$ GeV}},
  \href{https://doi.org/10.1007/JHEP03(2017)005}{\emph{JHEP} {\bfseries 03}
  (2017) 005} [\href{https://arxiv.org/abs/1611.05028}{{\ttfamily
  1611.05028}}].

\bibitem{Benakli:2017whb}
K.~Benakli, Y.~Chen, E.~Dudas and Y.~Mambrini, \emph{{Minimal model of
  gravitino dark matter}},
  \href{https://doi.org/10.1103/PhysRevD.95.095002}{\emph{Phys. Rev. D}
  {\bfseries 95} (2017) 095002}
  [\href{https://arxiv.org/abs/1701.06574}{{\ttfamily 1701.06574}}].

\bibitem{Monteux:2015qqa}
A.~Monteux and C.S.~Shin, \emph{{Thermal Goldstino Production with Low
  Reheating Temperatures}},
  \href{https://doi.org/10.1103/PhysRevD.92.035002}{\emph{Phys. Rev. D}
  {\bfseries 92} (2015) 035002}
  [\href{https://arxiv.org/abs/1505.03149}{{\ttfamily 1505.03149}}].

\bibitem{Kolda:2014ppa}
C.~Kolda and J.~Unwin, \emph{{X-ray lines from R-parity violating decays of keV
  sparticles}}, \href{https://doi.org/10.1103/PhysRevD.90.023535}{\emph{Phys.
  Rev. D} {\bfseries 90} (2014) 023535}
  [\href{https://arxiv.org/abs/1403.5580}{{\ttfamily 1403.5580}}].

\bibitem{Hut:1979xw}
P.~Hut and K.A.~Olive, \emph{{A Cosmological Upper Limit On The Mass Of Heavy
  Neutrinos}}, \href{https://doi.org/10.1016/0370-2693(79)90039-X}{\emph{Phys.
  Lett. B} {\bfseries 87} (1979) 144}.

\bibitem{Davoudiasl:2012uw}
H.~Davoudiasl and R.N.~Mohapatra, \emph{{On Relating the Genesis of Cosmic
  Baryons and Dark Matter}},
  \href{https://doi.org/10.1088/1367-2630/14/9/095011}{\emph{New J. Phys.}
  {\bfseries 14} (2012) 095011}
  [\href{https://arxiv.org/abs/1203.1247}{{\ttfamily 1203.1247}}].

\bibitem{Petraki:2013wwa}
K.~Petraki and R.R.~Volkas, \emph{{Review of asymmetric dark matter}},
  \href{https://doi.org/10.1142/S0217751X13300287}{\emph{Int. J. Mod. Phys.}
  {\bfseries A28} (2013) 1330028}
  [\href{https://arxiv.org/abs/1305.4939}{{\ttfamily 1305.4939}}].

\bibitem{Zurek:2013wia}
K.M.~Zurek, \emph{{Asymmetric Dark Matter: Theories, Signatures, and
  Constraints}},
  \href{https://doi.org/10.1016/j.physrep.2013.12.001}{\emph{Phys. Rept.}
  {\bfseries 537} (2014) 91} [\href{https://arxiv.org/abs/1308.0338}{{\ttfamily
  1308.0338}}].

\bibitem{Hisano:2006nn}
J.~Hisano, S.~Matsumoto, M.~Nagai, O.~Saito and M.~Senami,
  \emph{{Non-perturbative effect on thermal relic abundance of dark matter}},
  \href{https://doi.org/10.1016/j.physletb.2007.01.012}{\emph{Phys. Lett. B}
  {\bfseries 646} (2007) 34}
  [\href{https://arxiv.org/abs/hep-ph/0610249}{{\ttfamily hep-ph/0610249}}].

\bibitem{Arkani-Hamed:2008hhe}
N.~Arkani-Hamed, D.P.~Finkbeiner, T.R.~Slatyer and N.~Weiner, \emph{{A Theory
  of Dark Matter}},
  \href{https://doi.org/10.1103/PhysRevD.79.015014}{\emph{Phys. Rev. D}
  {\bfseries 79} (2009) 015014}
  [\href{https://arxiv.org/abs/0810.0713}{{\ttfamily 0810.0713}}].

\bibitem{vonHarling:2014kha}
B.~von Harling and K.~Petraki, \emph{{Bound-state formation for thermal relic
  dark matter and unitarity}},
  \href{https://doi.org/10.1088/1475-7516/2014/12/033}{\emph{JCAP} {\bfseries
  12} (2014) 033} [\href{https://arxiv.org/abs/1407.7874}{{\ttfamily
  1407.7874}}].

\bibitem{An:2016gad}
H.~An, M.B.~Wise and Y.~Zhang, \emph{{Effects of Bound States on Dark Matter
  Annihilation}}, \href{https://doi.org/10.1103/PhysRevD.93.115020}{\emph{Phys.
  Rev. D} {\bfseries 93} (2016) 115020}
  [\href{https://arxiv.org/abs/1604.01776}{{\ttfamily 1604.01776}}].

\bibitem{Asadi:2016ybp}
P.~Asadi, M.~Baumgart, P.J.~Fitzpatrick, E.~Krupczak and T.R.~Slatyer,
  \emph{{Capture and Decay of Electroweak WIMPonium}},
  \href{https://doi.org/10.1088/1475-7516/2017/02/005}{\emph{JCAP} {\bfseries
  02} (2017) 005} [\href{https://arxiv.org/abs/1610.07617}{{\ttfamily
  1610.07617}}].

\bibitem{Mitridate:2017izz}
A.~Mitridate, M.~Redi, J.~Smirnov and A.~Strumia, \emph{{Cosmological
  Implications of Dark Matter Bound States}},
  \href{https://doi.org/10.1088/1475-7516/2017/05/006}{\emph{JCAP} {\bfseries
  05} (2017) 006} [\href{https://arxiv.org/abs/1702.01141}{{\ttfamily
  1702.01141}}].

\bibitem{Binder:2020efn}
T.~Binder, B.~Blobel, J.~Harz and K.~Mukaida, \emph{{Dark matter bound-state
  formation at higher order: a non-equilibrium quantum field theory approach}},
  \href{https://doi.org/10.1007/JHEP09(2020)086}{\emph{JHEP} {\bfseries 09}
  (2020) 086} [\href{https://arxiv.org/abs/2002.07145}{{\ttfamily
  2002.07145}}].

\bibitem{Smirnov:2019ngs}
J.~Smirnov and J.F.~Beacom, \emph{{TeV-Scale Thermal WIMPs: Unitarity and its
  Consequences}},
  \href{https://doi.org/10.1103/PhysRevD.100.043029}{\emph{Phys. Rev. D}
  {\bfseries 100} (2019) 043029}
  [\href{https://arxiv.org/abs/1904.11503}{{\ttfamily 1904.11503}}].

\bibitem{Bottaro:2021snn}
S.~Bottaro, D.~Buttazzo, M.~Costa, R.~Franceschini, P.~Panci, D.~Redigolo
  et~al., \emph{{Closing the window on WIMP Dark Matter}},
  \href{https://doi.org/10.1140/epjc/s10052-021-09917-9}{\emph{Eur. Phys. J. C}
  {\bfseries 82} (2022) 31} [\href{https://arxiv.org/abs/2107.09688}{{\ttfamily
  2107.09688}}].

\bibitem{Hall:1997ah}
L.J.~Hall, T.~Moroi and H.~Murayama, \emph{{Sneutrino cold dark matter with
  lepton number violation}},
  \href{https://doi.org/10.1016/S0370-2693(98)00196-8}{\emph{Phys. Lett. B}
  {\bfseries 424} (1998) 305}
  [\href{https://arxiv.org/abs/hep-ph/9712515}{{\ttfamily hep-ph/9712515}}].

\bibitem{Tucker-Smith:2001myb}
D.~Tucker-Smith and N.~Weiner, \emph{{Inelastic dark matter}},
  \href{https://doi.org/10.1103/PhysRevD.64.043502}{\emph{Phys. Rev. D}
  {\bfseries 64} (2001) 043502}
  [\href{https://arxiv.org/abs/hep-ph/0101138}{{\ttfamily hep-ph/0101138}}].

\bibitem{Tucker-Smith:2004mxa}
D.~Tucker-Smith and N.~Weiner, \emph{{The Status of inelastic dark matter}},
  \href{https://doi.org/10.1103/PhysRevD.72.063509}{\emph{Phys. Rev. D}
  {\bfseries 72} (2005) 063509}
  [\href{https://arxiv.org/abs/hep-ph/0402065}{{\ttfamily hep-ph/0402065}}].

\bibitem{Chang:2008gd}
S.~Chang, G.D.~Kribs, D.~Tucker-Smith and N.~Weiner, \emph{{Inelastic Dark
  Matter in Light of DAMA/LIBRA}},
  \href{https://doi.org/10.1103/PhysRevD.79.043513}{\emph{Phys. Rev. D}
  {\bfseries 79} (2009) 043513}
  [\href{https://arxiv.org/abs/0807.2250}{{\ttfamily 0807.2250}}].

\bibitem{March-Russell:2008rkh}
J.~March-Russell, C.~McCabe and M.~McCullough, \emph{{Inelastic Dark Matter,
  Non-Standard Halos and the DAMA/LIBRA Results}},
  \href{https://doi.org/10.1088/1126-6708/2009/05/071}{\emph{JHEP} {\bfseries
  05} (2009) 071} [\href{https://arxiv.org/abs/0812.1931}{{\ttfamily
  0812.1931}}].

\bibitem{Cui:2009xq}
Y.~Cui, D.E.~Morrissey, D.~Poland and L.~Randall, \emph{{Candidates for
  Inelastic Dark Matter}},
  \href{https://doi.org/10.1088/1126-6708/2009/05/076}{\emph{JHEP} {\bfseries
  05} (2009) 076} [\href{https://arxiv.org/abs/0901.0557}{{\ttfamily
  0901.0557}}].

\bibitem{Alves:2009nf}
D.S.M.~Alves, S.R.~Behbahani, P.~Schuster and J.G.~Wacker, \emph{{Composite
  Inelastic Dark Matter}},
  \href{https://doi.org/10.1016/j.physletb.2010.08.006}{\emph{Phys. Lett. B}
  {\bfseries 692} (2010) 323}
  [\href{https://arxiv.org/abs/0903.3945}{{\ttfamily 0903.3945}}].

\bibitem{Chang:2010en}
S.~Chang, N.~Weiner and I.~Yavin, \emph{{Magnetic Inelastic Dark Matter}},
  \href{https://doi.org/10.1103/PhysRevD.82.125011}{\emph{Phys. Rev. D}
  {\bfseries 82} (2010) 125011}
  [\href{https://arxiv.org/abs/1007.4200}{{\ttfamily 1007.4200}}].

\bibitem{Barello:2014uda}
G.~Barello, S.~Chang and C.A.~Newby, \emph{{A Model Independent Approach to
  Inelastic Dark Matter Scattering}},
  \href{https://doi.org/10.1103/PhysRevD.90.094027}{\emph{Phys. Rev. D}
  {\bfseries 90} (2014) 094027}
  [\href{https://arxiv.org/abs/1409.0536}{{\ttfamily 1409.0536}}].

\bibitem{Bramante:2016rdh}
J.~Bramante, P.J.~Fox, G.D.~Kribs and A.~Martin, \emph{{Inelastic frontier:
  Discovering dark matter at high recoil energy}},
  \href{https://doi.org/10.1103/PhysRevD.94.115026}{\emph{Phys. Rev. D}
  {\bfseries 94} (2016) 115026}
  [\href{https://arxiv.org/abs/1608.02662}{{\ttfamily 1608.02662}}].

\bibitem{Berlin:2018jbm}
A.~Berlin and F.~Kling, \emph{{Inelastic Dark Matter at the LHC Lifetime
  Frontier: ATLAS, CMS, LHCb, CODEX-b, FASER, and MATHUSLA}},
  \href{https://doi.org/10.1103/PhysRevD.99.015021}{\emph{Phys. Rev. D}
  {\bfseries 99} (2019) 015021}
  [\href{https://arxiv.org/abs/1810.01879}{{\ttfamily 1810.01879}}].

\bibitem{Chung:2003fi}
D.J.H.~Chung, L.L.~Everett, G.L.~Kane, S.F.~King, J.D.~Lykken and L.-T.~Wang,
  \emph{{The Soft supersymmetry breaking Lagrangian: Theory and applications}},
  \href{https://doi.org/10.1016/j.physrep.2004.08.032}{\emph{Phys. Rept.}
  {\bfseries 407} (2005) 1}
  [\href{https://arxiv.org/abs/hep-ph/0312378}{{\ttfamily hep-ph/0312378}}].

\bibitem{Ellwanger:2009dp}
U.~Ellwanger, C.~Hugonie and A.M.~Teixeira, \emph{{The Next-to-Minimal
  Supersymmetric Standard Model}},
  \href{https://doi.org/10.1016/j.physrep.2010.07.001}{\emph{Phys. Rept.}
  {\bfseries 496} (2010) 1} [\href{https://arxiv.org/abs/0910.1785}{{\ttfamily
  0910.1785}}].

\bibitem{Roszkowski:2017nbc}
L.~Roszkowski, E.M.~Sessolo and S.~Trojanowski, \emph{{WIMP dark matter
  candidates and searches\textemdash{}current status and future prospects}},
  \href{https://doi.org/10.1088/1361-6633/aab913}{\emph{Rept. Prog. Phys.}
  {\bfseries 81} (2018) 066201}
  [\href{https://arxiv.org/abs/1707.06277}{{\ttfamily 1707.06277}}].

\bibitem{Delgado:2020url}
A.~Delgado and M.~Quir\'os, \emph{{Higgsino Dark Matter in the MSSM}},
  \href{https://doi.org/10.1103/PhysRevD.103.015024}{\emph{Phys. Rev. D}
  {\bfseries 103} (2021) 015024}
  [\href{https://arxiv.org/abs/2008.00954}{{\ttfamily 2008.00954}}].

\bibitem{Kowalska:2018toh}
K.~Kowalska and E.M.~Sessolo, \emph{{The discreet charm of higgsino dark matter
  - a pocket review}}, \href{https://doi.org/10.1155/2018/6828560}{\emph{Adv.
  High Energy Phys.} {\bfseries 2018} (2018) 6828560}
  [\href{https://arxiv.org/abs/1802.04097}{{\ttfamily 1802.04097}}].

\bibitem{Han:2013gba}
T.~Han, Z.~Liu and A.~Natarajan, \emph{{Dark matter and Higgs bosons in the
  MSSM}}, \href{https://doi.org/10.1007/JHEP11(2013)008}{\emph{JHEP} {\bfseries
  11} (2013) 008} [\href{https://arxiv.org/abs/1303.3040}{{\ttfamily
  1303.3040}}].

\bibitem{Cabrera:2016wwr}
M.E.~Cabrera, J.A.~Casas, A.~Delgado, S.~Robles and R.~Ruiz~de Austri,
  \emph{{Naturalness of MSSM dark matter}},
  \href{https://doi.org/10.1007/JHEP08(2016)058}{\emph{JHEP} {\bfseries 08}
  (2016) 058} [\href{https://arxiv.org/abs/1604.02102}{{\ttfamily
  1604.02102}}].

\bibitem{Baum:2017enm}
S.~Baum, M.~Carena, N.R.~Shah and C.E.M.~Wagner, \emph{{Higgs portals for
  thermal Dark Matter. EFT perspectives and the NMSSM}},
  \href{https://doi.org/10.1007/JHEP04(2018)069}{\emph{JHEP} {\bfseries 04}
  (2018) 069} [\href{https://arxiv.org/abs/1712.09873}{{\ttfamily
  1712.09873}}].

\bibitem{Panico:2006em}
G.~Panico, M.~Serone and A.~Wulzer, \emph{{Electroweak Symmetry Breaking and
  Precision Tests with a Fifth Dimension}},
  \href{https://doi.org/10.1016/j.nuclphysb.2006.10.032}{\emph{Nucl. Phys. B}
  {\bfseries 762} (2007) 189}
  [\href{https://arxiv.org/abs/hep-ph/0605292}{{\ttfamily hep-ph/0605292}}].

\bibitem{Servant:2002aq}
G.~Servant and T.M.P.~Tait, \emph{{Is the lightest Kaluza-Klein particle a
  viable dark matter candidate?}},
  \href{https://doi.org/10.1016/S0550-3213(02)01012-X}{\emph{Nucl. Phys. B}
  {\bfseries 650} (2003) 391}
  [\href{https://arxiv.org/abs/hep-ph/0206071}{{\ttfamily hep-ph/0206071}}].

\bibitem{Haba:2009xu}
N.~Haba, S.~Matsumoto, N.~Okada and T.~Yamashita, \emph{{Gauge-Higgs Dark
  Matter}}, \href{https://doi.org/10.1007/JHEP03(2010)064}{\emph{JHEP}
  {\bfseries 03} (2010) 064} [\href{https://arxiv.org/abs/0910.3741}{{\ttfamily
  0910.3741}}].

\bibitem{Panico:2008bx}
G.~Panico, E.~Ponton, J.~Santiago and M.~Serone, \emph{{Dark Matter and
  Electroweak Symmetry Breaking in Models with Warped Extra Dimensions}},
  \href{https://doi.org/10.1103/PhysRevD.77.115012}{\emph{Phys. Rev. D}
  {\bfseries 77} (2008) 115012}
  [\href{https://arxiv.org/abs/0801.1645}{{\ttfamily 0801.1645}}].

\bibitem{Maru:2018ocf}
N.~Maru, N.~Okada and S.~Okada, \emph{{$SU(2)_L$ doublet vector dark matter
  from gauge-Higgs unification}},
  \href{https://doi.org/10.1103/PhysRevD.98.075021}{\emph{Phys. Rev. D}
  {\bfseries 98} (2018) 075021}
  [\href{https://arxiv.org/abs/1803.01274}{{\ttfamily 1803.01274}}].

\bibitem{Regis:2006hc}
M.~Regis, M.~Serone and P.~Ullio, \emph{{A Dark Matter Candidate from an Extra
  (Non-Universal) Dimension}},
  \href{https://doi.org/10.1088/1126-6708/2007/03/084}{\emph{JHEP} {\bfseries
  03} (2007) 084} [\href{https://arxiv.org/abs/hep-ph/0612286}{{\ttfamily
  hep-ph/0612286}}].

\bibitem{Kribs:2009fy}
G.D.~Kribs, T.S.~Roy, J.~Terning and K.M.~Zurek, \emph{{Quirky Composite Dark
  Matter}}, \href{https://doi.org/10.1103/PhysRevD.81.095001}{\emph{Phys. Rev.
  D} {\bfseries 81} (2010) 095001}
  [\href{https://arxiv.org/abs/0909.2034}{{\ttfamily 0909.2034}}].

\bibitem{Appelquist:2015yfa}
T.~Appelquist et~al., \emph{{Stealth Dark Matter: Dark scalar baryons through
  the Higgs portal}},
  \href{https://doi.org/10.1103/PhysRevD.92.075030}{\emph{Phys. Rev. D}
  {\bfseries 92} (2015) 075030}
  [\href{https://arxiv.org/abs/1503.04203}{{\ttfamily 1503.04203}}].

\bibitem{Antipin:2015xia}
O.~Antipin, M.~Redi, A.~Strumia and E.~Vigiani, \emph{{Accidental Composite
  Dark Matter}}, \href{https://doi.org/10.1007/JHEP07(2015)039}{\emph{JHEP}
  {\bfseries 07} (2015) 039}
  [\href{https://arxiv.org/abs/1503.08749}{{\ttfamily 1503.08749}}].

\bibitem{Kribs:2016cew}
G.D.~Kribs and E.T.~Neil, \emph{{Review of strongly-coupled composite dark
  matter models and lattice simulations}},
  \href{https://doi.org/10.1142/S0217751X16430041}{\emph{Int. J. Mod. Phys. A}
  {\bfseries 31} (2016) 1643004}
  [\href{https://arxiv.org/abs/1604.04627}{{\ttfamily 1604.04627}}].

\bibitem{Dondi:2019olm}
N.A.~Dondi, F.~Sannino and J.~Smirnov, \emph{{Thermal history of composite dark
  matter}}, \href{https://doi.org/10.1103/PhysRevD.101.103010}{\emph{Phys. Rev.
  D} {\bfseries 101} (2020) 103010}
  [\href{https://arxiv.org/abs/1905.08810}{{\ttfamily 1905.08810}}].

\bibitem{Garani:2021zrr}
R.~Garani, M.~Redi and A.~Tesi, \emph{{Dark QCD matters}},
  \href{https://doi.org/10.1007/JHEP12(2021)139}{\emph{JHEP} {\bfseries 12}
  (2021) 139} [\href{https://arxiv.org/abs/2105.03429}{{\ttfamily
  2105.03429}}].

\bibitem{Cline:2021itd}
J.M.~Cline, \emph{{Dark atoms and composite dark matter}},  in \emph{{Les
  Houches summer school on Dark Matter}}, 8, 2021
  [\href{https://arxiv.org/abs/2108.10314}{{\ttfamily 2108.10314}}].

\bibitem{Hambye:2009fg}
T.~Hambye and M.H.G.~Tytgat, \emph{{Confined hidden vector dark matter}},
  \href{https://doi.org/10.1016/j.physletb.2009.11.050}{\emph{Phys. Lett. B}
  {\bfseries 683} (2010) 39} [\href{https://arxiv.org/abs/0907.1007}{{\ttfamily
  0907.1007}}].

\bibitem{SpierMoreiraAlves:2010err}
D.~Spier Moreira~Alves, S.R.~Behbahani, P.~Schuster and J.G.~Wacker, \emph{{The
  Cosmology of Composite Inelastic Dark Matter}},
  \href{https://doi.org/10.1007/JHEP06(2010)113}{\emph{JHEP} {\bfseries 06}
  (2010) 113} [\href{https://arxiv.org/abs/1003.4729}{{\ttfamily 1003.4729}}].

\bibitem{Antipin:2014qva}
O.~Antipin, M.~Redi and A.~Strumia, \emph{{Dynamical generation of the weak and
  Dark Matter scales from strong interactions}},
  \href{https://doi.org/10.1007/JHEP01(2015)157}{\emph{JHEP} {\bfseries 01}
  (2015) 157} [\href{https://arxiv.org/abs/1410.1817}{{\ttfamily 1410.1817}}].

\bibitem{Hochberg:2014kqa}
Y.~Hochberg, E.~Kuflik, H.~Murayama, T.~Volansky and J.G.~Wacker, \emph{{Model
  for Thermal Relic Dark Matter of Strongly Interacting Massive Particles}},
  \href{https://doi.org/10.1103/PhysRevLett.115.021301}{\emph{Phys. Rev. Lett.}
  {\bfseries 115} (2015) 021301}
  [\href{https://arxiv.org/abs/1411.3727}{{\ttfamily 1411.3727}}].

\bibitem{Carmona:2015haa}
A.~Carmona and M.~Chala, \emph{{Composite Dark Sectors}},
  \href{https://doi.org/10.1007/JHEP06(2015)105}{\emph{JHEP} {\bfseries 06}
  (2015) 105} [\href{https://arxiv.org/abs/1504.00332}{{\ttfamily
  1504.00332}}].

\bibitem{Lonsdale:2017mzg}
S.J.~Lonsdale, M.~Schroor and R.R.~Volkas, \emph{{Asymmetric Dark Matter and
  the hadronic spectra of hidden QCD}},
  \href{https://doi.org/10.1103/PhysRevD.96.055027}{\emph{Phys. Rev. D}
  {\bfseries 96} (2017) 055027}
  [\href{https://arxiv.org/abs/1704.05213}{{\ttfamily 1704.05213}}].

\bibitem{DeLuca:2018mzn}
V.~De~Luca, A.~Mitridate, M.~Redi, J.~Smirnov and A.~Strumia, \emph{{Colored
  Dark Matter}}, \href{https://doi.org/10.1103/PhysRevD.97.115024}{\emph{Phys.
  Rev. D} {\bfseries 97} (2018) 115024}
  [\href{https://arxiv.org/abs/1801.01135}{{\ttfamily 1801.01135}}].

\bibitem{Kribs:2018oad}
G.D.~Kribs, A.~Martin and T.~Tong, \emph{{Effective Theories of Dark Mesons
  with Custodial Symmetry}},
  \href{https://doi.org/10.1007/JHEP08(2019)020}{\emph{JHEP} {\bfseries 08}
  (2019) 020} [\href{https://arxiv.org/abs/1809.10183}{{\ttfamily
  1809.10183}}].

\bibitem{Tsai:2020vpi}
Y.-D.~Tsai, R.~McGehee and H.~Murayama, \emph{{Resonant Self-Interacting Dark
  Matter from Dark QCD}},  \href{https://arxiv.org/abs/2008.08608}{{\ttfamily
  2008.08608}}.

\bibitem{Cheng:2021kjg}
H.-C.~Cheng, L.~Li and E.~Salvioni, \emph{{A theory of dark pions}},
  \href{https://doi.org/10.1007/JHEP01(2022)122}{\emph{JHEP} {\bfseries 01}
  (2022) 122} [\href{https://arxiv.org/abs/2110.10691}{{\ttfamily
  2110.10691}}].

\bibitem{Buckley:2012ky}
M.R.~Buckley and E.T.~Neil, \emph{{Thermal dark matter from a confining
  sector}}, \href{https://doi.org/10.1103/PhysRevD.87.043510}{\emph{Phys. Rev.
  D} {\bfseries 87} (2013) 043510}
  [\href{https://arxiv.org/abs/1209.6054}{{\ttfamily 1209.6054}}].

\bibitem{LatticeStrongDynamicsLSD:2013elk}
{\scshape Lattice Strong Dynamics (LSD)} collaboration, \emph{{Lattice
  Calculation of Composite Dark Matter Form Factors}},
  \href{https://doi.org/10.1103/PhysRevD.88.014502}{\emph{Phys. Rev. D}
  {\bfseries 88} (2013) 014502}
  [\href{https://arxiv.org/abs/1301.1693}{{\ttfamily 1301.1693}}].

\bibitem{Cline:2016nab}
J.M.~Cline, W.~Huang and G.D.~Moore, \emph{{Challenges for models with
  composite states}},
  \href{https://doi.org/10.1103/PhysRevD.94.055029}{\emph{Phys. Rev. D}
  {\bfseries 94} (2016) 055029}
  [\href{https://arxiv.org/abs/1607.07865}{{\ttfamily 1607.07865}}].

\bibitem{Mitridate:2017oky}
A.~Mitridate, M.~Redi, J.~Smirnov and A.~Strumia, \emph{{Dark Matter as a
  weakly coupled Dark Baryon}},
  \href{https://doi.org/10.1007/JHEP10(2017)210}{\emph{JHEP} {\bfseries 10}
  (2017) 210} [\href{https://arxiv.org/abs/1707.05380}{{\ttfamily
  1707.05380}}].

\bibitem{Morrison:2020yeg}
L.~Morrison, S.~Profumo and D.J.~Robinson, \emph{{Large $N$-ightmare Dark
  Matter}}, \href{https://doi.org/10.1088/1475-7516/2021/05/058}{\emph{JCAP}
  {\bfseries 05} (2021) 058}
  [\href{https://arxiv.org/abs/2010.03586}{{\ttfamily 2010.03586}}].

\bibitem{Faraggi:2000pv}
A.E.~Faraggi and M.~Pospelov, \emph{{Selfinteracting dark matter from the
  hidden heterotic string sector}},
  \href{https://doi.org/10.1016/S0927-6505(01)00121-9}{\emph{Astropart. Phys.}
  {\bfseries 16} (2002) 451}
  [\href{https://arxiv.org/abs/hep-ph/0008223}{{\ttfamily hep-ph/0008223}}].

\bibitem{Juknevich:2009ji}
J.E.~Juknevich, D.~Melnikov and M.J.~Strassler, \emph{{A Pure-Glue Hidden
  Valley I. States and Decays}},
  \href{https://doi.org/10.1088/1126-6708/2009/07/055}{\emph{JHEP} {\bfseries
  07} (2009) 055} [\href{https://arxiv.org/abs/0903.0883}{{\ttfamily
  0903.0883}}].

\bibitem{Juknevich:2009gg}
J.E.~Juknevich, \emph{{Pure-glue hidden valleys through the Higgs portal}},
  \href{https://doi.org/10.1007/JHEP08(2010)121}{\emph{JHEP} {\bfseries 08}
  (2010) 121} [\href{https://arxiv.org/abs/0911.5616}{{\ttfamily 0911.5616}}].

\bibitem{Soni:2016gzf}
A.~Soni and Y.~Zhang, \emph{{Hidden SU(N) Glueball Dark Matter}},
  \href{https://doi.org/10.1103/PhysRevD.93.115025}{\emph{Phys. Rev. D}
  {\bfseries 93} (2016) 115025}
  [\href{https://arxiv.org/abs/1602.00714}{{\ttfamily 1602.00714}}].

\bibitem{Forestell:2016qhc}
L.~Forestell, D.E.~Morrissey and K.~Sigurdson, \emph{{Non-Abelian Dark Forces
  and the Relic Densities of Dark Glueballs}},
  \href{https://doi.org/10.1103/PhysRevD.95.015032}{\emph{Phys. Rev. D}
  {\bfseries 95} (2017) 015032}
  [\href{https://arxiv.org/abs/1605.08048}{{\ttfamily 1605.08048}}].

\bibitem{Morningstar:1999rf}
C.J.~Morningstar and M.J.~Peardon, \emph{{The Glueball spectrum from an
  anisotropic lattice study}},
  \href{https://doi.org/10.1103/PhysRevD.60.034509}{\emph{Phys. Rev. D}
  {\bfseries 60} (1999) 034509}
  [\href{https://arxiv.org/abs/hep-lat/9901004}{{\ttfamily hep-lat/9901004}}].

\bibitem{Mathieu:2008me}
V.~Mathieu, N.~Kochelev and V.~Vento, \emph{{The Physics of Glueballs}},
  \href{https://doi.org/10.1142/S0218301309012124}{\emph{Int. J. Mod. Phys. E}
  {\bfseries 18} (2009) 1} [\href{https://arxiv.org/abs/0810.4453}{{\ttfamily
  0810.4453}}].

\bibitem{Blinnikov:1983gh}
S.I.~Blinnikov and M.~Khlopov, \emph{{Possible astronomical effects of mirror
  particles}}, {\emph{Sov. Astron.} {\bfseries 27} (1983) 371}.

\bibitem{Goldberg:1986nk}
H.~Goldberg and L.J.~Hall, \emph{{A New Candidate for Dark Matter}},
  \href{https://doi.org/10.1016/0370-2693(86)90731-8}{\emph{Phys. Lett.}
  {\bfseries B174} (1986) 151}.

\bibitem{Chacko:2005pe}
Z.~Chacko, H.-S.~Goh and R.~Harnik, \emph{{The Twin Higgs: Natural electroweak
  breaking from mirror symmetry}},
  \href{https://doi.org/10.1103/PhysRevLett.96.231802}{\emph{Phys. Rev. Lett.}
  {\bfseries 96} (2006) 231802}
  [\href{https://arxiv.org/abs/hep-ph/0506256}{{\ttfamily hep-ph/0506256}}].

\bibitem{Chacko:2005vw}
Z.~Chacko, Y.~Nomura, M.~Papucci and G.~Perez, \emph{{Natural little hierarchy
  from a partially goldstone twin Higgs}},
  \href{https://doi.org/10.1088/1126-6708/2006/01/126}{\emph{JHEP} {\bfseries
  01} (2006) 126} [\href{https://arxiv.org/abs/hep-ph/0510273}{{\ttfamily
  hep-ph/0510273}}].

\bibitem{Chacko:2005un}
Z.~Chacko, H.-S.~Goh and R.~Harnik, \emph{{A Twin Higgs model from left-right
  symmetry}}, \href{https://doi.org/10.1088/1126-6708/2006/01/108}{\emph{JHEP}
  {\bfseries 01} (2006) 108}
  [\href{https://arxiv.org/abs/hep-ph/0512088}{{\ttfamily hep-ph/0512088}}].

\bibitem{Barbieri:2005ri}
R.~Barbieri, T.~Gregoire and L.J.~Hall, \emph{{Mirror world at the large hadron
  collider}},  \href{https://arxiv.org/abs/hep-ph/0509242}{{\ttfamily
  hep-ph/0509242}}.

\bibitem{Craig:2013fga}
N.~Craig and K.~Howe, \emph{{Doubling down on naturalness with a supersymmetric
  twin Higgs}}, \href{https://doi.org/10.1007/JHEP03(2014)140}{\emph{JHEP}
  {\bfseries 03} (2014) 140} [\href{https://arxiv.org/abs/1312.1341}{{\ttfamily
  1312.1341}}].

\bibitem{GarciaGarcia:2015fol}
I.~Garcia~Garcia, R.~Lasenby and J.~March-Russell, \emph{{Twin Higgs WIMP Dark
  Matter}}, \href{https://doi.org/10.1103/PhysRevD.92.055034}{\emph{Phys. Rev.
  D} {\bfseries 92} (2015) 055034}
  [\href{https://arxiv.org/abs/1505.07109}{{\ttfamily 1505.07109}}].

\bibitem{Craig:2015xla}
N.~Craig and A.~Katz, \emph{{The Fraternal WIMP Miracle}},
  \href{https://doi.org/10.1088/1475-7516/2015/10/054}{\emph{JCAP} {\bfseries
  10} (2015) 054} [\href{https://arxiv.org/abs/1505.07113}{{\ttfamily
  1505.07113}}].

\bibitem{Farina:2015uea}
M.~Farina, \emph{{Asymmetric Twin Dark Matter}},
  \href{https://doi.org/10.1088/1475-7516/2015/11/017}{\emph{JCAP} {\bfseries
  11} (2015) 017} [\href{https://arxiv.org/abs/1506.03520}{{\ttfamily
  1506.03520}}].

\bibitem{Farina:2016ndq}
M.~Farina, A.~Monteux and C.S.~Shin, \emph{{Twin mechanism for baryon and dark
  matter asymmetries}},
  \href{https://doi.org/10.1103/PhysRevD.94.035017}{\emph{Phys. Rev. D}
  {\bfseries 94} (2016) 035017}
  [\href{https://arxiv.org/abs/1604.08211}{{\ttfamily 1604.08211}}].

\bibitem{Prilepina:2016rlq}
V.~Prilepina and Y.~Tsai, \emph{{Reconciling Large And Small-Scale Structure In
  Twin Higgs Models}},
  \href{https://doi.org/10.1007/JHEP09(2017)033}{\emph{JHEP} {\bfseries 09}
  (2017) 033} [\href{https://arxiv.org/abs/1611.05879}{{\ttfamily
  1611.05879}}].

\bibitem{Craig:2016lyx}
N.~Craig, S.~Koren and T.~Trott, \emph{{Cosmological Signals of a Mirror Twin
  Higgs}}, \href{https://doi.org/10.1007/JHEP05(2017)038}{\emph{JHEP}
  {\bfseries 05} (2017) 038}
  [\href{https://arxiv.org/abs/1611.07977}{{\ttfamily 1611.07977}}].

\bibitem{Berger:2016vxi}
J.~Berger, K.~Jedamzik and D.G.E.~Walker, \emph{{Cosmological Constraints on
  Decoupled Dark Photons and Dark Higgs}},
  \href{https://doi.org/10.1088/1475-7516/2016/11/032}{\emph{JCAP} {\bfseries
  11} (2016) 032} [\href{https://arxiv.org/abs/1605.07195}{{\ttfamily
  1605.07195}}].

\bibitem{Chacko:2016hvu}
Z.~Chacko, N.~Craig, P.J.~Fox and R.~Harnik, \emph{{Cosmology in Mirror Twin
  Higgs and Neutrino Masses}},
  \href{https://doi.org/10.1007/JHEP07(2017)023}{\emph{JHEP} {\bfseries 07}
  (2017) 023} [\href{https://arxiv.org/abs/1611.07975}{{\ttfamily
  1611.07975}}].

\bibitem{Csaki:2017spo}
C.~Csaki, E.~Kuflik and S.~Lombardo, \emph{{Viable Twin Cosmology from Neutrino
  Mixing}}, \href{https://doi.org/10.1103/PhysRevD.96.055013}{\emph{Phys. Rev.
  D} {\bfseries 96} (2017) 055013}
  [\href{https://arxiv.org/abs/1703.06884}{{\ttfamily 1703.06884}}].

\bibitem{Chacko:2018vss}
Z.~Chacko, D.~Curtin, M.~Geller and Y.~Tsai, \emph{{Cosmological Signatures of
  a Mirror Twin Higgs}},
  \href{https://doi.org/10.1007/JHEP09(2018)163}{\emph{JHEP} {\bfseries 09}
  (2018) 163} [\href{https://arxiv.org/abs/1803.03263}{{\ttfamily
  1803.03263}}].

\bibitem{Elor:2018xku}
G.~Elor, H.~Liu, T.R.~Slatyer and Y.~Soreq, \emph{{Complementarity for Dark
  Sector Bound States}},
  \href{https://doi.org/10.1103/PhysRevD.98.036015}{\emph{Phys. Rev. D}
  {\bfseries 98} (2018) 036015}
  [\href{https://arxiv.org/abs/1801.07723}{{\ttfamily 1801.07723}}].

\bibitem{Hochberg:2018vdo}
Y.~Hochberg, E.~Kuflik and H.~Murayama, \emph{{Twin Higgs model with strongly
  interacting massive particle dark matter}},
  \href{https://doi.org/10.1103/PhysRevD.99.015005}{\emph{Phys. Rev. D}
  {\bfseries 99} (2019) 015005}
  [\href{https://arxiv.org/abs/1805.09345}{{\ttfamily 1805.09345}}].

\bibitem{Francis:2018xjd}
A.~Francis, R.J.~Hudspith, R.~Lewis and S.~Tulin, \emph{{Dark Matter from
  Strong Dynamics: The Minimal Theory of Dark Baryons}},
  \href{https://doi.org/10.1007/JHEP12(2018)118}{\emph{JHEP} {\bfseries 12}
  (2018) 118} [\href{https://arxiv.org/abs/1809.09117}{{\ttfamily
  1809.09117}}].

\bibitem{Harigaya:2019shz}
K.~Harigaya, R.~Mcgehee, H.~Murayama and K.~Schutz, \emph{{A predictive mirror
  twin Higgs with small Z$_{2}$ breaking}},
  \href{https://doi.org/10.1007/JHEP05(2020)155}{\emph{JHEP} {\bfseries 05}
  (2020) 155} [\href{https://arxiv.org/abs/1905.08798}{{\ttfamily
  1905.08798}}].

\bibitem{Ibe:2019ena}
M.~Ibe, A.~Kamada, S.~Kobayashi, T.~Kuwahara and W.~Nakano, \emph{{Baryon-Dark
  Matter Coincidence in Mirrored Unification}},
  \href{https://doi.org/10.1103/PhysRevD.100.075022}{\emph{Phys. Rev. D}
  {\bfseries 100} (2019) 075022}
  [\href{https://arxiv.org/abs/1907.03404}{{\ttfamily 1907.03404}}].

\bibitem{Dunsky:2019upk}
D.~Dunsky, L.J.~Hall and K.~Harigaya, \emph{{Dark Matter, Dark Radiation and
  Gravitational Waves from Mirror Higgs Parity}},
  \href{https://doi.org/10.1007/JHEP02(2020)078}{\emph{JHEP} {\bfseries 02}
  (2020) 078} [\href{https://arxiv.org/abs/1908.02756}{{\ttfamily
  1908.02756}}].

\bibitem{Csaki:2019qgb}
C.~Cs\'aki, C.-S.~Guan, T.~Ma and J.~Shu, \emph{{Twin Higgs with exact
  Z$_{2}$}}, \href{https://doi.org/10.1007/JHEP12(2020)005}{\emph{JHEP}
  {\bfseries 12} (2020) 005}
  [\href{https://arxiv.org/abs/1910.14085}{{\ttfamily 1910.14085}}].

\bibitem{Koren:2019iuv}
S.~Koren and R.~McGehee, \emph{{Freezing-in twin dark matter}},
  \href{https://doi.org/10.1103/PhysRevD.101.055024}{\emph{Phys. Rev. D}
  {\bfseries 101} (2020) 055024}
  [\href{https://arxiv.org/abs/1908.03559}{{\ttfamily 1908.03559}}].

\bibitem{Terning:2019hgj}
J.~Terning, C.B.~Verhaaren and K.~Zora, \emph{{Composite Twin Dark Matter}},
  \href{https://doi.org/10.1103/PhysRevD.99.095020}{\emph{Phys. Rev. D}
  {\bfseries 99} (2019) 095020}
  [\href{https://arxiv.org/abs/1902.08211}{{\ttfamily 1902.08211}}].

\bibitem{Johns:2020rtp}
L.~Johns and S.~Koren, \emph{{The Hydrogen Mixing Portal, Its Origins, and Its
  Cosmological Effects}},  \href{https://arxiv.org/abs/2012.06591}{{\ttfamily
  2012.06591}}.

\bibitem{Roux:2020wkp}
J.-S.~Roux and J.M.~Cline, \emph{{Constraining galactic structures of mirror
  dark matter}}, \href{https://doi.org/10.1103/PhysRevD.102.063518}{\emph{Phys.
  Rev. D} {\bfseries 102} (2020) 063518}
  [\href{https://arxiv.org/abs/2001.11504}{{\ttfamily 2001.11504}}].

\bibitem{Ritter:2021hgu}
A.C.~Ritter and R.R.~Volkas, \emph{{Implementing asymmetric dark matter and
  dark electroweak baryogenesis in a mirror two-Higgs-doublet model}},
  \href{https://doi.org/10.1103/PhysRevD.104.035032}{\emph{Phys. Rev. D}
  {\bfseries 104} (2021) 035032}
  [\href{https://arxiv.org/abs/2101.07421}{{\ttfamily 2101.07421}}].

\bibitem{Curtin:2021alk}
D.~Curtin and S.~Gryba, \emph{{Twin Higgs portal dark matter}},
  \href{https://doi.org/10.1007/JHEP08(2021)009}{\emph{JHEP} {\bfseries 08}
  (2021) 009} [\href{https://arxiv.org/abs/2101.11019}{{\ttfamily
  2101.11019}}].

\bibitem{Curtin:2021spx}
D.~Curtin, S.~Gryba, J.~Setford, D.~Hooper and J.~Scholtz, \emph{{Resurrecting
  the Fraternal Twin WIMP Miracle}},
  \href{https://arxiv.org/abs/2106.12578}{{\ttfamily 2106.12578}}.

\bibitem{Foot:2002iy}
R.~Foot and S.~Mitra, \emph{{Ordinary atom mirror atom bound states: A New
  window on the mirror world}},
  \href{https://doi.org/10.1103/PhysRevD.66.061301}{\emph{Phys. Rev.}
  {\bfseries D66} (2002) 061301}
  [\href{https://arxiv.org/abs/hep-ph/0204256}{{\ttfamily hep-ph/0204256}}].

\bibitem{Foot:2003jt}
R.~Foot and R.R.~Volkas, \emph{{Was ordinary matter synthesized from mirror
  matter? An Attempt to explain why Omega(Baryon) approximately equal to 0.2
  Omega(Dark)}}, \href{https://doi.org/10.1103/PhysRevD.68.021304}{\emph{Phys.
  Rev. D} {\bfseries 68} (2003) 021304}
  [\href{https://arxiv.org/abs/hep-ph/0304261}{{\ttfamily hep-ph/0304261}}].

\bibitem{Foot:2004pa}
R.~Foot, \emph{{Mirror matter-type dark matter}},
  \href{https://doi.org/10.1142/S0218271804006449}{\emph{Int. J. Mod. Phys.}
  {\bfseries D13} (2004) 2161}
  [\href{https://arxiv.org/abs/astro-ph/0407623}{{\ttfamily
  astro-ph/0407623}}].

\bibitem{Foot:2004wz}
R.~Foot and R.R.~Volkas, \emph{{Spheroidal galactic halos and mirror dark
  matter}}, \href{https://doi.org/10.1103/PhysRevD.70.123508}{\emph{Phys. Rev.}
  {\bfseries D70} (2004) 123508}
  [\href{https://arxiv.org/abs/astro-ph/0407522}{{\ttfamily
  astro-ph/0407522}}].

\bibitem{Khlopov:2008ty}
M.Y.~Khlopov and C.~Kouvaris, \emph{{Composite dark matter from a model with
  composite Higgs boson}},
  \href{https://doi.org/10.1103/PhysRevD.78.065040}{\emph{Phys. Rev. D}
  {\bfseries 78} (2008) 065040}
  [\href{https://arxiv.org/abs/0806.1191}{{\ttfamily 0806.1191}}].

\bibitem{Kaplan:2009de}
D.E.~Kaplan, G.Z.~Krnjaic, K.R.~Rehermann and C.M.~Wells, \emph{{Atomic Dark
  Matter}}, \href{https://doi.org/10.1088/1475-7516/2010/05/021}{\emph{JCAP}
  {\bfseries 1005} (2010) 021}
  [\href{https://arxiv.org/abs/0909.0753}{{\ttfamily 0909.0753}}].

\bibitem{Kaplan:2011yj}
D.E.~Kaplan, G.Z.~Krnjaic, K.R.~Rehermann and C.M.~Wells, \emph{{Dark Atoms:
  Asymmetry and Direct Detection}},
  \href{https://doi.org/10.1088/1475-7516/2011/10/011}{\emph{JCAP} {\bfseries
  1110} (2011) 011} [\href{https://arxiv.org/abs/1105.2073}{{\ttfamily
  1105.2073}}].

\bibitem{Behbahani:2010xa}
S.R.~Behbahani, M.~Jankowiak, T.~Rube and J.G.~Wacker, \emph{{Nearly
  Supersymmetric Dark Atoms}},
  \href{https://doi.org/10.1155/2011/709492}{\emph{Adv. High Energy Phys.}
  {\bfseries 2011} (2011) 709492}
  [\href{https://arxiv.org/abs/1009.3523}{{\ttfamily 1009.3523}}].

\bibitem{Cline:2012is}
J.M.~Cline, Z.~Liu and W.~Xue, \emph{{Millicharged Atomic Dark Matter}},
  \href{https://doi.org/10.1103/PhysRevD.85.101302}{\emph{Phys. Rev. D}
  {\bfseries 85} (2012) 101302}
  [\href{https://arxiv.org/abs/1201.4858}{{\ttfamily 1201.4858}}].

\bibitem{Cyr-Racine:2012tfp}
F.-Y.~Cyr-Racine and K.~Sigurdson, \emph{{Cosmology of atomic dark matter}},
  \href{https://doi.org/10.1103/PhysRevD.87.103515}{\emph{Phys. Rev. D}
  {\bfseries 87} (2013) 103515}
  [\href{https://arxiv.org/abs/1209.5752}{{\ttfamily 1209.5752}}].

\bibitem{Cline:2013pca}
J.M.~Cline, Z.~Liu, G.~Moore and W.~Xue, \emph{{Scattering properties of dark
  atoms and molecules}},
  \href{https://doi.org/10.1103/PhysRevD.89.043514}{\emph{Phys. Rev. D}
  {\bfseries 89} (2014) 043514}
  [\href{https://arxiv.org/abs/1311.6468}{{\ttfamily 1311.6468}}].

\bibitem{Pearce:2015zca}
L.~Pearce, K.~Petraki and A.~Kusenko, \emph{{Signals from dark atom formation
  in halos}}, \href{https://doi.org/10.1103/PhysRevD.91.083532}{\emph{Phys.
  Rev. D} {\bfseries 91} (2015) 083532}
  [\href{https://arxiv.org/abs/1502.01755}{{\ttfamily 1502.01755}}].

\bibitem{Choquette:2015mca}
J.~Choquette and J.M.~Cline, \emph{{Minimal non-Abelian model of atomic dark
  matter}}, \href{https://doi.org/10.1103/PhysRevD.92.115011}{\emph{Phys. Rev.
  D} {\bfseries 92} (2015) 115011}
  [\href{https://arxiv.org/abs/1509.05764}{{\ttfamily 1509.05764}}].

\bibitem{Petraki:2014uza}
K.~Petraki, L.~Pearce and A.~Kusenko, \emph{{Self-interacting asymmetric dark
  matter coupled to a light massive dark photon}},
  \href{https://doi.org/10.1088/1475-7516/2014/07/039}{\emph{JCAP} {\bfseries
  07} (2014) 039} [\href{https://arxiv.org/abs/1403.1077}{{\ttfamily
  1403.1077}}].

\bibitem{Cirelli:2016rnw}
M.~Cirelli, P.~Panci, K.~Petraki, F.~Sala and M.~Taoso, \emph{{Dark Matter's
  secret liaisons: phenomenology of a dark U(1) sector with bound states}},
  \href{https://doi.org/10.1088/1475-7516/2017/05/036}{\emph{JCAP} {\bfseries
  05} (2017) 036} [\href{https://arxiv.org/abs/1612.07295}{{\ttfamily
  1612.07295}}].

\bibitem{Petraki:2016cnz}
K.~Petraki, M.~Postma and J.~de~Vries, \emph{{Radiative bound-state-formation
  cross-sections for dark matter interacting via a Yukawa potential}},
  \href{https://doi.org/10.1007/JHEP04(2017)077}{\emph{JHEP} {\bfseries 04}
  (2017) 077} [\href{https://arxiv.org/abs/1611.01394}{{\ttfamily
  1611.01394}}].

\bibitem{Ciarcelluti:2004ik}
P.~Ciarcelluti, \emph{{Cosmology with mirror dark matter. 1. Linear evolution
  of perturbations}},
  \href{https://doi.org/10.1142/S0218271805006213}{\emph{Int. J. Mod. Phys. D}
  {\bfseries 14} (2005) 187}
  [\href{https://arxiv.org/abs/astro-ph/0409630}{{\ttfamily
  astro-ph/0409630}}].

\bibitem{Ciarcelluti:2004ip}
P.~Ciarcelluti, \emph{{Cosmology with mirror dark matter. 2. Cosmic microwave
  background and large scale structure}},
  \href{https://doi.org/10.1142/S0218271805006225}{\emph{Int. J. Mod. Phys. D}
  {\bfseries 14} (2005) 223}
  [\href{https://arxiv.org/abs/astro-ph/0409633}{{\ttfamily
  astro-ph/0409633}}].

\bibitem{Ciarcelluti:2008vs}
P.~Ciarcelluti and A.~Lepidi, \emph{{Thermodynamics of the early Universe with
  mirror dark matter}},
  \href{https://doi.org/10.1103/PhysRevD.78.123003}{\emph{Phys. Rev. D}
  {\bfseries 78} (2008) 123003}
  [\href{https://arxiv.org/abs/0809.0677}{{\ttfamily 0809.0677}}].

\bibitem{Ciarcelluti:2010zz}
P.~Ciarcelluti, \emph{{Cosmology with mirror dark matter}},
  \href{https://doi.org/10.1142/S0218271810018438}{\emph{Int. J. Mod. Phys. D}
  {\bfseries 19} (2010) 2151}
  [\href{https://arxiv.org/abs/1102.5530}{{\ttfamily 1102.5530}}].

\bibitem{Ciarcelluti:2012zz}
P.~Ciarcelluti and Q.~Wallemacq, \emph{{Is dark matter made of mirror matter?
  Evidence from cosmological data}},
  \href{https://doi.org/10.1016/j.physletb.2013.12.057}{\emph{Phys. Lett. B}
  {\bfseries 729} (2014) 62} [\href{https://arxiv.org/abs/1211.5354}{{\ttfamily
  1211.5354}}].

\bibitem{Ciarcelluti:2014scd}
P.~Ciarcelluti and Q.~Wallemacq, \emph{{Cosmological constraints on mirror
  matter parameters}}, \href{https://doi.org/10.1155/2014/148319}{\emph{Adv.
  High Energy Phys.} {\bfseries 2014} (2014) 148319}
  [\href{https://arxiv.org/abs/1401.4763}{{\ttfamily 1401.4763}}].

\bibitem{Cudell:2014wca}
J.-R.~Cudell, M.Y.~Khlopov and Q.~Wallemacq, \emph{{Effects of dark atom
  excitations}}, \href{https://doi.org/10.1142/S0217732314400069}{\emph{Mod.
  Phys. Lett. A} {\bfseries 29} (2014) 1440006}
  [\href{https://arxiv.org/abs/1411.1655}{{\ttfamily 1411.1655}}].

\bibitem{HOLDOM198665}
B.~Holdom, \emph{Searching for $\epsilon$ charges and a new u(1)},
  \href{https://doi.org/http://dx.doi.org/10.1016/0370-2693(86)90470-3}{\emph{Physics
  Letters B} {\bfseries 178} (1986) 65 }.

\bibitem{HOLDOM1986196}
B.~Holdom, \emph{Two u(1)'s and $\epsilon$ charge shifts},
  \href{https://doi.org/http://dx.doi.org/10.1016/0370-2693(86)91377-8}{\emph{Physics
  Letters B} {\bfseries 166} (1986) 196 }.

\bibitem{Agrawal:2017rvu}
P.~Agrawal, F.-Y.~Cyr-Racine, L.~Randall and J.~Scholtz, \emph{{Dark
  Catalysis}}, \href{https://doi.org/10.1088/1475-7516/2017/08/021}{\emph{JCAP}
  {\bfseries 08} (2017) 021}
  [\href{https://arxiv.org/abs/1702.05482}{{\ttfamily 1702.05482}}].

\bibitem{Ackerman:2008gi}
L.~Ackerman, M.R.~Buckley, S.M.~Carroll and M.~Kamionkowski, \emph{{Dark Matter
  and Dark Radiation}},
  \href{https://doi.org/10.1103/PhysRevD.79.023519}{\emph{Phys. Rev. D}
  {\bfseries 79} (2009) 023519}
  [\href{https://arxiv.org/abs/0810.5126}{{\ttfamily 0810.5126}}].

\bibitem{Feng:2009mn}
J.L.~Feng, M.~Kaplinghat, H.~Tu and H.-B.~Yu, \emph{{Hidden Charged Dark
  Matter}}, \href{https://doi.org/10.1088/1475-7516/2009/07/004}{\emph{JCAP}
  {\bfseries 07} (2009) 004} [\href{https://arxiv.org/abs/0905.3039}{{\ttfamily
  0905.3039}}].

\bibitem{Agrawal:2016quu}
P.~Agrawal, F.-Y.~Cyr-Racine, L.~Randall and J.~Scholtz, \emph{{Make Dark
  Matter Charged Again}},
  \href{https://doi.org/10.1088/1475-7516/2017/05/022}{\emph{JCAP} {\bfseries
  1705} (2017) 022} [\href{https://arxiv.org/abs/1610.04611}{{\ttfamily
  1610.04611}}].

\bibitem{Kahn:2021ttr}
Y.~Kahn and T.~Lin, \emph{{Searches for light dark matter using condensed
  matter systems}},  \href{https://arxiv.org/abs/2108.03239}{{\ttfamily
  2108.03239}}.

\bibitem{Mitridate:2022tnv}
A.~Mitridate, T.~Trickle, Z.~Zhang and K.M.~Zurek, \emph{{Snowmass White Paper:
  Light Dark Matter Direct Detection at the Interface With Condensed Matter
  Physics}},  in \emph{{2022 Snowmass Summer Study}}, 3, 2022
  [\href{https://arxiv.org/abs/2203.07492}{{\ttfamily 2203.07492}}].

\bibitem{PhysRevLett.39.165}
B.W.~Lee and S.~Weinberg, \emph{Cosmological lower bound on heavy-neutrino
  masses}, \href{https://doi.org/10.1103/PhysRevLett.39.165}{\emph{Phys. Rev.
  Lett.} {\bfseries 39} (1977) 165}.

\bibitem{McDermott:2010pa}
S.D.~McDermott, H.-B.~Yu and K.M.~Zurek, \emph{{Turning off the Lights: How
  Dark is Dark Matter?}},
  \href{https://doi.org/10.1103/PhysRevD.83.063509}{\emph{Phys. Rev. D}
  {\bfseries 83} (2011) 063509}
  [\href{https://arxiv.org/abs/1011.2907}{{\ttfamily 1011.2907}}].

\bibitem{Holdom:1985ag}
B.~Holdom, \emph{{Two U(1)'s and Epsilon Charge Shifts}},
  \href{https://doi.org/10.1016/0370-2693(86)91377-8}{\emph{Phys. Lett. B}
  {\bfseries 166} (1986) 196}.

\bibitem{Okun:1982xi}
L.B.~Okun, \emph{{Limits Of Electrodynamics: Paraphotons?}}, {\emph{Sov. Phys.
  JETP} {\bfseries 56} (1982) 502}.

\bibitem{Battaglieri:2017aum}
M.~Battaglieri et~al., \emph{{US Cosmic Visions: New Ideas in Dark Matter 2017:
  Community Report}},  in \emph{{U.S. Cosmic Visions: New Ideas in Dark Matter
  College Park, MD, USA, March 23-25, 2017}}, 2017
  [\href{https://arxiv.org/abs/1707.04591}{{\ttfamily 1707.04591}}].

\bibitem{Preskill:1982cy}
J.~Preskill, M.B.~Wise and F.~Wilczek, \emph{{Cosmology of the Invisible
  Axion}}, \href{https://doi.org/10.1016/0370-2693(83)90637-8}{\emph{Phys.
  Lett. B} {\bfseries 120} (1983) 127}.

\bibitem{Dine:1982ah}
M.~Dine and W.~Fischler, \emph{{The Not So Harmless Axion}},
  \href{https://doi.org/10.1016/0370-2693(83)90639-1}{\emph{Phys. Lett. B}
  {\bfseries 120} (1983) 137}.

\bibitem{Abbott:1982af}
L.F.~Abbott and P.~Sikivie, \emph{{A Cosmological Bound on the Invisible
  Axion}}, \href{https://doi.org/10.1016/0370-2693(83)90638-X}{\emph{Phys.
  Lett. B} {\bfseries 120} (1983) 133}.

\bibitem{Vilenkin:1984ib}
A.~Vilenkin, \emph{{Cosmic Strings and Domain Walls}},
  \href{https://doi.org/10.1016/0370-1573(85)90033-X}{\emph{Phys. Rept.}
  {\bfseries 121} (1985) 263}.

\bibitem{Davis:1986xc}
R.L.~Davis, \emph{{Cosmic Axions from Cosmic Strings}},
  \href{https://doi.org/10.1016/0370-2693(86)90300-X}{\emph{Phys. Lett. B}
  {\bfseries 180} (1986) 225}.

\bibitem{Hiramatsu:2012gg}
T.~Hiramatsu, M.~Kawasaki, K.~Saikawa and T.~Sekiguchi, \emph{{Production of
  dark matter axions from collapse of string-wall systems}},
  \href{https://doi.org/10.1103/PhysRevD.85.105020}{\emph{Phys. Rev. D}
  {\bfseries 85} (2012) 105020}
  [\href{https://arxiv.org/abs/1202.5851}{{\ttfamily 1202.5851}}].

\bibitem{Kawasaki:2014sqa}
M.~Kawasaki, K.~Saikawa and T.~Sekiguchi, \emph{{Axion dark matter from
  topological defects}},
  \href{https://doi.org/10.1103/PhysRevD.91.065014}{\emph{Phys. Rev. D}
  {\bfseries 91} (2015) 065014}
  [\href{https://arxiv.org/abs/1412.0789}{{\ttfamily 1412.0789}}].

\bibitem{Fleury:2015aca}
L.~Fleury and G.D.~Moore, \emph{{Axion dark matter: strings and their cores}},
  \href{https://doi.org/10.1088/1475-7516/2016/01/004}{\emph{JCAP} {\bfseries
  01} (2016) 004} [\href{https://arxiv.org/abs/1509.00026}{{\ttfamily
  1509.00026}}].

\bibitem{Klaer:2017ond}
V.B..~Klaer and G.D.~Moore, \emph{{The dark-matter axion mass}},
  \href{https://doi.org/10.1088/1475-7516/2017/11/049}{\emph{JCAP} {\bfseries
  11} (2017) 049} [\href{https://arxiv.org/abs/1708.07521}{{\ttfamily
  1708.07521}}].

\bibitem{Gorghetto:2018myk}
M.~Gorghetto, E.~Hardy and G.~Villadoro, \emph{{Axions from Strings: the
  Attractive Solution}},
  \href{https://doi.org/10.1007/JHEP07(2018)151}{\emph{JHEP} {\bfseries 07}
  (2018) 151} [\href{https://arxiv.org/abs/1806.04677}{{\ttfamily
  1806.04677}}].

\bibitem{Vaquero:2018tib}
A.~Vaquero, J.~Redondo and J.~Stadler, \emph{{Early seeds of axion
  miniclusters}},
  \href{https://doi.org/10.1088/1475-7516/2019/04/012}{\emph{JCAP} {\bfseries
  04} (2019) 012} [\href{https://arxiv.org/abs/1809.09241}{{\ttfamily
  1809.09241}}].

\bibitem{Buschmann:2019icd}
M.~Buschmann, J.W.~Foster and B.R.~Safdi, \emph{{Early-Universe Simulations of
  the Cosmological Axion}},
  \href{https://doi.org/10.1103/PhysRevLett.124.161103}{\emph{Phys. Rev. Lett.}
  {\bfseries 124} (2020) 161103}
  [\href{https://arxiv.org/abs/1906.00967}{{\ttfamily 1906.00967}}].

\bibitem{Hindmarsh:2019csc}
M.~Hindmarsh, J.~Lizarraga, A.~Lopez-Eiguren and J.~Urrestilla, \emph{{Scaling
  Density of Axion Strings}},
  \href{https://doi.org/10.1103/PhysRevLett.124.021301}{\emph{Phys. Rev. Lett.}
  {\bfseries 124} (2020) 021301}
  [\href{https://arxiv.org/abs/1908.03522}{{\ttfamily 1908.03522}}].

\bibitem{Gorghetto:2020qws}
M.~Gorghetto, E.~Hardy and G.~Villadoro, \emph{{More axions from strings}},
  \href{https://doi.org/10.21468/SciPostPhys.10.2.050}{\emph{SciPost Phys.}
  {\bfseries 10} (2021) 050}
  [\href{https://arxiv.org/abs/2007.04990}{{\ttfamily 2007.04990}}].

\bibitem{Hindmarsh:2021vih}
M.~Hindmarsh, J.~Lizarraga, A.~Lopez-Eiguren and J.~Urrestilla, \emph{{Approach
  to scaling in axion string networks}},
  \href{https://doi.org/10.1103/PhysRevD.103.103534}{\emph{Phys. Rev. D}
  {\bfseries 103} (2021) 103534}
  [\href{https://arxiv.org/abs/2102.07723}{{\ttfamily 2102.07723}}].

\bibitem{Buschmann:2021sdq}
M.~Buschmann, J.W.~Foster, A.~Hook, A.~Peterson, D.E.~Willcox, W.~Zhang et~al.,
  \emph{{Dark Matter from Axion Strings with Adaptive Mesh Refinement}},
  \href{https://arxiv.org/abs/2108.05368}{{\ttfamily 2108.05368}}.

\bibitem{Sikivie:1982qv}
P.~Sikivie, \emph{{Of Axions, Domain Walls and the Early Universe}},
  \href{https://doi.org/10.1103/PhysRevLett.48.1156}{\emph{Phys. Rev. Lett.}
  {\bfseries 48} (1982) 1156}.

\bibitem{Chang:1998tb}
S.~Chang, C.~Hagmann and P.~Sikivie, \emph{{Studies of the motion and decay of
  axion walls bounded by strings}},
  \href{https://doi.org/10.1103/PhysRevD.59.023505}{\emph{Phys. Rev.}
  {\bfseries D59} (1999) 023505}
  [\href{https://arxiv.org/abs/hep-ph/9807374}{{\ttfamily hep-ph/9807374}}].

\bibitem{Hiramatsu:2010yn}
T.~Hiramatsu, M.~Kawasaki and K.~Saikawa, \emph{{Evolution of String-Wall
  Networks and Axionic Domain Wall Problem}},
  \href{https://doi.org/10.1088/1475-7516/2011/08/030}{\emph{JCAP} {\bfseries
  1108} (2011) 030} [\href{https://arxiv.org/abs/1012.4558}{{\ttfamily
  1012.4558}}].

\bibitem{Hiramatsu:2012sc}
T.~Hiramatsu, M.~Kawasaki, K.~Saikawa and T.~Sekiguchi, \emph{{Axion cosmology
  with long-lived domain walls}},
  \href{https://doi.org/10.1088/1475-7516/2013/01/001}{\emph{JCAP} {\bfseries
  1301} (2013) 001} [\href{https://arxiv.org/abs/1207.3166}{{\ttfamily
  1207.3166}}].

\bibitem{Ringwald:2015dsf}
A.~Ringwald and K.~Saikawa, \emph{{Axion dark matter in the post-inflationary
  Peccei-Quinn symmetry breaking scenario}},
  \href{https://doi.org/10.1103/PhysRevD.93.085031,
  10.1103/PhysRevD.94.049908}{\emph{Phys. Rev.} {\bfseries D93} (2016) 085031}
  [\href{https://arxiv.org/abs/1512.06436}{{\ttfamily 1512.06436}}].

\bibitem{Harigaya:2018ooc}
K.~Harigaya and M.~Kawasaki, \emph{{QCD axion dark matter from long-lived
  domain walls during matter domination}},
  \href{https://doi.org/10.1016/j.physletb.2018.04.056}{\emph{Phys. Lett.}
  {\bfseries B782} (2018) 1}
  [\href{https://arxiv.org/abs/1802.00579}{{\ttfamily 1802.00579}}].

\bibitem{Caputo:2019wsd}
A.~Caputo and M.~Reig, \emph{{Cosmic implications of a low-scale solution to
  the axion domain wall problem}},
  \href{https://doi.org/10.1103/PhysRevD.100.063530}{\emph{Phys. Rev.}
  {\bfseries D100} (2019) 063530}
  [\href{https://arxiv.org/abs/1905.13116}{{\ttfamily 1905.13116}}].

\bibitem{Nilles:1981py}
H.P.~Nilles and S.~Raby, \emph{{Supersymmetry and the strong CP problem}},
  \href{https://doi.org/10.1016/0550-3213(82)90547-8}{\emph{Nucl. Phys. B}
  {\bfseries 198} (1982) 102}.

\bibitem{Hall:1995eq}
L.J.~Hall and S.~Raby, \emph{{A Complete supersymmetric SO(10) model}},
  \href{https://doi.org/10.1103/PhysRevD.51.6524}{\emph{Phys. Rev. D}
  {\bfseries 51} (1995) 6524}
  [\href{https://arxiv.org/abs/hep-ph/9501298}{{\ttfamily hep-ph/9501298}}].

\bibitem{Co:2018mho}
R.T.~Co, E.~Gonzalez and K.~Harigaya, \emph{{Axion Misalignment Driven to the
  Hilltop}}, \href{https://doi.org/10.1007/JHEP05(2019)163}{\emph{JHEP}
  {\bfseries 05} (2019) 163}
  [\href{https://arxiv.org/abs/1812.11192}{{\ttfamily 1812.11192}}].

\bibitem{Takahashi:2019pqf}
F.~Takahashi and W.~Yin, \emph{{QCD axion on hilltop by a phase shift of
  $\pi$}}, \href{https://doi.org/10.1007/JHEP10(2019)120}{\emph{JHEP}
  {\bfseries 10} (2019) 120}
  [\href{https://arxiv.org/abs/1908.06071}{{\ttfamily 1908.06071}}].

\bibitem{Huang:2020etx}
J.~Huang, A.~Madden, D.~Racco and M.~Reig, \emph{{Maximal axion misalignment
  from a minimal model}},
  \href{https://doi.org/10.1007/JHEP10(2020)143}{\emph{JHEP} {\bfseries 10}
  (2020) 143} [\href{https://arxiv.org/abs/2006.07379}{{\ttfamily
  2006.07379}}].

\bibitem{Dvali:1995ce}
G.R.~Dvali, \emph{{Removing the cosmological bound on the axion scale}},
  \href{https://arxiv.org/abs/hep-ph/9505253}{{\ttfamily hep-ph/9505253}}.

\bibitem{Banks:1996ea}
T.~Banks and M.~Dine, \emph{{The Cosmology of string theoretic axions}},
  \href{https://doi.org/10.1016/S0550-3213(97)00413-6}{\emph{Nucl. Phys.}
  {\bfseries B505} (1997) 445}
  [\href{https://arxiv.org/abs/hep-th/9608197}{{\ttfamily hep-th/9608197}}].

\bibitem{Choi:1996fs}
K.~Choi, H.B.~Kim and J.E.~Kim, \emph{{Axion cosmology with a stronger QCD in
  the early universe}},
  \href{https://doi.org/10.1016/S0550-3213(97)00066-7}{\emph{Nucl. Phys. B}
  {\bfseries 490} (1997) 349}
  [\href{https://arxiv.org/abs/hep-ph/9606372}{{\ttfamily hep-ph/9606372}}].

\bibitem{Co:2018phi}
R.T.~Co, E.~Gonzalez and K.~Harigaya, \emph{{Axion Misalignment Driven to the
  Bottom}}, \href{https://doi.org/10.1007/JHEP05(2019)162}{\emph{JHEP}
  {\bfseries 05} (2019) 162}
  [\href{https://arxiv.org/abs/1812.11186}{{\ttfamily 1812.11186}}].

\bibitem{Dimopoulos:1988pw}
S.~Dimopoulos and L.J.~Hall, \emph{{Inflation and Invisible Axions}},
  \href{https://doi.org/10.1103/PhysRevLett.60.1899}{\emph{Phys. Rev. Lett.}
  {\bfseries 60} (1988) 1899}.

\bibitem{Graham:2018jyp}
P.W.~Graham and A.~Scherlis, \emph{{Stochastic axion scenario}},
  \href{https://doi.org/10.1103/PhysRevD.98.035017}{\emph{Phys. Rev. D}
  {\bfseries 98} (2018) 035017}
  [\href{https://arxiv.org/abs/1805.07362}{{\ttfamily 1805.07362}}].

\bibitem{Takahashi:2018tdu}
F.~Takahashi, W.~Yin and A.H.~Guth, \emph{{QCD axion window and low-scale
  inflation}}, \href{https://doi.org/10.1103/PhysRevD.98.015042}{\emph{Phys.
  Rev. D} {\bfseries 98} (2018) 015042}
  [\href{https://arxiv.org/abs/1805.08763}{{\ttfamily 1805.08763}}].

\bibitem{Kitajima:2019ibn}
N.~Kitajima, Y.~Tada and F.~Takahashi, \emph{{Stochastic inflation with an
  extremely large number of $e$-folds}},
  \href{https://doi.org/10.1016/j.physletb.2019.135097}{\emph{Phys. Lett. B}
  {\bfseries 800} (2020) 135097}
  [\href{https://arxiv.org/abs/1908.08694}{{\ttfamily 1908.08694}}].

\bibitem{Dienes:2022zbh}
K.R.~Dienes and B.~Thomas, \emph{{More is Different: Non-Minimal Dark Sectors
  and their Implications for Particle Physics, Astrophysics, and Cosmology --
  13 Take-Away Lessons for Snowmass 2021}},  in \emph{{2022 Snowmass Summer
  Study}}, 3, 2022 [\href{https://arxiv.org/abs/2203.17258}{{\ttfamily
  2203.17258}}].

\bibitem{planck}
{\scshape Planck} collaboration, \emph{{Planck 2018 results. VI. Cosmological
  parameters}},  \href{https://arxiv.org/abs/1807.06209}{{\ttfamily
  1807.06209}}.

\bibitem{sakharov}
A.D.~Sakharov, \emph{{Violation of CP Invariance, c Asymmetry, and Baryon
  Asymmetry of the Universe}},
  \href{https://doi.org/10.1070/PU1991v034n05ABEH002497}{\emph{Pisma Zh. Eksp.
  Teor. Fiz.} {\bfseries 5} (1967) 32}.

\bibitem{Arnold:1987mh}
P.B.~Arnold and L.D.~McLerran, \emph{{Sphalerons, Small Fluctuations and Baryon
  Number Violation in Electroweak Theory}},
  \href{https://doi.org/10.1103/PhysRevD.36.581}{\emph{Phys. Rev. D} {\bfseries
  36} (1987) 581}.

\bibitem{Arnold:1996dy}
P.B.~Arnold, D.~Son and L.G.~Yaffe, \emph{{The Hot baryon violation rate is O
  (alpha-w**5 T**4)}},
  \href{https://doi.org/10.1103/PhysRevD.55.6264}{\emph{Phys. Rev. D}
  {\bfseries 55} (1997) 6264}
  [\href{https://arxiv.org/abs/hep-ph/9609481}{{\ttfamily hep-ph/9609481}}].

\bibitem{Abe:2019vii}
{\scshape T2K} collaboration, \emph{{Constraint on the matter--antimatter
  symmetry-violating phase in neutrino oscillations}},
  \href{https://doi.org/10.1038/s41586-020-2177-0}{\emph{Nature} {\bfseries
  580} (2020) 339} [\href{https://arxiv.org/abs/1910.03887}{{\ttfamily
  1910.03887}}].

\bibitem{Fukugita:1986hr}
M.~Fukugita and T.~Yanagida, \emph{{Baryogenesis Without Grand Unification}},
  \href{https://doi.org/10.1016/0370-2693(86)91126-3}{\emph{Phys. Lett.}
  {\bfseries B174} (1986) 45}.

\bibitem{Cline:2006ts}
J.M.~Cline, \emph{{Baryogenesis}},  in \emph{{Les Houches Summer School -
  Session 86: Particle Physics and Cosmology: The Fabric of Spacetime}}, 9,
  2006 [\href{https://arxiv.org/abs/hep-ph/0609145}{{\ttfamily
  hep-ph/0609145}}].

\bibitem{Morrissey:2012db}
D.E.~Morrissey and M.J.~Ramsey-Musolf, \emph{{Electroweak baryogenesis}},
  \href{https://doi.org/10.1088/1367-2630/14/12/125003}{\emph{New J. Phys.}
  {\bfseries 14} (2012) 125003}
  [\href{https://arxiv.org/abs/1206.2942}{{\ttfamily 1206.2942}}].

\bibitem{Konstandin:2013caa}
T.~Konstandin, \emph{{Quantum Transport and Electroweak Baryogenesis}},
  \href{https://doi.org/10.3367/UFNe.0183.201308a.0785}{\emph{Phys. Usp.}
  {\bfseries 56} (2013) 747} [\href{https://arxiv.org/abs/1302.6713}{{\ttfamily
  1302.6713}}].

\bibitem{Garbrecht:2018mrp}
B.~Garbrecht, \emph{{Why is there more matter than antimatter? Calculational
  methods for leptogenesis and electroweak baryogenesis}},
  \href{https://doi.org/10.1016/j.ppnp.2019.103727}{\emph{Prog. Part. Nucl.
  Phys.} {\bfseries 110} (2020) 103727}
  [\href{https://arxiv.org/abs/1812.02651}{{\ttfamily 1812.02651}}].

\bibitem{Bodeker:2020ghk}
D.~Bodeker and W.~Buchmuller, \emph{{Baryogenesis from the weak scale to the
  grand unification scale}},
  \href{https://doi.org/10.1103/RevModPhys.93.035004}{\emph{Rev. Mod. Phys.}
  {\bfseries 93} (2021) 035004}
  [\href{https://arxiv.org/abs/2009.07294}{{\ttfamily 2009.07294}}].

\bibitem{Froggatt:1978nt}
C.D.~Froggatt and H.B.~Nielsen, \emph{{Hierarchy of Quark Masses, Cabibbo
  Angles and CP Violation}},
  \href{https://doi.org/10.1016/0550-3213(79)90316-X}{\emph{Nucl. Phys. B}
  {\bfseries 147} (1979) 277}.

\bibitem{Randall:1999vf}
L.~Randall and R.~Sundrum, \emph{{An Alternative to compactification}},
  \href{https://doi.org/10.1103/PhysRevLett.83.4690}{\emph{Phys. Rev. Lett.}
  {\bfseries 83} (1999) 4690}
  [\href{https://arxiv.org/abs/hep-th/9906064}{{\ttfamily hep-th/9906064}}].

\bibitem{Randall:1999ee}
L.~Randall and R.~Sundrum, \emph{{A Large mass hierarchy from a small extra
  dimension}}, \href{https://doi.org/10.1103/PhysRevLett.83.3370}{\emph{Phys.
  Rev. Lett.} {\bfseries 83} (1999) 3370}
  [\href{https://arxiv.org/abs/hep-ph/9905221}{{\ttfamily hep-ph/9905221}}].

\bibitem{Contino:2006qr}
R.~Contino, L.~Da~Rold and A.~Pomarol, \emph{{Light custodians in natural
  composite Higgs models}},
  \href{https://doi.org/10.1103/PhysRevD.75.055014}{\emph{Phys. Rev. D}
  {\bfseries 75} (2007) 055014}
  [\href{https://arxiv.org/abs/hep-ph/0612048}{{\ttfamily hep-ph/0612048}}].

\bibitem{Weinberg:1972ws}
S.~Weinberg, \emph{{Electromagnetic and weak masses}},
  \href{https://doi.org/10.1103/PhysRevLett.29.388}{\emph{Phys. Rev. Lett.}
  {\bfseries 29} (1972) 388}.

\end{thebibliography}\endgroup

\end{document}